\documentclass[a4paper,11pt]{article}
\pdfoutput=1 
\usepackage{jheppub}
\usepackage{color} 
\usepackage{float}
\usepackage{ulem}

\usepackage{amsmath,graphicx,amssymb,subfigure,bbm}
\usepackage[all]{xy}

\begin{document}

\title{\boldmath Nonlinear realisation of chiral symmetry breaking in  holographic soft wall models}

\author[a]{Alfonso Ballon-Bayona}
\author[b]{and Luis A. H. Mamani}
\affiliation[a]{Instituto de F\'{i}sica, Universidade
Federal do Rio de Janeiro, \\
Caixa Postal 68528, RJ 21941-972, Brazil.}       
\affiliation[b]{Centro de Ci\^encias Naturais e Humanas,
Universidade Federal do ABC,\\
Rua Santa Ad\'elia 166, 09210-170, Santo Andr\'e, SP, Brazil}
\emailAdd{aballonb@if.ufrj.br}
\emailAdd{luis.mamani@ufabc.edu.br}

\abstract{
We investigate nonlinear extensions of the holographic soft wall model proposed by Karch, Katz, Son and Stephanov \cite{Karch:2006pv} with a positive quadratic dilaton. We consider a Higgs potential for the tachyonic field that brings a more natural realisation of chiral symmetry breaking in the infrared regime. Utilising the AdS/CFT dictionary and holographic renormalisation we find the chiral condensate as a function of the quark mass. The nonlinearity of the Higgs potential leads to a nonlinear relation between the chiral condensate and the quark mass. Solving the effective Schrödinger equations for the field perturbations we estimate meson masses and decay constants and evaluate their dependence on the quark mass. 
In the axial and pseudoscalar sector we find an interesting behaviour for the decay constants as the quark mass increases. We also investigate the effect of a 5d running mass for the tachyonic field. We conclude that nonlinear soft wall models with a  Higgs potential for the tachyon and a positive quadratic dilaton do not provide spontaneous symmetry breaking in the chiral limit.}

\maketitle
\flushbottom

\section{Introduction}

The dynamical breaking of chiral symmetry in Quantum Chromodynamics (QCD) is a fascinating phenomenon that occurs in the strongly coupled (non-perturbative) regime. Chiral symmetry breaking is the mechanism responsible for dynamical generation of quark masses at low energies and it is crucial for the description of mesons and baryons. The well established order parameter associated with chiral symmetry breaking is the so-called chiral condensate $\langle \bar q q \rangle$, which is the VEV of the quark mass operator. A nonzero chiral condensate signalises chiral symmetry breaking and it can be generated by a nontrivial QCD vacuum. The phenomenon of chiral symmetry breaking is closely related to the phenomenon of quark confinement \cite{Coleman:1980mx}.

The traditional approaches used to investigating the non-perturbative regime of QCD are lattice QCD and the Schwinger-Dyson equations. They have provided plenty of insights on the mechanism for dynamical chiral symmetry breaking as well as quantitative predictions for hadronic phenomenology.  More recently, a complementary approach to non-perturbative QCD has been developed. The so-called holographic QCD approach builds 5d gravitational models dual to 4d non-perturbative quantum field theories similar to QCD. This approach is based on the AdS/CFT correspondence and the gauge/gravity duality. 

One of the main achievements of the holographic QCD approach is the discovery of a universal criterion for confinement \cite{Kinar:1998vq}, based on the holographic map between 5d classical strings and 4d Wilson loops \cite{Maldacena:1998im}. The problem of chiral symmetry breaking is also fascinating in holographic QCD. Since the chiral symmetry group $U(N_f)_L \times U(N_f)_R$ is a 4d global symmetry group, it maps to 5d (local) gauge symmetry group. This implies that 4d chiral symmetry breaking corresponds to  5d (non-Abelian) gauge symmetry breaking. A minimal 5d holographic QCD model for chiral symmetry breaking was proposed in \cite{Erlich:2005qh,DaRold:2005mxj}, based on the so-called hard wall model \cite{Polchinski:2001tt}. In that approach the breaking of the 5d Non-Abelian gauge symmetry is induced by a 5d scalar field $X$, dual to the quark mass operator $\bar q q$. The AdS/CFT dictionary maps the 5d squared mass $m_X^2$ to the conformal dimension of  $\bar q q$ implying a tachyonic mass $m_X^2=-3$.

Although the model in \cite{Erlich:2005qh,DaRold:2005mxj} successfully introduces the quark mass and chiral condensate, it does not provide a dynamical mechanism for chiral symmetry in the sense that the chiral condensate is fixed by boundary conditions in the infrared regime. An alternative approach was proposed in  \cite{Karch:2006pv}, which is known as the soft wall model. That approach was physically motivated by experimental data on the meson spectrum indicating a set of linear Regge trajectories. In holographic QCD  meson masses arise as eigenvalues of effective Schrödinger potentials for 5d field perturbations. The soft wall model introduces a smooth cut off in the form a background scalar field $\Phi(z)$, known as the dilaton. As shown in \cite{Karch:2006pv}, a positive quadratic dilaton $\Phi(z) \sim z^2$ in the IR leads to approximate linear Regge trajectories for the mesons. In the soft wall model a  nontrivial dynamics for the tachyonic field $X$ is driven by the presence of the dilaton field $\Phi(z)$ and instead of imposing a boundary condition at an arbitrary energy scale the chiral condensate is obtained from a regularity condition.

Despite providing a more realistic meson spectrum, the original soft wall model leads to a linear realisation of chiral symmetry breaking driven by an IR regularity condition.  Indeed, the chiral condensate  $\langle \bar q q \rangle$ is just proportional to the quark mass $m_q$ as a consequence of the linearity of the tachyonic field equation. This implies, in particular, that taking the chiral limit $m_q \to 0$ in the soft wall model does not lead to spontaneous chiral symmetry breaking (the chiral condensate vanishes). This differs significantly from QCD where spontaneous symmetry breaking always occurs in the chiral limit. 

In this work we investigate nonlinear extensions of the holographic soft wall model. We consider a 5d Higgs potential for the tachyonic field $X$ leading to a nonlinear differential equation for the tachyonic field. The nonlinearity allows us to find a family of solutions in the IR depending on only one parameter, that we call $C_0$. For fixed $C_0$ we will be able to solve numerically the tachyon differential equation. Extracting the source and VEV coefficients for the tachyon solution near the boundary, we utilise the AdS/CFT correspondence and map those parameters to the quark mass and chiral condensate respectively. We will show that the nonlinearity of the tachyon differential equation implies a nonlinear relation between the chiral condensate  $\langle \bar q q \rangle$ and the quark mass $m_q$. We will show, in particular, that for the original Higgs potential the chiral condensate always grows with the quark mass, as expected in QCD , see i.e. \cite{McNeile:2012xh}.  We will find, however, that near the chiral limit $m_q \to 0$ the nonlinear soft wall models based on a Higgs potential behave in the same way as the original soft wall model and therefore the chiral condensate vanishes. A similar result was recently found in \cite{Chelabi:2015gpc}. 

In this work we perform a systematic analysis of the background and field perturbations for nonlinear soft wall models based on a positive quadratic dilaton $\Phi$ and a Higgs potential for the tachyonic field $X$. For completeness we consider the cases of negative and positive couplings for the Higgs potential $U(|X|) = m_X^2 |X|^2 + \lambda |X|^4$. A negative coupling $\lambda<0$ never provides a  minimum whereas the positive case $\lambda>0$ corresponds to the original Mexican hat potential leading to a minimum. 
Our results for the meson spectrum and decay constants support our main conclusions and also bring some pleasant surprises. In particular, we find an interesting behaviour for the meson decay constants in the axial sector as functions of the quark mass. Our approach was a bit inspired on the holographic QCD model of \cite{Iatrakis:2010jb} where chiral symmetry breaking maps to tachyon condensation in string theory. Our results agree qualitatively with \cite{Iatrakis:2010jb} in the regime of large $m_q$ but differ in the chiral limit. The holographic model of \cite{Gherghetta:2009ac} was also a useful guide in our approach. 

The paper is organised as follows. In section \ref{Sec:HWSW} we briefly review chiral symmetry breaking in the hard wall and soft wall models.  In section \ref{Sec:NLSW} we present the nonlinear extensions of the soft wall model based on a Higgs potential. We perform a systematic analysis of the background including holographic renormalisation for the chiral condensate. Solving the effective Schrödinger equations for the 5d field perturbations we obtain the meson masses and decay constants.  In section \ref{Sec:NLSWRM} we introduce a running mass  for the 5d tachyonic field and investigate its effect on the chiral condensate, the meson spectrum and decay constants. We present our conclusions in section \ref{Sec:Conclusions}. Appendix \ref{Sec:4daction} describes the equations for the 5d field perturbations and the dictionary for the meson decay constants. Appendix \ref{Sec:GKK} briefly reviews the holographic model of \cite{Gherghetta:2009ac} whilst appendix \ref{Sec:Num} describes  some complementary numerical results.  

\section{Hard and soft wall models for chiral symmetry breaking}
\label{Sec:HWSW}

Holographic models for QCD at zero temperature satisfy Poincaré invariance. The 5d metric for those models is usually written as
\begin{equation}
ds^2= e^{2A_s(z)} [ - dt^2 + d \vec{x}^2 + dz^2 ]    \, , \label{Eq:hQCDmetric}
\end{equation}
where $A_s(z)$ is the (string-frame) warp-factor given as a function of the conformal coordinate $z$. In this work we are interested in the simplest backgrounds where the 5d metric is just Lorentzian AdS, i.e. $A_s(z)= - \ln z$ \footnote{As it is usual in holographic QCD, we work in units where the AdS radius is set to $1$.}. Simplifying the background as much as possible allows us to focus on the dynamics of chiral symmetry breaking. 

\subsection{The hard wall model}

In the hard wall model \cite{Polchinski:2001tt} the 5d background is given by 
\begin{equation}
ds^2 = \frac{1}{z^2} [ - dt^2 + d \vec{x}^2 + dz^2 ] \quad , \quad 
0<z\leq z_0 \, ,
\end{equation}
which corresponds to a 5d Lorentzian AdS space ending in an infrared (IR) hard wall at $z=z_0$. A nice property of the hard wall model is that it satisfies the confinement criterion, found in \cite{Kinar:1998vq}. Soon it was realised that the hard wall model allows us to estimate the spectrum of glueballs \cite{BoschiFilho:2002vd}. 

The flavour sector of the hard wall model was introduced in \cite{Erlich:2005qh,DaRold:2005mxj}. The key ingredients were the 4d quark mass operator $\bar q_R q_L$ as well as the left and right  4d currents $J^{\mu,a}_{(L/R)}= \bar q_{L/R} \gamma^{\mu} T^a q_{L/R}$ associated with the chiral symmetry group $SU(N_f)_L \times SU(N_f)_R$. According to the AdS/CFT dictionary, the 4d quark mass operator maps to a 5d scalar $X$ field with mass given by the relation $m_X^2= \Delta (\Delta -4)$ with $\Delta$ the conformal dimension of the quark mass operator. In the extreme ultraviolet (UV) we have $\Delta=3$ and this corresponds to $m_X^2= -3$. This 5d mass would indicate an instability in Minkowski space and therefore the 5d field $X$ is usually called the tachyon. The left and right 4d currents, on the other hand, are conserved in the extreme UV so they are mapped to 5d gauge fields $A^{\mu,a}_{L/R}$. 

In QCD chiral symmetry is broken dynamically in the IR due to a nontrivial vacuum and the dynamics of the quark mass operator. An important quantity is the so-called chiral condensate $\sigma$, which is the VEV of the quark mass operator, i.e. $\sigma = \langle \bar q q \rangle$. The authors in \cite{Erlich:2005qh,DaRold:2005mxj} realised that chiral symmetry breaking in 4d corresponds to 5d gauge symmetry breaking of the fields  $A^{\mu,a}_{L/R}$ driven by a nontrivial scalar field $X$. 
In this work we use the conventions of \cite{Erlich:2005qh} and write the 5d action as
\begin{equation}
S = - \int d^5 x \sqrt{-g} {\rm Tr} \Big [| D_m X |^2 + m_X^2 |X|^2 + \frac{1}{g_5^2} {F_{mn}^{(L)}}^2 + \frac{1}{4g_5^2} {F_{mn}^{(R)}}^2 \Big ] \, , \label{ErlichAction}
\end{equation}
 where
 \begin{align}
F_{mn}^{(L/R)} &= \partial_m A_n^{(L/R)} - \partial_n A_m^{(L/R)} - i [A_m^{(L/R)},A_n^{(L/R)}] \, , \nonumber \\
D_m X &= \partial_m X - i A_m^{(L)} X + i X A_m^{(R)} \, . 
\end{align}
Note that under the non-Abelian gauge group $SU(N_f)_L \times SU(N_f)_R$ the left and right vector fields $A_m^{(L/R)}$ transform as adjoint fields whereas the scalar field $X$ behaves as a bifundamental. The 5d gauge coupling is fixed as $g_5^2=(12 \pi^2)/N_c$ in order to reproduce the expected large $Q^2$ behaviour  for the current correlator in QCD \cite{Erlich:2005qh,DaRold:2005mxj}.

From now on we take $N_f=2$ and make the assumption of isospin (flavour) symmetry, i.e. $m_u=m_d$ and $\sigma_u=\sigma_d$, which is a good approximation for the light quark sector in QCD. We therefore take the following ansatz for the 5d tachyonic field:
\begin{equation}
X(z)=\frac12 v(z)  I_{2 \times 2} \, . \label{TachyonBckgd}      
\end{equation}
In QCD at zero temperature (and zero density) we do not expect any vectorial condensate so we take the following ansatz for the 5d gauge fields
\begin{equation}
A_m^{L/R}=0 \, .     \label{GaugeBckgd}
\end{equation}
The 5d action in \eqref{ErlichAction} reduces to a 1d action for the field v(z) and the Euler-Lagrange equation takes the form 
\begin{equation}
\Big [ (z \partial_z)^2 - 4 z \partial_z - m_X^2 \Big ] v = 0 \,, 
\label{veqHW}
\end{equation}
with exact solution 
\begin{equation}
v(z) = c_1 z^{\Delta_{-}} + c_3 z^{\Delta_{+}} \quad , \quad 
\Delta_{\pm}= 2 \pm \sqrt{4 + m_X^2} \, . 
\end{equation}
This has the expected scaling behaviour for a 5d scalar field dual to an operator of dimension $\Delta_+$ and a coupling of conformal dimension $\Delta_{-}=4-\Delta_{+}$. In order to match to the quark mass operator in the extreme UV we choose $\Delta_{+}=3$. The source coefficient $c_1$ is related to the quark mass $m_q$ by $c_1 = m_q \zeta$ where $\zeta$ is a normalisation constant. This constant is usually fixed as $\zeta = \sqrt{N_c}/(2 \pi)$ to be consistent with counting rules of large-$N_c$ QCD \cite{Cherman:2008eh}.  The VEV coefficient $c_3$ will be related to the quark condensate $\sigma_u$. The precise relation is scheme-dependent and will be obtained in the next section. 

In the hard wall model the VEV coefficient $c_3$ is fixed by boundary conditions at the IR wall $z=z_0$. This differs from what we expect in QCD where the chiral condensate is generated dynamically due to a nontrivial vacuum. The hard wall model, however, has the nice feature of being the simplest model that describes at the same time confinement and chiral symmetry breaking. 

\subsection{The (linear) soft wall model}

A shortcoming of the hard wall model is that it leads to hadronic masses that grow too fast as we increase the radial number. For example, the  resonances of the $\rho$ meson have squared masses that grow as $m_{\rho^{(n)}}^2 \sim n^2$ for large $n$. Experimental data, on the other hand, indicates an approximate linear dependence for the squared masses, i.e. $m_{\rho^{(n)}}^2 \sim n$. 

The quadratic dependence in the hard wall model can be thought as a consequence of having a metric that abruptly ends at a cutoff $z=z_0$. Motivated by this problem, the authors of \cite{Karch:2006pv} proposed the idea of a smooth cutoff driven by a background scalar field $\Phi(z)$  so that the geometry does not abruptly end.  This was inspired by string theory where the field $\Phi$ is known as the dilaton. Later works have explored the backreaction of this background field on the AdS metric and how it maps to the Yang-Mills operator ${\rm Tr} F^2$ \cite{Csaki:2006ji,Gursoy:2007er,Ballon-Bayona:2017sxa} \footnote{The AdS deformation by a scalar field may also be interpreted in terms of supersymmetry breaking, e.g. \cite{Ghoroku:2005vt}. }.

In the soft wall model the background is given by 
\begin{equation}
ds^2 = \frac{1}{z^2} [ - dt^2 + d \vec{x}^2 + dz^2 ] \quad , \quad 
\Phi(z) = \phi_{\infty} z^2 \, ,
\end{equation}
where $\phi_{\infty}$ is a constant providing an IR mass scale. The flavour sector of the soft wall model is described by the action 
\begin{equation}
S = - \int d^5 x \sqrt{-g} \, e^{-\Phi(z)} {\rm Tr} \Big [| D_m X |^2 + m_X^2 |X|^2 + \frac{1}{g_5^2} {F_{mn}^{(L)}}^2 + \frac{1}{4 g_5^2} {F_{mn}^{(R)}}^2 \Big ] \, , \label{KarchAction}
\end{equation}
The positive quadratic z dependence of $\Phi$ leads, at large $z$,  to a harmonic oscillator form for the effective Schrödinger potentials  for the field perturbations. This guarantees a linear dependence on the radial number for the meson squared masses, i.e. $m_{(n)}^2 \sim n$.

In the flavour sector we take again the ansatz \eqref{TachyonBckgd}-\eqref{GaugeBckgd} for the background fields and this time we obtain the equation
\begin{equation}
\Big [ z^2 \partial_z^2 - (3 + 2 \phi_{\infty} z^2)  z \partial_z  - m_X^2 \Big ] v = 0 \,. 
\label{veqSW}
\end{equation}
In order to find an exact solution we rewrite \eqref{veqSW} in terms of a new variable $x \equiv \phi_{\infty} z^2$ and redefine the field $v(x)$ as $x^{\beta} \tilde v (x)$. We arrive at the 2nd order differential equation
\begin{equation}
\Big \{ x^2 \partial_x^2  + [ 2 \beta - 1 - x ] x \partial_x 
+ \beta^2 - 2 \beta - \frac{m_X^2}{4} - \beta x \Big \} \tilde v = 0 \,. 
\end{equation}
This equation becomes Kummer's equation if 
\begin{equation}
\beta^2 - 2 \beta - \frac{m_X^2}{4} = 0 \quad \to \quad 
\beta = 1 \pm \sqrt{1 + \frac{m_X^2}{4}} = \frac{ \Delta_{\pm}}{2} \,.
\end{equation}
The solutions for $\tilde v(z)$ are of the form $M(\beta , 2 \beta -1 ; x)$ (Kummer) and $U(\beta , 2 \beta -1 ; x)$ (Tricomi). From this analysis we conclude that the exact solution for $v(z)$ can be written as 
\begin{equation}
v(z) =  \tilde c_1 z^{\Delta_{-}} U \left ( \frac{\Delta_{-}}{2} ,  \Delta_{-} - 1 ; \phi_{\infty} z^2  \right ) + \tilde c_3 z^{\Delta_{+}} M( \frac{\Delta_{+}}{2} ,  \Delta_{+} - 1 ; \phi_{\infty} z^2 ) \, , 
\label{vsolSW}    
\end{equation}
where $\tilde c_1$ and $\tilde c_3$ are constant coefficients. 
For $m_X^2=-3$ we have $\Delta_{-}=1$ and $\Delta_{+}=3$. This exact solution was first obtained in \cite{Colangelo:2008us}. In the limit $x \to \infty$  we have $M(a, b;x) \sim e^x x^{a-b}$ and $U(a,b;x) \sim x^{-a}$; therefore we need to set $\tilde c_3=0$ in order to avoid a divergent solution in the IR. The tachyon solution in \eqref{vsolSW} becomes
\begin{equation}
v(z) = \frac{\sqrt{\pi}}{2}  c_1 z  \, U \left ( \frac12 ,  0 ; \phi_{\infty} z^2  \right )  \, , 
\label{vsolSWv2}    
\end{equation}
where $\tilde c_1$ was rewritten as $(\sqrt{\pi}/2) c_1$.  In the UV (small $z$), the solution in \eqref{vsolSWv2} behaves as \begin{equation}
v^{UV}(z) = c_1 z + d_3(c_1) z^3 \ln z + c_3(c_1) z^3 + \dots     
\end{equation}
The logarithmic term could have been anticipated from an asymptotic analysis using the Frobenius method. Both coefficients $d_3$ and $c_3$ are actually proportional to $c_1$ and then vanish in the chiral limit $c_1 \to 0$. 
In the IR the tachyon solution \eqref{vsolSWv2} behaves as 
\begin{equation}
v^{IR}(z)  
= C_0\Big [ 1 + {\cal O} (z^{-2}) \Big ] \quad , \quad 
C_0 = \frac{\sqrt{\pi}}{2} \frac{c_1}{\sqrt{\phi_{\infty}}} \,. 
\end{equation}
The parameter $C_0$ characterises the tachyon solution at large $z$. As expected, the solution is regular in the IR. This regular solution was, however, chosen by setting $\tilde c_3=0$. In the next section we will introduce a nonlinear potential for the tachyonic field. The nonlinearity of the new differential equation will naturally lead to regular solutions in the IR without the need of fixing any integration constant. 

The soft wall model represents an important step towards the construction of a minimal model for describing the flavour sector in holographic QCD. The positive quadratic behaviour of the dilaton $\Phi(z)$ leads naturally to a linear spectrum for the mesons. Moreover, when taking backreaction into account, the same positive quadratic dilaton leads to a linear spectrum for glueballs \cite{Gursoy:2007er,Ballon-Bayona:2017sxa}. The soft wall model has also served as a useful guide to other approaches such as light-front holography \cite{Brodsky:2014yha}.

\section{Nonlinear soft wall models}
\label{Sec:NLSW}

The authors in \cite{Karch:2006pv} realised that the linear dependence of $c_3$ on $c_1$ differed significantly from what one expects in QCD for the chiral condensate $\sigma_q$ as a function of the quark mass $m_q$. They suggested that the model could be improved by adding nonlinear terms to the tachyonic potential. The nonlinearity would lead to solutions in the IR characterised by only one parameter, $C_0$,  and the relation between the IR parameter $C_0$ and the UV parameter $c_1$ would also be nonlinear. 

In this work we investigate a nonlinear potential of the Higgs form 
\begin{equation}
U(|X|) = m_X^2 |X|^2 + \lambda |X|^4 \,.    
\label{Eq:Higgspot}
\end{equation}
This time the 5d action for the flavour sector becomes
\begin{equation}
S = - \int d^5 x \sqrt{-g} \, e^{-\Phi(z)} {\rm Tr} \Big [| D_m X |^2 + m_X^2 |X|^2 + \lambda |X|^4 + \frac{1}{g_5^2} {F_{mn}^{(L)}}^2 + \frac{1}{4 g_5^2} {F_{mn}^{(R)}}^2 \Big ] \, . \label{HiggsAction}
\end{equation}
We will consider both cases $\lambda <0$ and $\lambda>0$ and also recover the original soft wall model in the limit $\lambda \to 0$. As in the previous cases, we take $m_X^2=-3$ for the tachyonic field.

For the background fields we take again the ansatz \eqref{TachyonBckgd}-\eqref{GaugeBckgd} and this time we obtain a nonlinear equation for the tachyon:
\begin{equation}
\Big [ z^2 \partial_z^2 - (3 + 2 \phi_{\infty} z^2)  z \partial_z  + 3 \Big ] v - \frac{\lambda}{2} v^3 = 0 \,. 
\label{Eq:Tachyon}
\end{equation}
Due to the nonlinearity of the differential equation,  we can not obtain an analytic solution. In the next subsections we describe the asymptotic solutions for the tachyon in the UV and IR regimes and present our numerical results for the tachyon profiles. Then we investigate the meson spectrum for the cases 
$\lambda <0$ and $\lambda>0$.  We finish the section describing the meson decay constants. 

\subsection{Asymptotic analysis}

In the UV we consider the Frobenius ansatz
\noindent
\begin{equation}
v(z)=c_1 z + d_3 z^3 \ln z + c_3 z^3 + d_5 z^5 \ln z + c_5 z^5 + \dots
\label{Eq:UVansatz}
\end{equation}
Plugging this ansatz into the tachyon equation \eqref{Eq:Tachyon} we find 
the UV coefficients
\begin{align}
d_3 &=  \frac14 {c_1} \left (c_1^2 \lambda + 4 \phi_{\infty}  \right )
\quad , \quad 
d_5 = \frac{3}{64} c_1 \left (c_1^2 \lambda + 4 \phi_{\infty}  \right )^2 + \, , \nonumber \\
c_5 &=   \frac{1}{256}   \left (c_1^2 \lambda + 4 \phi_{\infty}  \right )  \left (-9 c_1^3 \lambda - 20 c_1 \phi_{\infty} + 48 c_3 \right )   \quad , \quad \dots 
\end{align}
 Note that $d_3$ and $d_5$ depend only on $c_1$ whereas $c_5$ depends on $c_1$ and $c_3$. In the special case $c_1^2 \lambda + 4 \phi_{\infty}=0$ all the subleading terms vanish and the linear solution $v(z) = c_1 z$ becomes exact. This is only possible for $\lambda<0$ because  in the soft wall model we have $\phi_{\infty}>0$. 

The VEV parameter $c_3$ appears to be independent of $c_1$. We will see, however, that in the IR the tachyon solution is characterised by a single parameter $C_0$. The UV parameters $c_1$ and $c_3$ then can be thought as functions of the IR parameter $C_0$. As a consequence $c_3$ will depend on $c_1$ in a nonlinear fashion, as predicted in \cite{Karch:2006pv}. We remind the reader that the UV parameters $c_1$ and $c_3$ are related to the quark mass $m_q$ and chiral condensate $\sigma_q$ respectively. As far as we are concerned, the first nonlinear realisation of chiral symmetry breaking via tachyon dynamics in terms of $c_3$ as a function of $c_1$  was developed in the holographic QCD model of \cite{Iatrakis:2010jb} \footnote{For earlier works on the dictionary between tachyon dynamics and the chiral condensate from a top-down approach see \cite{Casero:2007ae,Bergman:2007pm}.}. More sophisticated models that take into account the backreaction of the tachyon were developed in 
\cite{Jarvinen:2011qe,Arean:2012mq,Jarvinen:2015ofa,Arean:2013tja}.

In the IR it is convenient to work with the variable $y=1/z$. The differential equation \eqref{Eq:Tachyon} may be written as
\noindent
\begin{equation}
 \Big [ (y\partial_y)^2+2\left(2+\phi_{\infty}y^{-2}\right)(y\partial_y) +
3 \Big ] v - \frac{\lambda}{2}v^3 = 0 \, .
\end{equation}\label{Eq:Tachyonv2}
\noindent
We first consider the power ansatz
\begin{equation}
v^{IR}(y) = C_0 y^{\alpha} \,. \label{Eq:IRAnsatz}    
\end{equation}
Plugging this ansatz into \eqref{Eq:Tachyonv2} we obtain the polynomial equation
\begin{equation}
C_0 \, y^{\alpha} ( \alpha^2 + 4 \alpha +3 ) + C_0 \, y^{\alpha-2} (2 \alpha \phi_{\infty}) - \frac{\lambda}{2} C_0^3 \, y^{3 \alpha} = 0 \,.  
\label{Eq:IRPolyn}
\end{equation}
The first term in \eqref{Eq:IRPolyn} is subleading compared to the second. The third term compete with the second term only when $\alpha<0$. We distinguish 3 cases: $\alpha>-1$, $\alpha=-1$ and $\alpha<-1$. The latter case is trivial because it leads to $C_0=0$.

In the case $\alpha>-1$ the second term always dominate and we find $\alpha=0$. This is the regular solution we are looking for and it admits the following expansion
\begin{equation}
v^{IR}(y)= C_0 + C_2 \, y^2 + C_4 \, y^4 + \dots \, ,  \label{Eq:IRRegsol}   
\end{equation}
with the subleading IR coefficients given by 
\begin{equation}
C_2 =  \frac{C_0}{8 \phi_{\infty}} ( C_0^2 \lambda - 6   ) \quad , \quad 
C_4 = \frac{3 C_0}{128 \phi_{\infty}^2} ( C_0^2 \lambda -  6) ( C_0^2 \lambda - 10) \,. \label{IRcoeffs}
\end{equation}
To guarantee the convergence of the series \eqref{Eq:IRRegsol} we need a condition of the form $|C_0^2 \lambda|/\phi_{\infty}<1$. In the special case $ C_0^2 \lambda - 6=0$ all the subleading coefficients vanish and the constant solution is exact. This can only obtained in the case $\lambda>0$ and, interestingly, it corresponds to the minimum of the Higgs potential $\partial U/\partial v =0$.  

In the special case $\alpha=-1$ we see from \eqref{Eq:IRPolyn} that the first term vanishes whereas the second and third terms are of order $y^{-3}$ leading to the condition $C_0^2 \lambda + 4 \phi_{\infty}=0$. This divergent solution is linear, i.e. $v(z) = C_0 \, z$ and it is valid only for $\lambda<0$. This linear solution appears to be exact and it may be related to previous approaches to  nonlinear soft wall models, such as \cite{Gherghetta:2009ac}. The model of \cite{Gherghetta:2009ac} is briefly discussed in appendix \ref{Sec:GKK}.

\subsection{Numerical solution}

Once the asymptotic analysis has been done, we  proceed to solve numerically the nonlinear differential equation for the tachyon field \eqref{Eq:Tachyon}. We may integrate \eqref{Eq:Tachyon} from the UV to the IR using the UV asymptotic solution \eqref{Eq:UVansatz} to extract initial conditions. By matching the numerical solution to the IR analytic solution \eqref{Eq:IRRegsol} we find a relation between the UV parameters $c_1$ and $c_3$ as well as the corresponding IR parameter $C_0$. Alternatively, we can integrate numerically \eqref{Eq:Tachyon} from the IR to the UV using the IR analytic solution \eqref{Eq:IRRegsol} and this time the matching procedure allows us to extract the UV parameters $c_1$ and $c_3$ in terms of the IR parameter $C_0$.

\begin{figure}[ht]
\centering
\includegraphics[width=7cm]{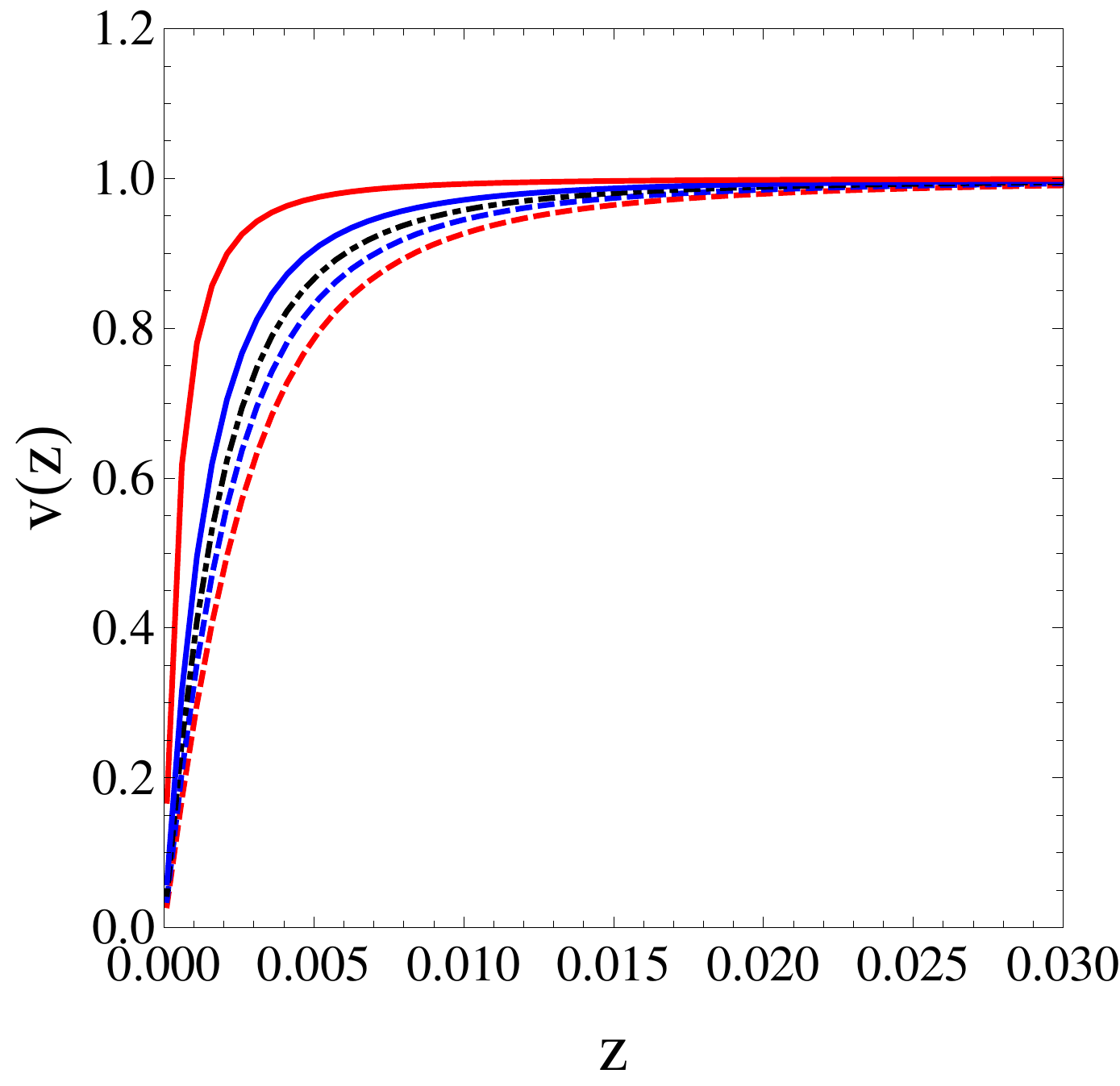}
\caption{Tachyon profiles for $C_0=1$ and $\lambda$ varying from -5 (red dashed) to 5 (red solid). The black dotdashed profile in the middle corresponds to $\lambda=0$ (linear soft wall model). }
\label{Fig:PlotTachyon}
\end{figure}

We show in Fig.~\ref{Fig:PlotTachyon} typical profiles for the tachyon field $v(z)$ for fixed $C_0$ and different values of $\lambda$. The linear soft wall model, corresponding to $\lambda=0$, is depicted by the black dotdashed line and we see that the effect of a nonlinear negative (positive) quartic coupling $\lambda$ is to deform the tachyon profile to the right (left). In the case of $\lambda>0$ we have an upper bound $C_0^2 \lambda =6$ for the tachyon solution. As we approach that bound the tachyon profile becomes constant in $z$, which is consistent with the asymptotic analysis done in the previous subsection. 
\noindent
\noindent

\begin{figure}[ht]
\centering
\includegraphics[width=7cm]{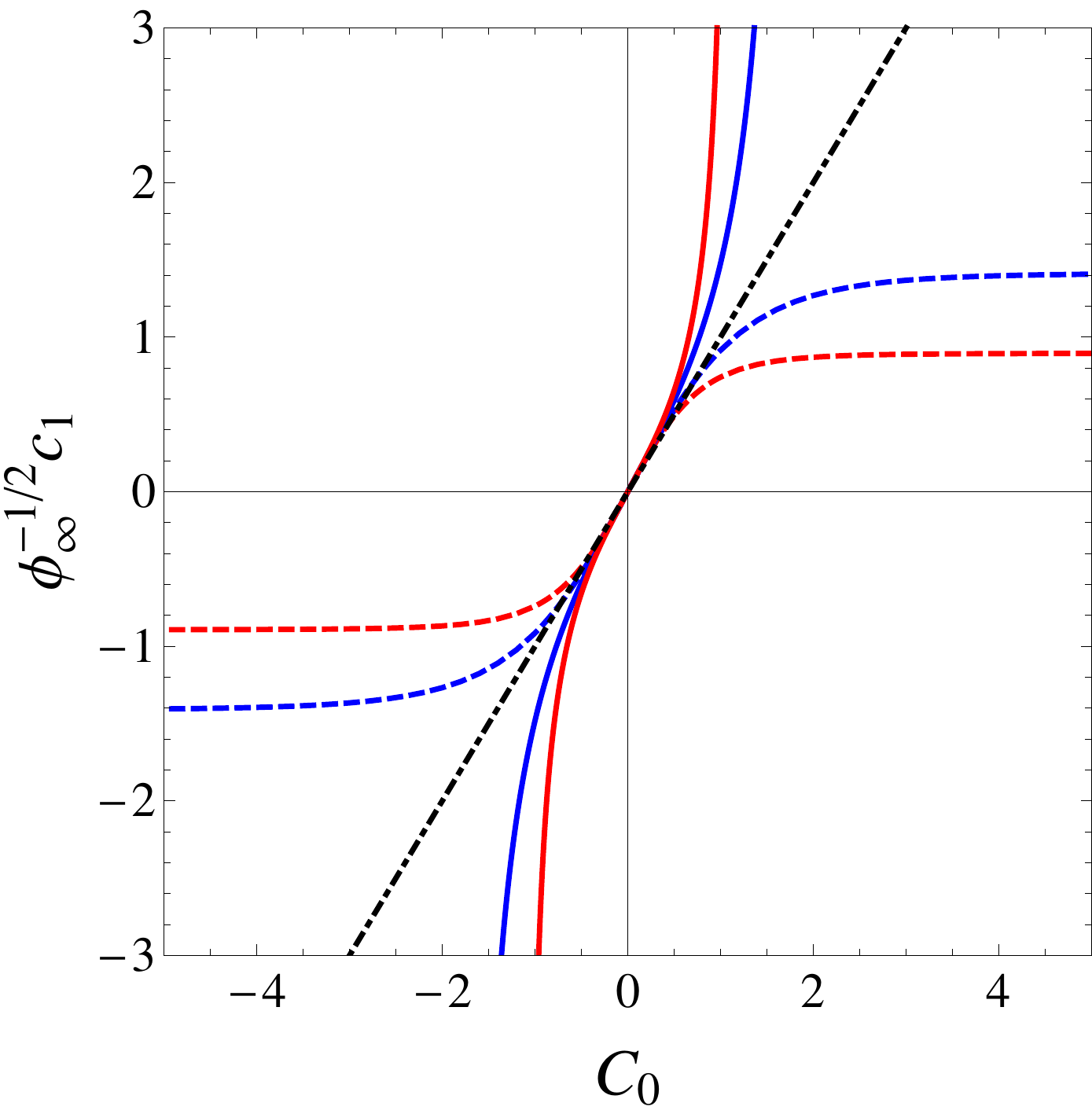}
\hfill
\includegraphics[width=7cm]{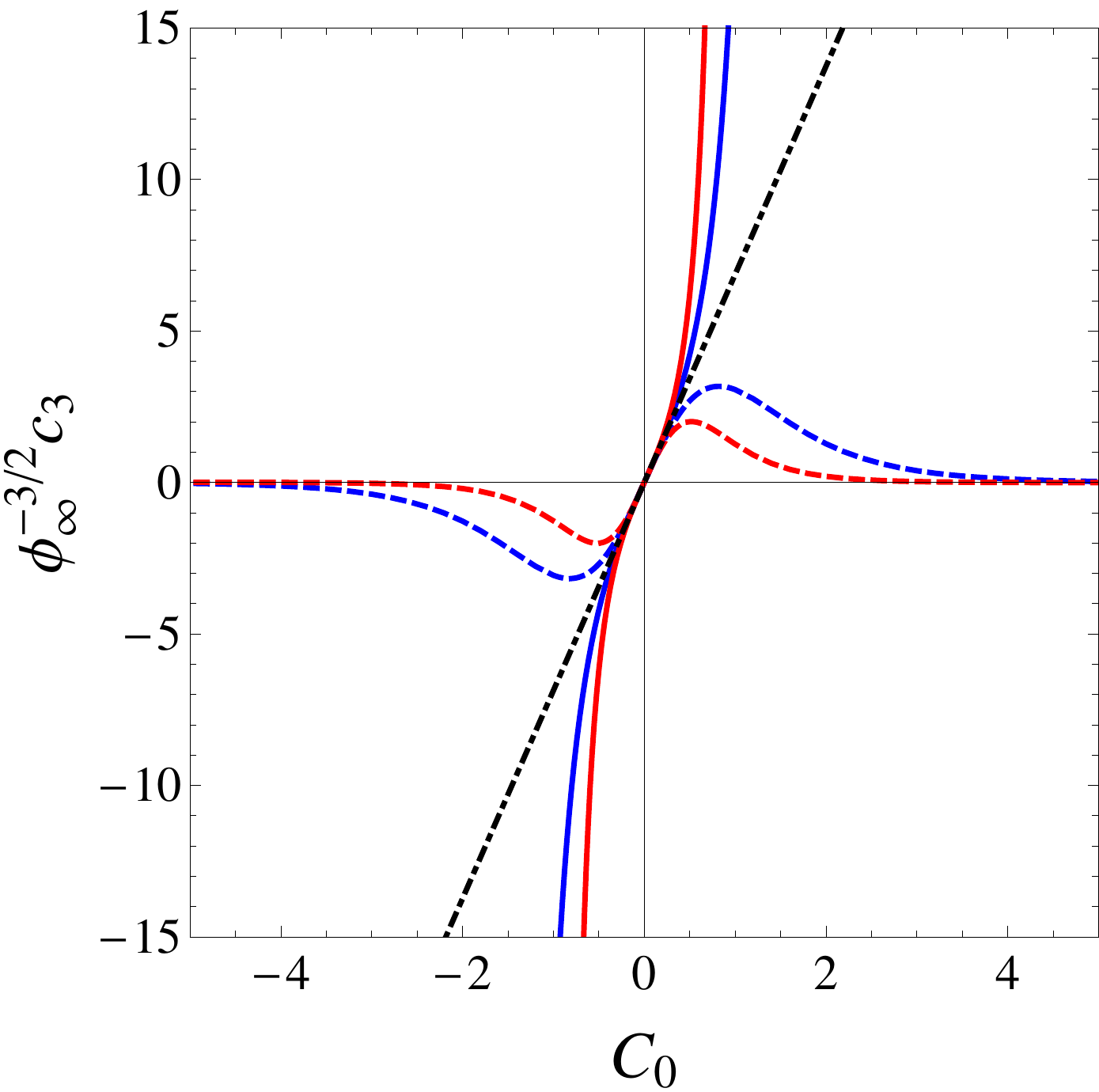}
\caption{Numerical results for $c_1$ (left panel) and $c_3$ (right panel) as functions of $C_0$ in units of $\sqrt{\phi_{\infty}}$. The blue and red solid lines (dashed lines) correspond to $\lambda=2$ and $\lambda=5$ ($\lambda=-2$ and $\lambda=-5$). The black dotdashed line corresponds to $\lambda=0$ (linear soft wall model). }
\label{Fig:Plotc1c3vsC0}
\end{figure}

Our numerical results for the UV parameters $c_1$ and $c_3$ as functions of the IR parameter $C_0$ are displayed in Fig.~\ref{Fig:Plotc1c3vsC0} for different values of $\lambda$. The linear soft wall model ($\lambda=0$) is represented by the black dotdashed line. All solutions enjoy the symmetry $(c_1,c_3)\leftrightarrow (-c_1,-c_3)$, correspoding to the symmetry $v \leftrightarrow -v$, present in the differential equation \eqref{Eq:Tachyon}. The physical regime, of course, corresponds to $c_1>0$. We see that the nonlinearity of the tachyon differential equation leads to nonlinear relations $c_1(C_0)$ and $c_3(C_0)$, as expected. However, in the chiral limit (corresponding to $c_1 \to 0$) all parameters go to zero. In particular, the VEV parameter $c_3$ (associated with the 4d chiral condensate) vanishes. This differs significantly from QCD where chiral symmetry is spontaneously broken in the chiral limit and therefore the chiral condensate is non-vanishing. Note that nonlinearity of the tachyon differential equation brings also saturation effects. In the case $\lambda>0$ there is an upper bound in $C_0$ (described in the previous paragraph) whereas in the case $\lambda<0$ we have an upper bound in $c_1$. An upper bound in $c_1$ implies a cutoff for the quark mass, which is not expected in QCD. 

\noindent
\noindent
\noindent
\begin{figure}[ht]
\centering
\includegraphics[width=7cm]{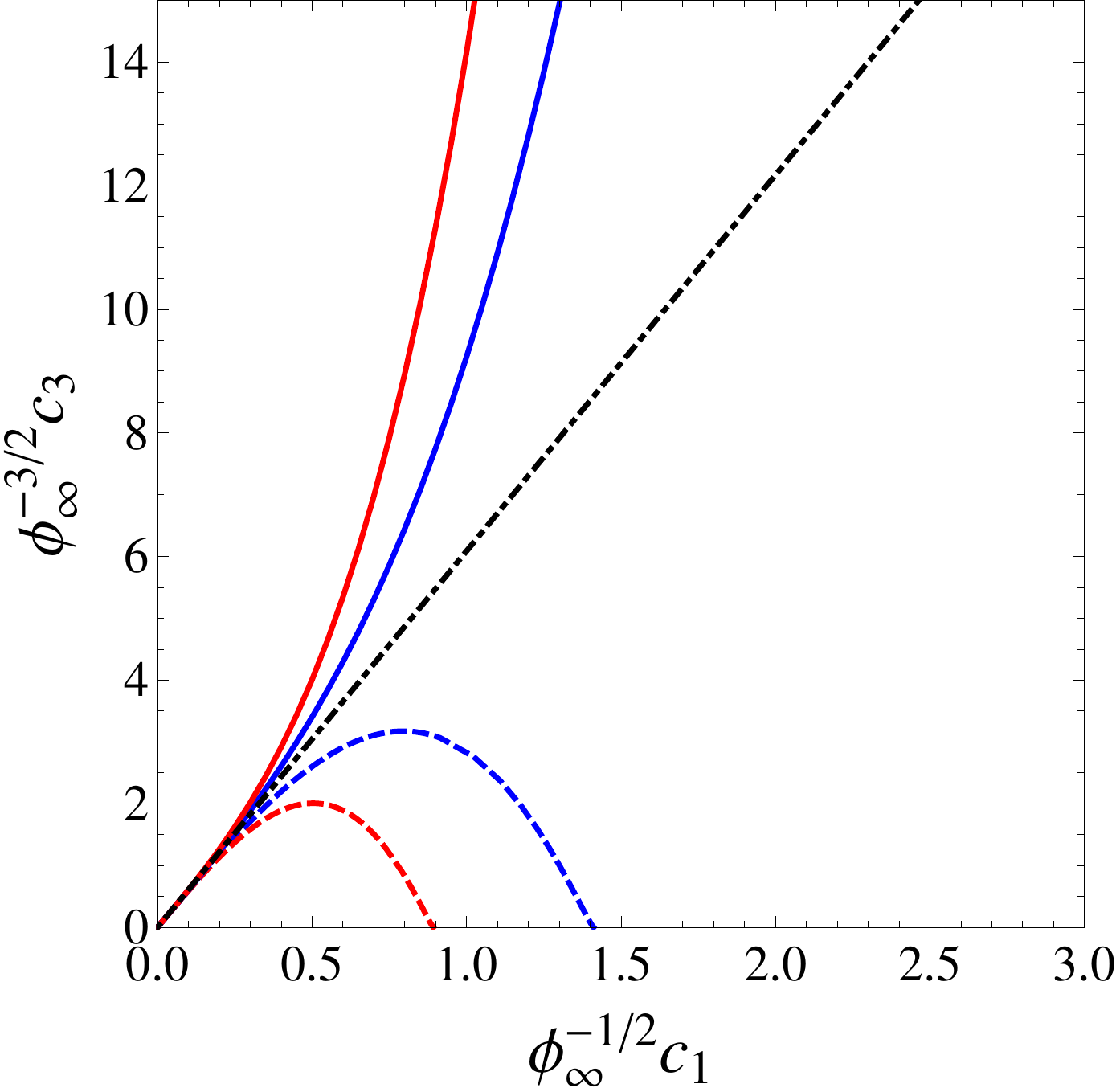}
\caption{The VEV parameter $c_3$ as a function of the source parameter $c_1$, in units of $\sqrt{\phi_{\infty}}$,  for different values of $\lambda$. The blue and red solid lines (dashed lines) correspond to $\lambda=2$ and $\lambda=5$ ($\lambda=-2$ and $\lambda=-5$). The black dotdashed line corresponds to $\lambda=0$ (linear soft wall model).}
\label{Fig:Plotc3vsc1}
\end{figure}
\noindent
In Fig~\ref{Fig:Plotc3vsc1} we show the VEV parameter $c_3$ as a function of the source parameter $c_1$ in the physical regime. 
This result will be interpreted in terms of the 4d chiral condensate $\langle \bar q q \rangle$ as a function of the quark mass $m_q$, as described in the next subsection. The linear relation $c_3 \sim c_1$, characteristic of the linear soft wall model, is actually a good approximation at small $c_1$. For positive $\lambda$ the VEV parameter $c_3$ grows monotonically with $c_1$, which is the expected behaviour for $\langle \bar q q \rangle$. A similar behaviour was obtained in the holographic QCD model developed in \cite{Iatrakis:2010jb}. 
The case of negative $\lambda$ leads to a non-monotonic function $c_3(c_1)$.  In that case the VEV parameter $c_3$ vanishes when $c_1$ reaches its upper bound.  We conclude that the case $\lambda>0$ leads to the more realistic scenario. This could have been antecipated by the fact that $\lambda>0$ in \eqref{Eq:Higgspot} corresponds to the Mexican hat potential for the Higgs field.

\subsection{Holographic renormalisation and the chiral condensate}
\label{subsec:Hologrenorm}

In this subsection we describe the procedure of holographic renormalisation for the nonlinear soft wall models proposed in this work. The case at hand is similar to the case of probe branes in a fixed background, see e.g. \cite{Karch:2005ms,Iatrakis:2010jb}. 
The starting point is the on-shell action. For the background tachyonic field the action in \eqref{HiggsAction}, when taken on-shell, can be written as
\begin{equation}
S_{{\rm os}}=S_{{\rm Bdy}} + S_{{\rm Int}} \, ,    
\end{equation}
where
\begin{equation}
S_{{\rm Bdy}} = - \frac12 \int d^4 x \, \Pi_z(z_0) v(z_0) \quad, \quad 
\Pi_z = - z^{-3} e^{- \Phi(z)} \partial_z v \, , 
\label{Eq:Sbdy}
\end{equation}
\begin{equation}
S_{{\rm Int}}=  \frac{\lambda}{8} \int d^4 x \, dz \, z^{-5}  e^{- \Phi(z)} v^4 \,.  
\label{Eq:Sint}
\end{equation}
$S_{{\rm Bdy}}$ is a boundary term (the AdS boundary is located at $z=z_0$ with $z_0 \to 0$) and $\Pi_z = \partial {\cal L}/ \partial (\partial_z v)$ is the conjugate momentum in $z$. The bulk term $S_{{\rm Int}}$ appears because of the nonlinear term in the tachyon potential. 

\subsubsection{Counterterms and covariant subtraction}

Plugging the UV asympotic solution \eqref{Eq:UVansatz} into the surface term \eqref{Eq:Sbdy} we see that it splits into divergent and finite pieces
\begin{align}
S_{{\rm Bdy}} &= S_{{\rm Bdy}}^{{\rm Div}} + S_{{\rm Bdy}}^{{\rm Fin}}  \, , \\  
S_{{\rm Bdy}}^{{\rm Div}} &= \frac12 \int d^4 x \, \Big [ c_1^2 z_0^{-2} + c_1^2 (c_1^2 \lambda + 4 \phi_{\infty}) \ln z_0 \Big ]    \, , \label{Eq:SbdyDiv} \\
S_{{\rm Bdy}}^{{\rm Fin}} &=  \frac12 \int d^4 x \, \Big [ \frac14 c_1^4 \lambda + 4 c_1 c_3 \Big ] \,. \label{Eq:SbdyFin}
\end{align}
We have omitted in \eqref{Eq:SbdyFin} the terms that vanish in the limit $z_0 \to 0$. 
The bulk term \eqref{Eq:Sint} can not be split in a simple way but,  from \eqref{Eq:UVansatz}, we find the divergent piece 
\begin{equation}
S_{{\rm Int}}^{{\rm Div}}= - \frac{\lambda}{8} \int d^4 x \, c_1^2 \ln z_0 \, , 
\label{Eq:SintDiv}
\end{equation}
and simply define the finite piece as 
\begin{equation}
S_{{\rm Int}}^{{\rm Fin}}= S_{{\rm Int}}- S_{{\rm Int}}^{{\rm Div}} \,.    
\end{equation}
In order to cancel the UV divergences \eqref{Eq:SbdyDiv}-\eqref{Eq:SintDiv} in a consistent way we introduce the covariant counterterms
\begin{align}
S_{{\rm ct}} &= - \int d^4 x \sqrt{- \gamma} \, \Big \{ a_1 v^2(z_0) + \ln z_0 \Big [ a_2 \Phi(z_0) v^2(z_0) - a_3 \lambda v^4(z_0) \Big ]\nonumber \\
&+ a_4 \Phi(z_0) v^2(z_0)   - a_5 \lambda \, v^4(z_0)   \Big \}  \, . \label{Eq:Sct}
\end{align}
The first three terms in \eqref{Eq:Sct} are the minimal required to cancel the UV divergences. The last two terms in \eqref{Eq:Sct} are finite counterterms associated with the renormalisation scheme dependence of the 4d dual theory.
The counterterms action in \eqref{Eq:Sct} can be split into divergent and finite pieces
\begin{align}
S_{{\rm ct}} &= S_{{\rm ct}}^{{\rm Div}} + S_{{\rm ct}}^{{\rm Fin}}  \, , \\  
S_{{\rm ct}}^{{\rm Div}} &= - \int d^4 x \, \Big \{ a_1 c_1^2 z_0^{-2} + \Big [ (2a_1 +a_2) c_1^2 \phi_{\infty} + \left (\frac{a_1}{2} - a_3  \right ) c_1^4 \lambda \Big ] \ln z_0 \Big \}    \, , \label{Eq:SctDiv} \\
S_{{\rm ct}}^{{\rm Fin}} &=  - \int d^4 x \, \Big [ 2 a_1 c_1 c_3 + a_4 c_1^2 \phi_{\infty} - a_5 c_1^4 \lambda  \Big ] \,. \label{Eq:SctFin}
\end{align}
The renormalised action is then defined as 
\begin{equation}
S_{{\rm Ren}} = S_{{\rm Bdy}} + S_{{\rm Int}} + S_{{\rm ct}} \, . \label{Eq:SRen}    
\end{equation}
 The UV divergences cancel for 
\begin{equation}
a_1 = \frac12  \quad , \quad a_2 = 1 \quad  , \quad a_3 = - \frac18 \, ,    
\end{equation}
and the renormalised action takes the form
\begin{equation}
S_{{\rm Ren}} = S_{{\rm Bdy}}^{{\rm Fin}} + S_{{\rm Int}}^{{\rm Fin}} + S_{{\rm ct}}^{{\rm Fin}} \,.    
\end{equation}

\subsubsection{The chiral condensate and the renormalised Hamiltonian}

\begin{figure}[ht]
\centering
\includegraphics[width=7cm]{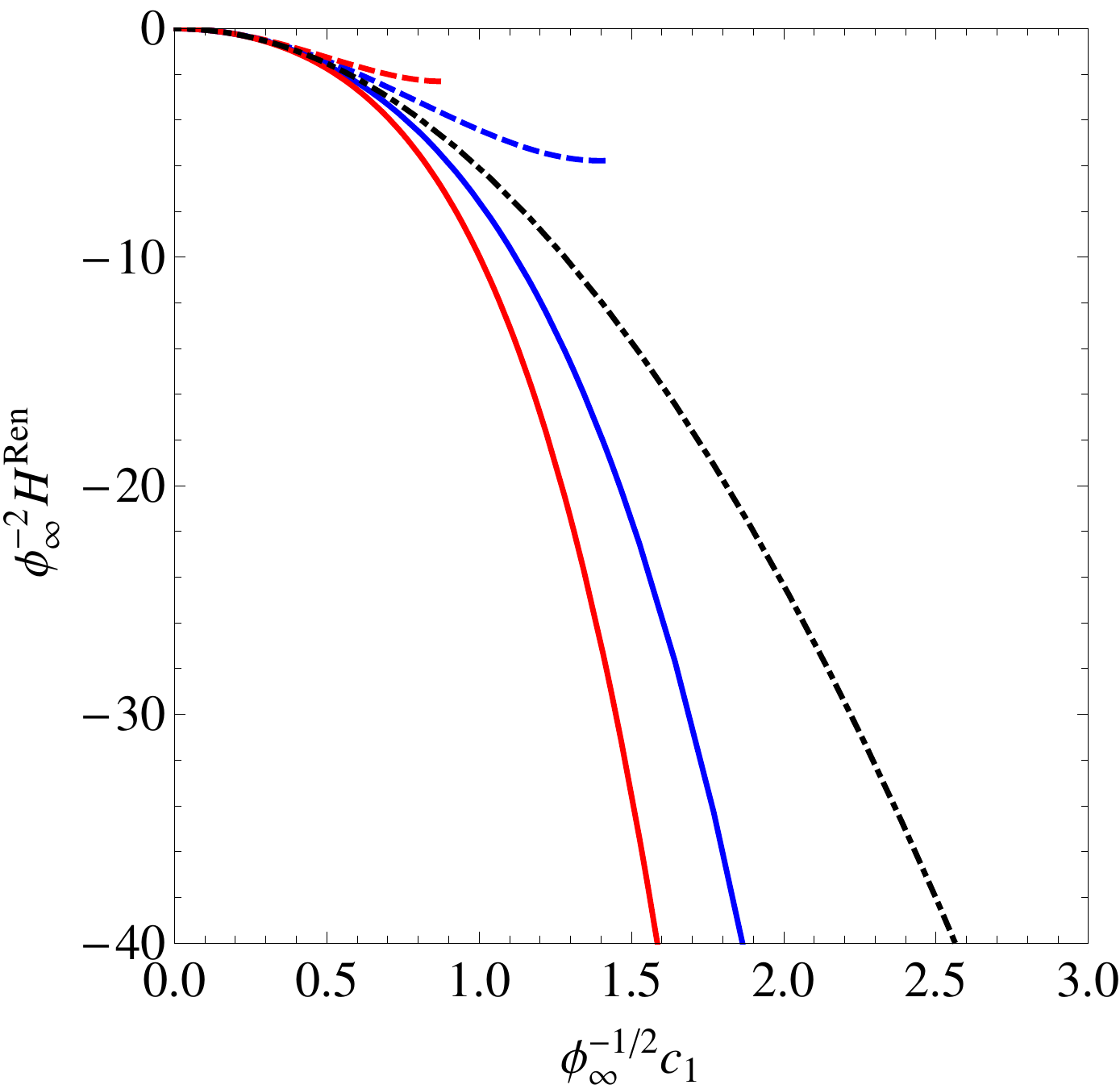}
\hfill
\includegraphics[width=7cm]{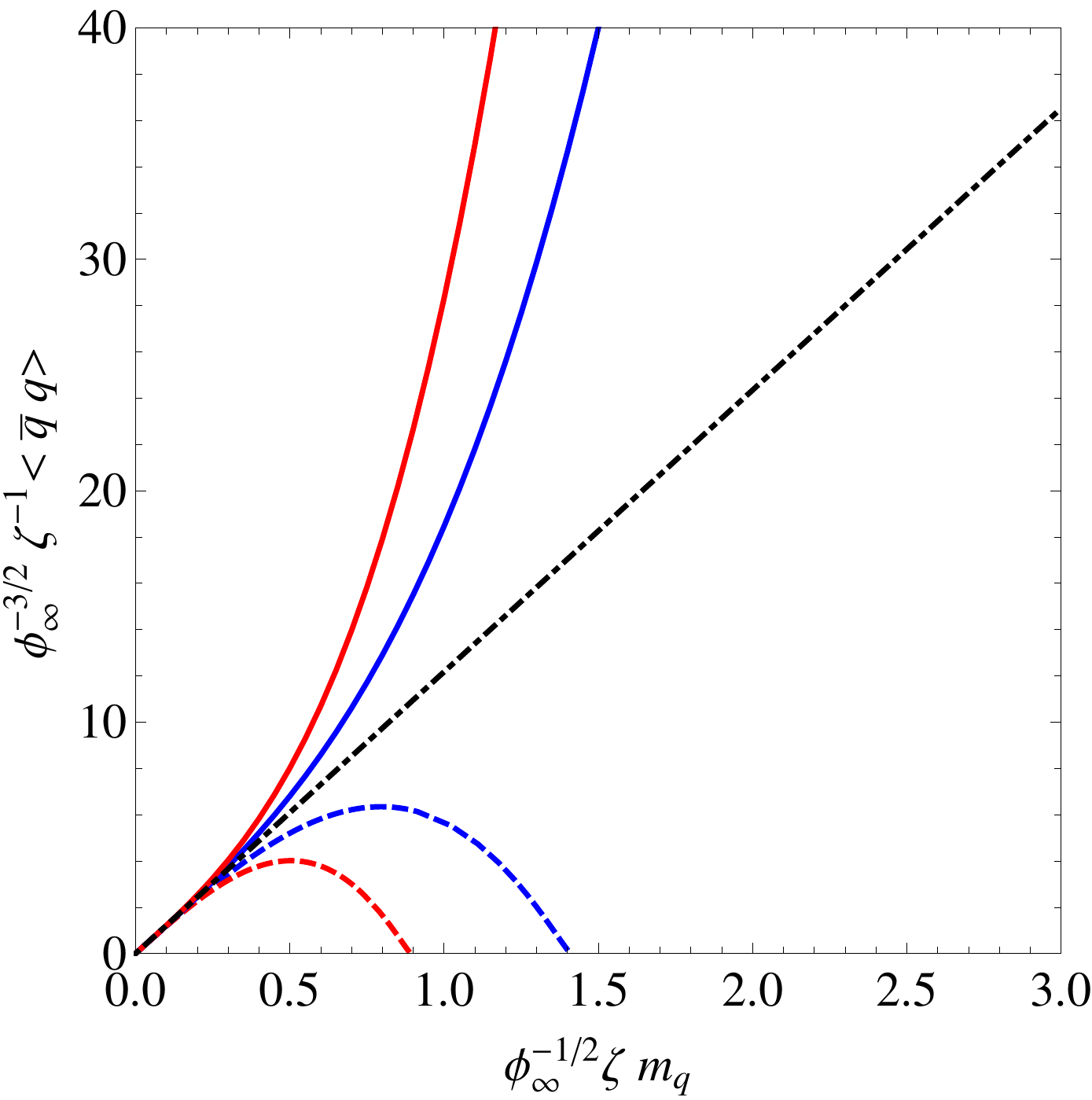}
\caption{{\bf Left (right) panel:} The renormalised Hamiltonian (chiral condensate) as a function of the UV parameter $c_1$ (quark mass) in units $\sqrt{\phi_{\infty}}$. The scheme was fixed setting $a_4=0$ and $a_5 = -1/16$. The blue and red solid (dashed) lines correspond to $\lambda=2$ and $\lambda=5$ ($\lambda=-2$ and $\lambda=-5$) respectively. The black dotdashed line corresponds to $\lambda=0$ (linear case).}
\label{Fig:HamAndCond}
\end{figure}

The CFT deformation due to the quark mass operator has the form $\int d^4 x \, m_q \langle \bar q q \rangle$. The holographic dictionary for the chiral condensate then takes the form
\begin{equation}
\langle \bar q q \rangle =  \zeta \Big [\frac{\delta S_{{\rm os}}}{\delta c_1} + \frac{\delta S_{{\rm ct}}}{\delta c_1} \Big ] \,, \label{Eq:chiralcond}
\end{equation}
where $\zeta = \sqrt{N_c}/(2 \pi)$ is the normalisation constant introduced in the previous section. We identify the first term
\begin{align}
\frac{\delta S_{{\rm os}}}{\delta c_1}  &= - \frac{\partial v(z_0)}{c_1} \Pi_z(z_0)  \nonumber \\
&= c_1 z_0^{-2} + \left ( 4 c_1 \phi_{\infty} - \frac32 c_1^3 \lambda \right ) \ln z_0 + 3 c_3 + c_1 \partial_{c_1} c_3 + \frac14 c_1^3 \lambda \, ,
\label{Eq:Barecond}
\end{align}
as the bare contribution and the second term
\begin{equation}
\frac{\delta S_{{\rm ct}}}{\delta c_1}  = - c_1 z_0^{-2} - \left ( 4 c_1 \phi_{\infty} + \frac32 c_1^3 \lambda \right ) \ln z_0 - c_3 
- c_1 \partial_{c_1} c_3 - 2 a_4 c_1 \phi_{\infty} + 4 a_5 c_1^3 \lambda \, ,
\label{Eq:ctcond}
\end{equation}
as the counterterms contribution. From \eqref{Eq:chiralcond}-\eqref{Eq:ctcond} we see that the UV divergences cancel and we arrive at the final expression for the  chiral condensate
\begin{equation}
  \langle \bar q q \rangle =  \zeta \Big [ 2 c_3 - 2 a_4 c_1 \phi_{\infty} + \frac14 c_1^3 \lambda (1 + 16 a_5) \Big ] \,. 
\end{equation}
The coefficients $a_4$ and $a_5$ reflect the scheme dependence of the chiral condensate. Since we already know that  $c_3$ has a very similar behaviour to the QCD chiral condensate, we suspect that the natural choice for fixing the scheme would be choosing $a_4=0$ and $a_5 = -1/16$. In Fig.~\ref{Fig:HamAndCond} we plot the renormalised Hamiltonian and the chiral condensate, in units $\sqrt{\phi_{\infty}}$, for that particular scheme. As promised, we find a nonlinear relation between the chiral condensate and the quark mass. As described in the previous subsection, the model with $\lambda>0$ provides a more realistic description of the chiral condensate. This conclusion will be supported by the analysis of the meson spectrum, performed in the next subsection. It can be checked that the chiral condensate could have been obtained directly using the relation $\langle \bar q q \rangle = - \partial H^{Ren}/\partial m_q$.

We finish this subsection with a general conclusion for the nonlinear soft wall models presented in this work. As described in the previous subsection, the VEV parameter $c_3$ always vanish in the limit $c_1 \to 0$. Regardless the renormalisation scheme, this result implies that the chiral condensate $\langle \bar q q \rangle$ always vanish in the chiral limit, in sharp contrast with QCD. This means that the nonlinear models at hand never lead to spontaneous symmetry breaking. This result will be confirmed later when we conclude that pseudo Nambu-Goldstone (NG) modes are absent in the spectrum of pseudoscalar mesons. An original solution to this problem was found by in \cite{Gherghetta:2009ac,Kelley:2010mu} where the dilaton profile was modified drastically near the boundary to allow for a description of spontaneous symmetry breaking. The authors in \cite{Gherghetta:2009ac,Kelley:2010mu} realised that the dilaton profile had to be negative in the UV to guarantee spontaneous symmetry breaking. In this work we take a more conservative approach of a positive quadratic dilaton to avoid instabilities in the fluctuations associated with the meson spectrum. Nevertheless, the model in \cite{Gherghetta:2009ac,Kelley:2010mu} deserves a further systematic study. That model is briefly reviewed in appendix \ref{Sec:GKK}.

\noindent
\noindent

\subsection{Meson spectrum ($\lambda<0$)}

The non-Abelian Higgs action in \eqref{HiggsAction} describes the dynamics of the scalar tachyonic field $X$ as well as the left and right gauge fields $A_m^{(L/R)}$. At the beginning of this section we analysed the background tachyonic field $X= \frac12 v(z)$ (the background gauge fields were set to zero). Now we consider the perturbations of  the fields $X$ and $A_m^{(L/R)}$. As described in appendix \ref{Sec:4daction}, the perturbations associated with the tachyonic field are the scalar field $S$ and the pseudoscalar fields $\pi^a$. The former will describe a tower of  scalar mesons whereas the latter will be related to the pseudoscalar mesons (pions in the case $N_f=2$). On the other hand, the gauge field fluctuations will be written as $A_m^{(L/R)}= V_m \pm A_m$ with $V_m = V_m^a T^a$ and $A_m=A_m^a T^a$ identified as the 5d fields dual to the vectorial and axial currents in the chiral symmetry group. 

As described in appendix \ref{Sec:4daction}, the vector field $V_m^a$ decompose as $(V_z^a,V_{\mu}^a)$ and $V_{\mu}^a$ will describe a tower of 4d vector  mesons. The axial-vector field $A_m^a$ decompose as $(A_z^a,A_{\mu}^a)$ and in turn $A_{\mu}^a$ decompose into transverse $A_{\mu}^{\perp,a}$ and  longitudinal parts $\partial_{\mu} \varphi^a$. The fields $A_{\mu}^{\perp,a}$ will describe a tower of axial-vector mesons whereas the fields $\varphi^a$ will couple to the fields $\pi^a$ and describe a tower of pseudoscalar mesons. The equations for the vector sector $V_{\mu}^{\perp,a}$, scalar sector $S$, axial-vector sector  $A_{\mu}^{\perp,a}$ and pseudoscalar sector ($\pi^a, \varphi^a$) are obtained from a second order expansion of the action \eqref{HiggsAction}, as described in appendix \ref{Sec:4daction}.

\subsubsection{Spectrum of the vector sector}
\label{Subsec:VM}

We start with the equation of motion \eqref{Eq:VectorMesonsEq} (the flavour index is hidden in the following analysis). After performing the Fourier transform $V_{\mu}(x^{\mu},z)\to V_{\mu}(k^{\mu},z)$ on \eqref{Eq:VectorMesonsEq}, where we have set $\square\to m_V^2$, the equation may be written in the Schrödinger form through the transformation $V_{\mu}= \xi_{\mu} e^{-B_{V}} \psi_{v_n}$, where $B_{V}=(A_s-\Phi)/2$ and $\xi_{\mu}$ is a (transverse) polarisation vector \footnote{We are introducing again the string-frame warp factor $A_s = - \ln z.$}. The effective Schrödinger equation reads 
\noindent
\begin{equation}\label{Eq:SchrodingerVector}
[ -\partial^2_z +V_{V}] \,\psi_{v_n} =m_{V}^2\, \psi_{v_n},
\end{equation}
\noindent
where the potential is given by
\begin{equation}
V_{V}=\left(\partial_z B_{V}\right)^2+\partial^2_{z} B_{V}.
\end{equation}
\noindent
In this case the problem has an exact solution \cite{Karch:2006pv}
\noindent
\begin{equation}
m^{2}_{V}=4\phi_{\infty}(1+n),\qquad n=0,1,2,\cdots
\end{equation}
\noindent
At this point the free parameter is $\phi_{\infty}$, we may fix the value of this parameter by comparing the first vector state with the corresponding experimental value of the $\rho$ meson, as was done in e.g.  ~\cite{Herzog:2006ra}. We obtain the value $\phi_{\infty}=(388\,\text{MeV})^2$. 

The spectrum obtained is shown in Table~\ref{Taba:VectorSF}, labelled as SW, compared against the holographic model of  \cite{Gherghetta:2009ac} and experimental data \cite{Tanabashi:2018oca}. The details of \cite{Gherghetta:2009ac} are given in appendix \ref{Sec:GKK}.
\begin{table}[ht]
\centering
\begin{tabular}{l |c|c|c}
\hline 
\hline
 $n$ & SW \cite{Karch:2006pv}& GKK \cite{Gherghetta:2009ac} & $\rho$ experimental \cite{Tanabashi:2018oca} \\
\hline 
 $1$ & 776  & 475  & $776\pm 1$  \\
 $2$ & 1097 & 1129 & $1282\pm 37$  \\
 $3$ & 1344 & 1429 & $1465\pm 25$ \\
 $4$ & 1552 & 1674 & $1720\pm 20$  \\
 $5$ & 1735 & 1884 & $1909\pm 30$  \\
 $6$ & 1901 & 2072 & $2149\pm 17$ \\
 $7$ & 2053 & 2243 & $2265\pm 40$ \\
\hline\hline
\end{tabular}
\caption{
The mass of the vector mesons (in MeV) obtained in the the soft wall model, compared against the holographic model \cite{Gherghetta:2009ac} and experimental results from PDG \cite{Tanabashi:2018oca}.
}
\label{Taba:VectorSF}
\end{table}

\subsubsection{Spectrum of the scalar sector}
\label{Sec:ScalarL}

\begin{figure}[ht]
\centering
\includegraphics[width=7cm]{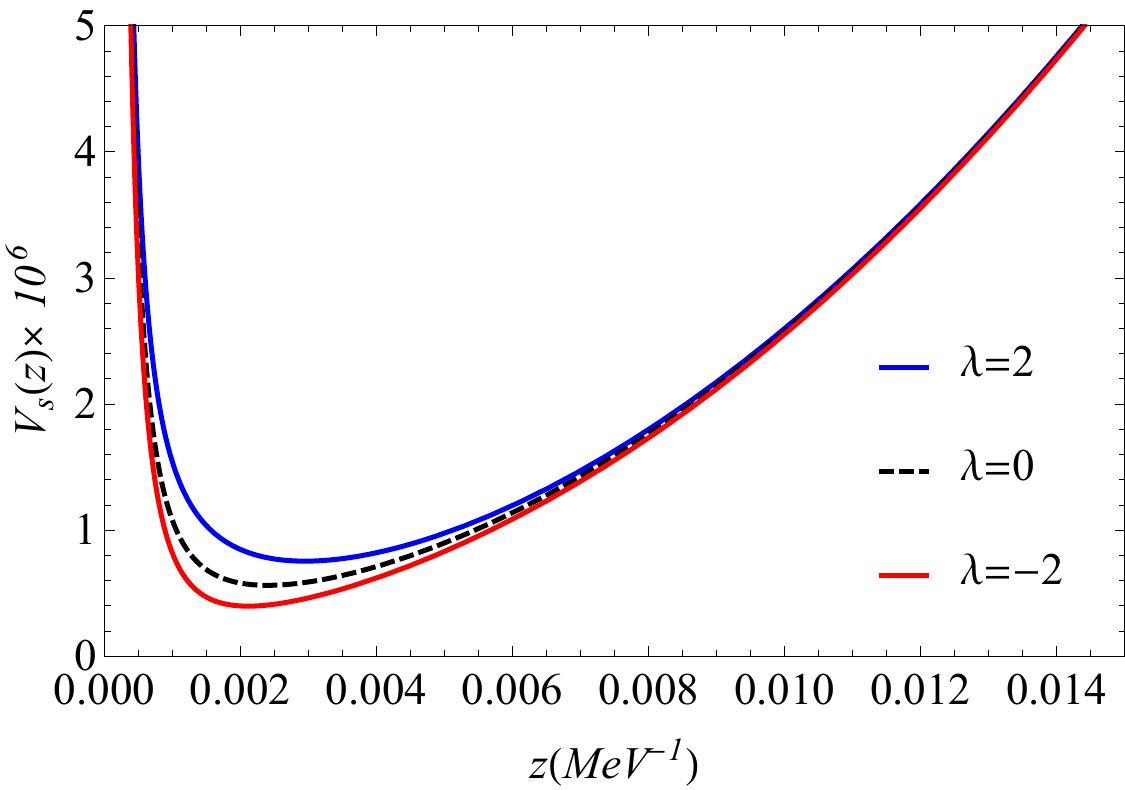}
\caption{The potential of the Schrödinger equation for 
$\phi_{\infty}=(388\,\text{MeV})^2$ and three different values of the parameter $\lambda$.
}
\label{Fig:PotScalarIGKK}
\end{figure}

\begin{figure}[ht]
\centering
\includegraphics[width=7cm]{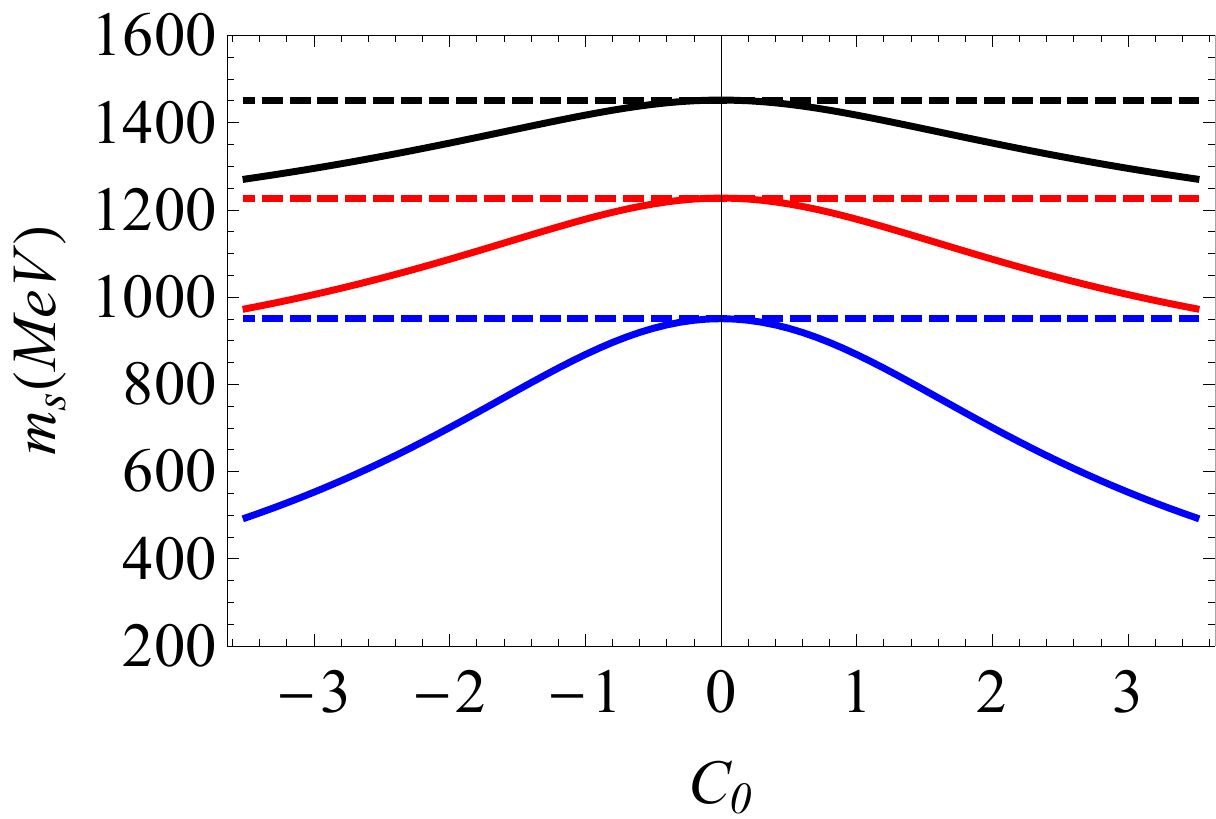}
\hfill
\includegraphics[width=7cm]{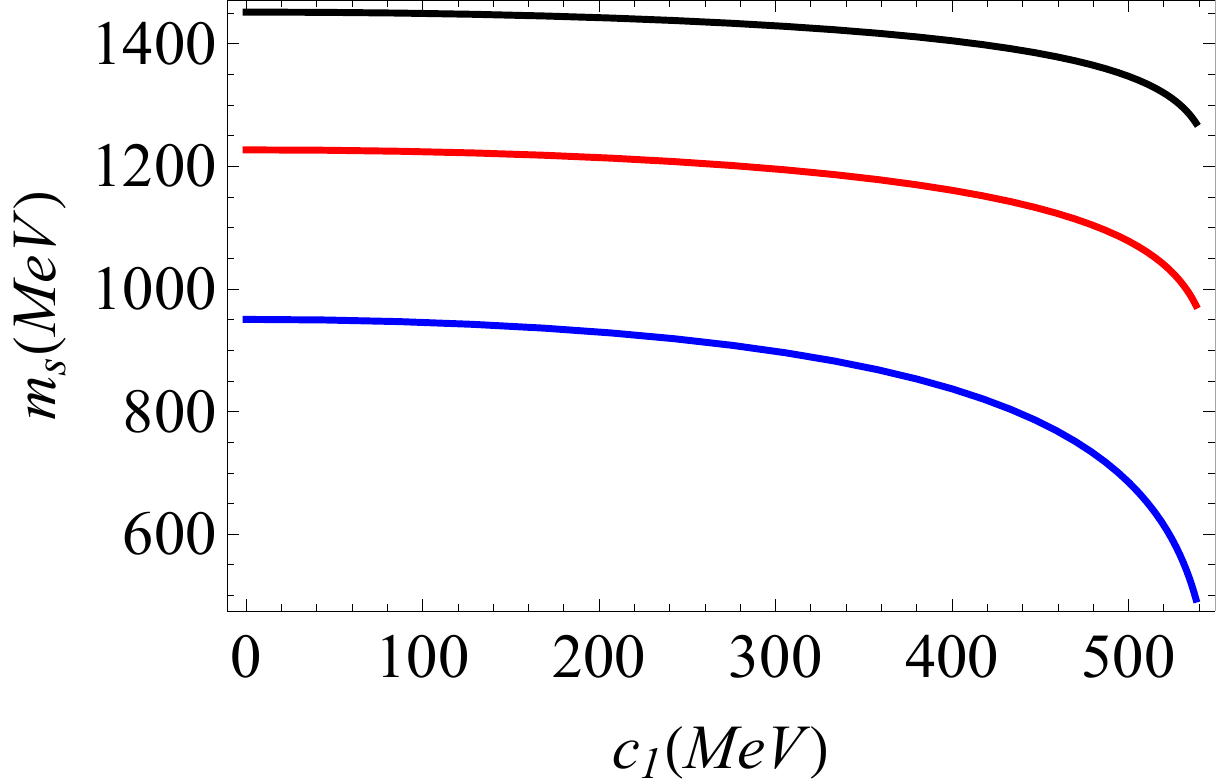}
\caption{
The mass of the scalar mesons as a function of $C_0$ (left) and $c_1$ (right). The source parameter $c_1$ is related to the quark mass by $c_1 = m_q \zeta$ with $\zeta=\sqrt{N_c}/(2\pi)$. 
Solid lines represent the results for $\lambda=-2$, while dashed lines for $\lambda=0$. The results were obtained setting $\phi_{\infty}=(388\,\text{MeV})^2$.
}
\label{Fig:ScalarMassIGKK}
\end{figure}

Now we proceed to calculate the spectrum of the scalar mesons. This sector is obtained from the fluctuations of the tachyon field, cf. Eq. \eqref{Eq:ScalarFlucts}, where $S(x,z)$ represents the scalar field related to the scalar mesons. After performing the Fourier transform $S(x^{\mu},z)\to S(k^{\mu},z)$ on \eqref{Eq:ScalarMesonsEq}, where we  set $\square\to m_s^2$ and , we arrive at the following equation
\noindent
\begin{equation}\label{Eq:ScalarEq}
e^{-3A_s+\Phi}\partial_z\left(e^{3A_s-\Phi}\partial_z S(k,z)\right)
+m_s^2 S(k,z)
-e^{2A_s}\left(m_X^{2}(z)+\frac32\lambda v^{2}(z)\right)S(k,z)=0,
\end{equation}
\noindent
where $m^2_{X}(z)=-3$. We rewrite the last equation in a Schrödinger form, redefining the scalar modes as  $S_n=e^{-B_S}\psi_{s_{n}}(z)$, where $B_S=3A_s/2-\Phi/2$. Thus, we get
\noindent
\begin{equation}\label{Eq:SchroScalarEq}
-\partial_z^2 \psi_{s_{n}}+V_S\,\psi_{s_{n}}=m_s^2\,\psi_{s_{n}},
\end{equation}
\noindent
with the Schrödinger potential given by
\noindent
\begin{equation}\label{Eq:Schro/potScalarL}
V_S=(\partial_z B_S)^2+\partial_z^2B_S
+e^{2A_s}\left(m_X^{2}+\frac{3\,\lambda}{2}v^2(z)\right).
\end{equation}
\noindent
For $\lambda=0$ the potential \eqref{Eq:Schro/potScalarL} reduces to the one obtained in \cite{Colangelo:2008us}. Notice how the parameter $\lambda$ controls the minimum value of the potential \eqref{Eq:Schro/potScalarL}, as shown in Fig.~\ref{Fig:PotScalarIGKK}. As we increase $\lambda$ the minimum increases and hence the masses of scalar mesons. 
For $\lambda<0$, the potential allows us to describe very light states. This statement is supported by the results displayed in Fig.~\ref{Fig:ScalarMassIGKK}, where we can see the evolution of the meson with the parameter $C_0$ (left panel) and $c_1$ (right panel). Those results were obtained for $\lambda=-2$ (solid lines) and $\lambda=0$ (dashed lines).

\noindent
\noindent

\noindent
\noindent

Those results indicate the possibility of fixing the parameter $C_0$, for given  $\lambda$, requiring the first eigenvalue of the Schrödinger equation to match the mass of the  scalar meson $f_0(550\text{MeV})$ \footnote{There were some attempts to obtain a light scalar meson in the top-down approach, see e.g. \cite{Kaplunovsky:2010eh,Ihl:2010zg}.}.  However, the status of the  $f_0(550\text{MeV})$ as a scalar meson is not established \cite{Tanabashi:2018oca}. We  follow a more conservative approach and consider  $f_0(980\text{MeV})$ as the first scalar meson, as in  ~\cite{DaRold:2005vr}. As the upper limit for the scalar mass in our model is the one obtained for $\lambda=0$, $m_s=950$ MeV, see Fig.~\ref{Fig:ScalarMassIGKK}, it is not possible to reach the state $f_0(980\text{MeV})$ when $\lambda<0$.  

It is worth mentioning that the behaviour of the scalar meson mass as a function of the quark mass, i.e., $c_1$, displayed in Fig.~\ref{Fig:ScalarMassIGKK} is opposite to the one expected in QCD.  Hence, the model with $\lambda<0$ is pathological in the scalar sector.

\subsubsection{Spectrum of the axial-vector sector}
\label{Sec:AV}

\begin{figure}[ht]
\centering\centering

\includegraphics[width=7cm]{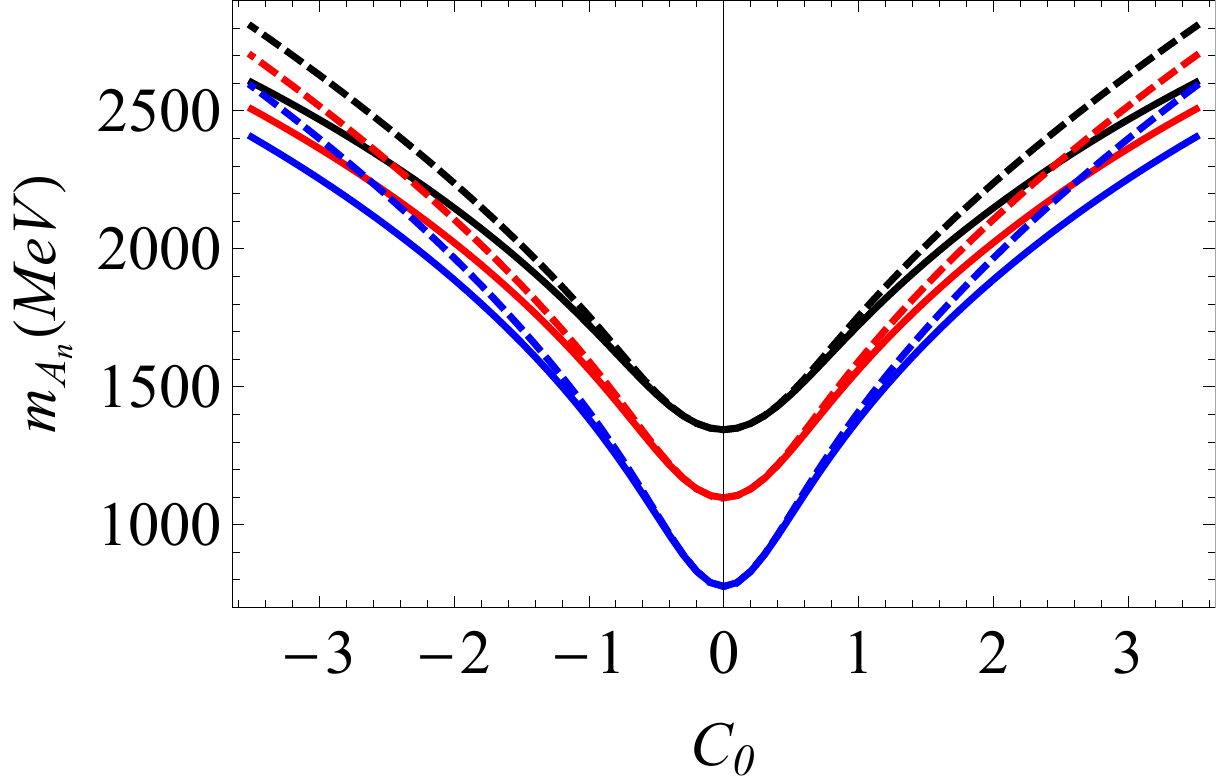}
\hfill
\includegraphics[width=7cm]{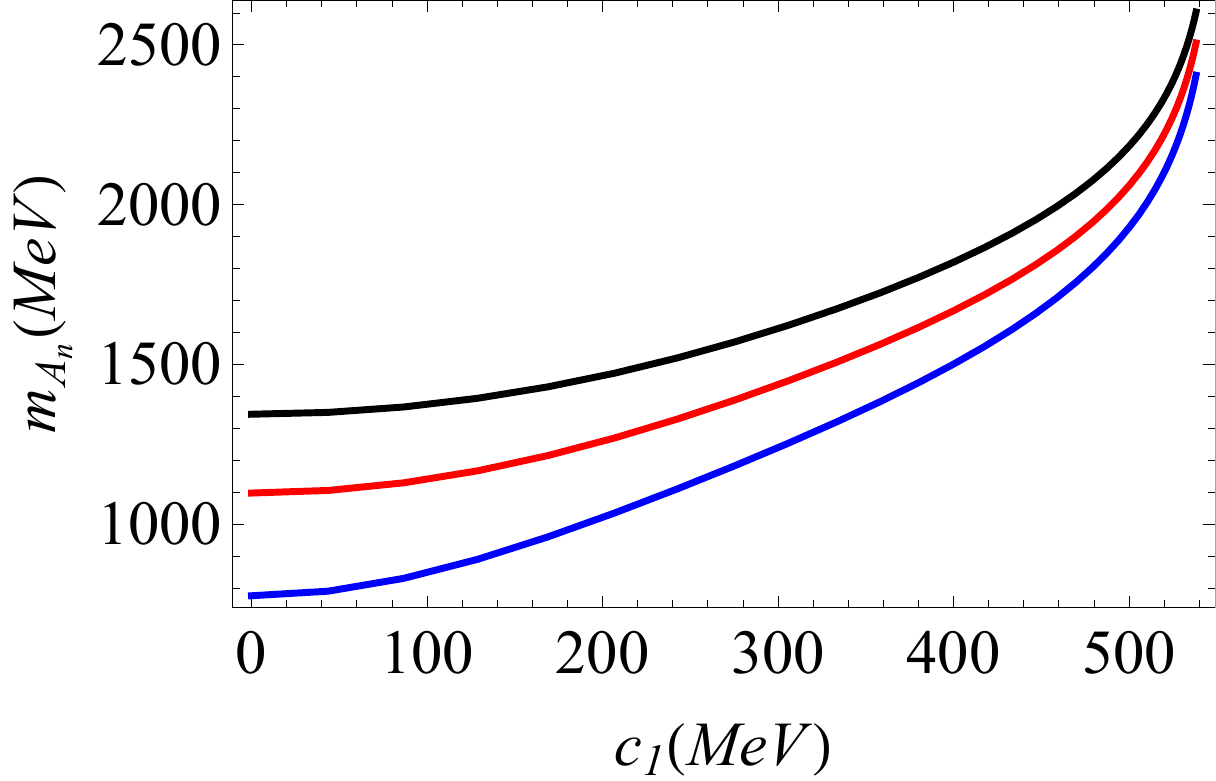}
\caption{
Masses of axial-vector mesons as functions of $C_0$ (left) and $c_1$ (right). The source parameter $c_1$ is related to the quark mass by $c_1 = m_q \zeta$ with $\zeta=\sqrt{N_c}/(2\pi)$. Solid lines represent the results for 
$\lambda=-2$, while dashed lines for $\lambda=0$. The 
results were obtained setting $\phi_{\infty}=(388\,\text{MeV})^2$.
}
\label{Fig:AVMassFig}
\end{figure}

 After performing the Fourier transform  $A^{\mu}_{\perp}(x^{\mu},z)\to A^{\mu}_{\perp}(k^{\mu},z)$  on \eqref{Eq:AxialMesonsEq}, with  $\square\to m_A^2$ and, redefining the axial-vector mode as $A_{\mu}= \xi_{\mu} e^{-B_{A}} \psi_{a_n}$, where $B_{A}=(A_s-\Phi)/2$, we arrive at the Schrödinger equation
\noindent
\begin{equation}\label{Eq:AVSchroEq}
-\partial^2_z \psi_{a_n}+V_{A}\,\psi_{a_n}=m_{A}^2\,\psi_{a_n},
\end{equation}
\noindent
with the Schrödinger potential given by
\begin{equation}\label{Eq:AVPotential}
V_{A}=\left(\partial_z B_{A}\right)^2+\partial^2_{z} B_{A}
+g_5^2\,e^{2A_s}v^2(z).
\end{equation}
\noindent
The left panel of Fig.~\ref{Fig:AVMassFig} shows our results for the masses of the axial-vector states as a function of the parameter $C_0$, compared against the corresponding values in the linear soft wall model, i.e., $\lambda=0$. The right panel of the figure shows how the masses depend on the parameter the quark mass parameter  $c_1$. We see that in the axial sector the masses increase with $c_1$, which is the expected behaviour for mesons. 
These results suggest the possibility of using the mass of the first axial-vector meson $a_1(1230)$ to fix the parameter $C_0$ for given $\lambda$. Note, however, that the axial-vector meson masses grow very fast and tend to diverge at a finite value of $c_1$. That value corresponds to the unphysical upper bound for the quark mass, described in the previous subsection. 
\noindent
\noindent

\subsubsection{Spectrum of the pseudoscalar sector}
\label{Sec:Pion}
\begin{figure}[ht]
\centering
\includegraphics[width=7cm]{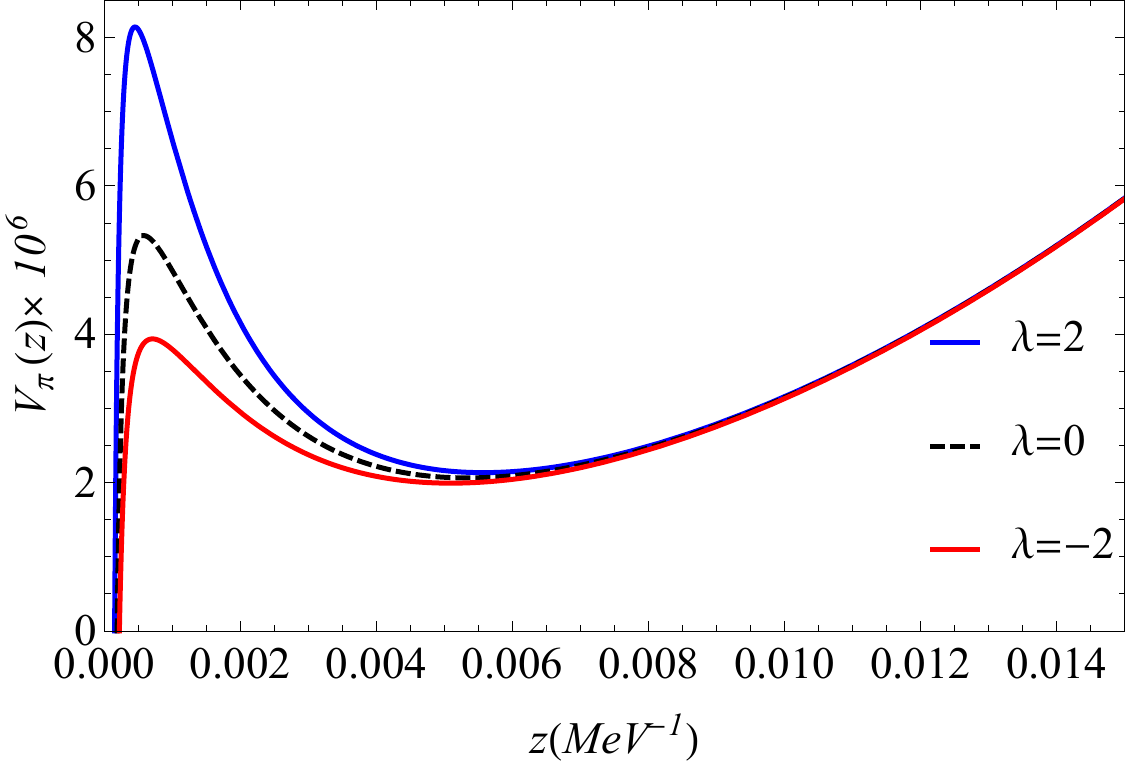}
\hfill
\includegraphics[width=7cm]{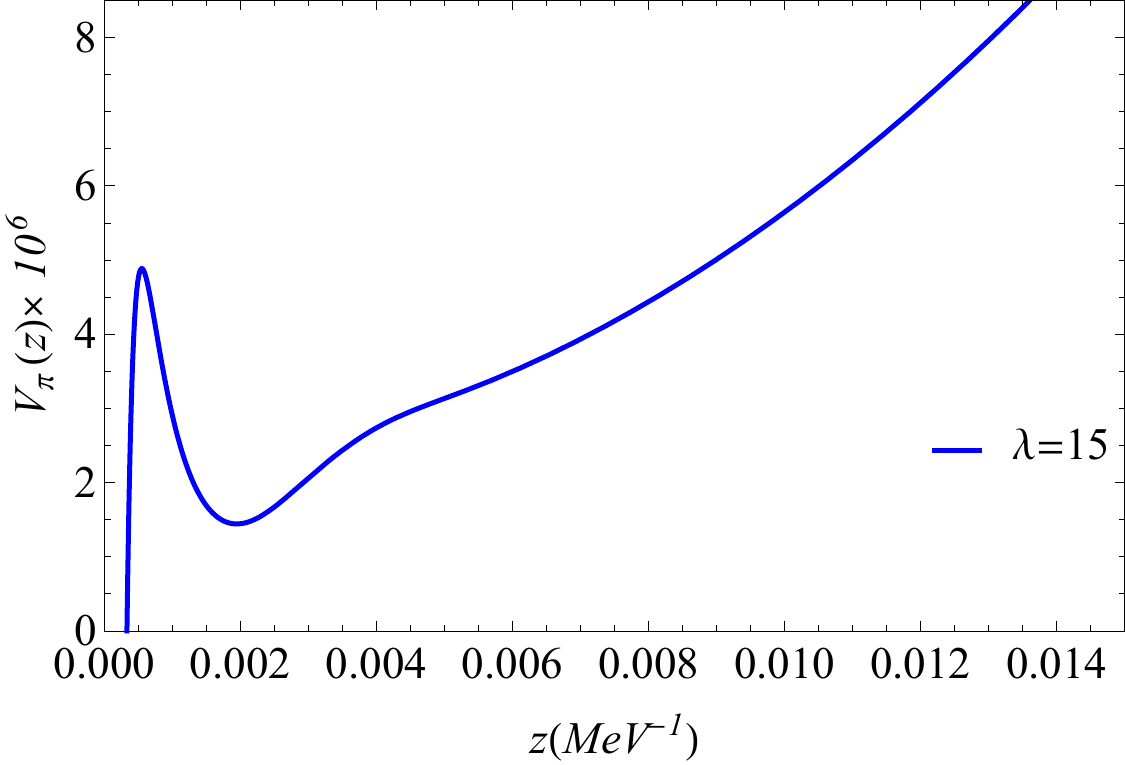}
\caption{
Left: The potential of the pseudoscalar Schrödinger equation in the NLSW model for $\lambda=2$ (blue line), $\lambda=0$ (dashed black line) and $\lambda=-2$ (red line), we have set $\phi_{\infty}=(388\,\text{MeV})^2$. Right: The potential of GKK model, this figure was obtained using the same parameters used in  \cite{Kelley:2010mu}.
}
\label{Fig:PionPotential}
\end{figure}

The pseudoscalar sector is special because it is described by a coupled system of differential equations \eqref{Eq:PionMesonsEq1}- \eqref{Eq:PionMesonsEq2}. After performing the Fourier transform $\pi(x^{\mu},z)\to \pi(k^{\mu},z)$ and $\varphi(x^{\mu},z)\to \varphi(k^{\mu},z)$ in both equations, where we have set $\square\to m_{\pi}^2$, we get the following equations
\noindent
\begin{eqnarray}
&e^{-A_s+\Phi}\partial_z\left(e^{A_s-\Phi}\partial_z\varphi\right)
+g_5^2e^{2A_s+2\log{v}}\left(\pi-\varphi\right)=0, \label{Eq:EqPion1}  \\
& -m_{\pi}^{2}\partial_z\varphi
+g_5^2e^{2A_s+2\log{v}}\partial_z\pi=0. \label{Eq:EqPion2}
\end{eqnarray}
\noindent
We follow \cite{Ballon-Bayona:2014oma} and decouple this system of equations. The decoupled equation is second order in the auxiliary field  $\Pi=\partial_z\pi_n$ and takes the form
\noindent
\begin{equation}\label{Eq:DecopledPionEq}
-\partial_z^2\Pi+\partial_z\left(\Phi-A_s
-\ln{\beta}\right)\partial_z\Pi
+\left(\partial_z^2\left(\Phi-A_s
-\ln{\beta}\right)-m_{\pi}^{2}+\beta\right)\Pi=0,
\end{equation}
\noindent
where we have introduced the function $\beta(z)=g_5^2e^{2A_s}v^2$. Defining the function $2B_{\pi}=A_s-\Phi+\log{\beta}$, then, introducing the transformation $\Pi=e^{-B_{\pi}}\psi_{\pi_n}$ we arrive at the Schrödinger equation
\noindent
\begin{equation}\label{Eq:SchroPion}
-\partial_z^2\psi_{\pi_n}+V_{\pi}\psi_{\pi_n}=m_{\pi}^{2}\psi_{\pi_n},
\end{equation}
\noindent
with the potential given by
\noindent
\begin{equation}\label{Eq:PionPotential}
V_{\pi}=\left(\partial_zB_{\pi}\right)^2-\partial_z^2B_{\pi}+\beta.
\end{equation}
\noindent
At this point it is interesting to display a plot of the potential \eqref{Eq:PionPotential}. This is shown on the left panel of Fig. \ref{Fig:PionPotential} where we observe potential wells emerging for different values of $\lambda$. These potential wells, however, are not deep enough to allow a light state in the spectrum. This is related to the fact that the soft wall backgrounds considered in this work do not provide spontaneous symmetry breaking in the chiral limit and therefore we would not expect pseudo NG modes. 
In the right panel of Fig. \ref{Fig:PionPotential} we display the potential obtained in the holographic model investigated in  \cite{Gherghetta:2009ac} where the potential well allows a light state in the spectrum, as described in  \cite{Kelley:2010mu}.
 
\noindent
\noindent

\begin{figure}[ht]
\centering
\includegraphics[width=7cm]{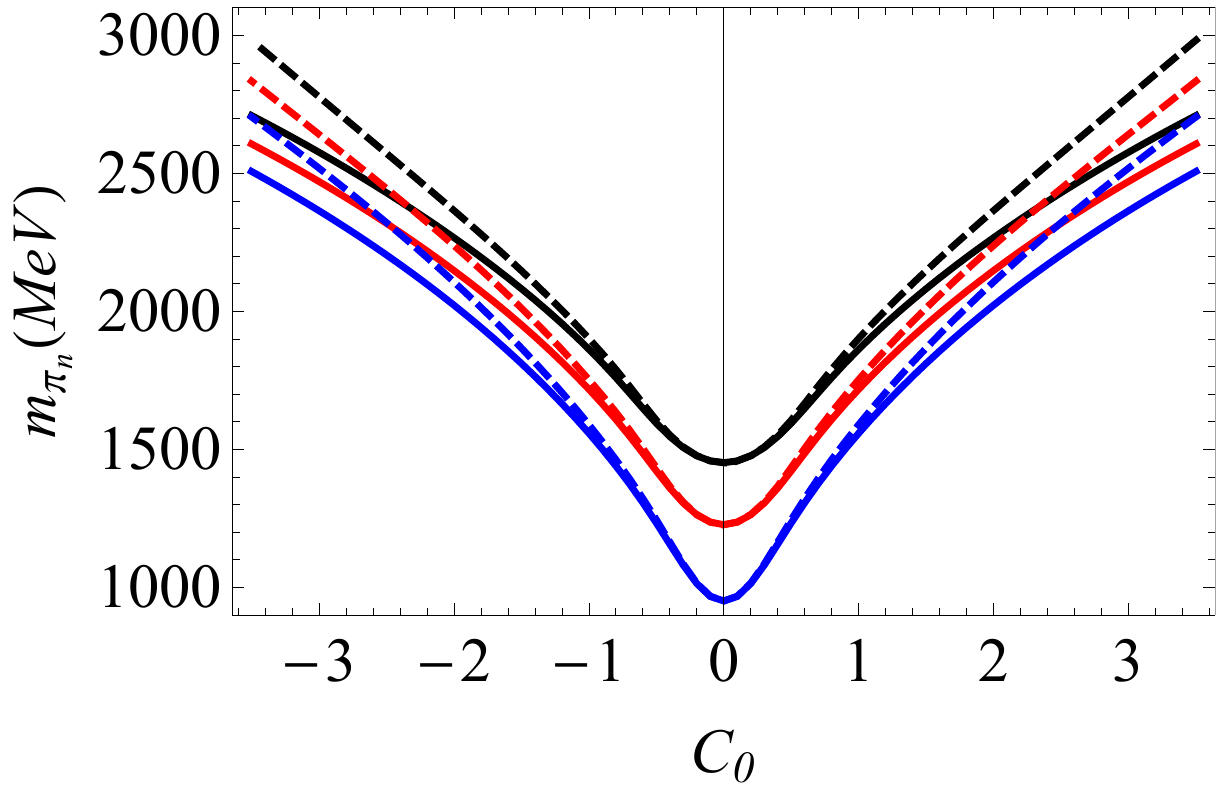}
\hfill
\includegraphics[width=7cm]{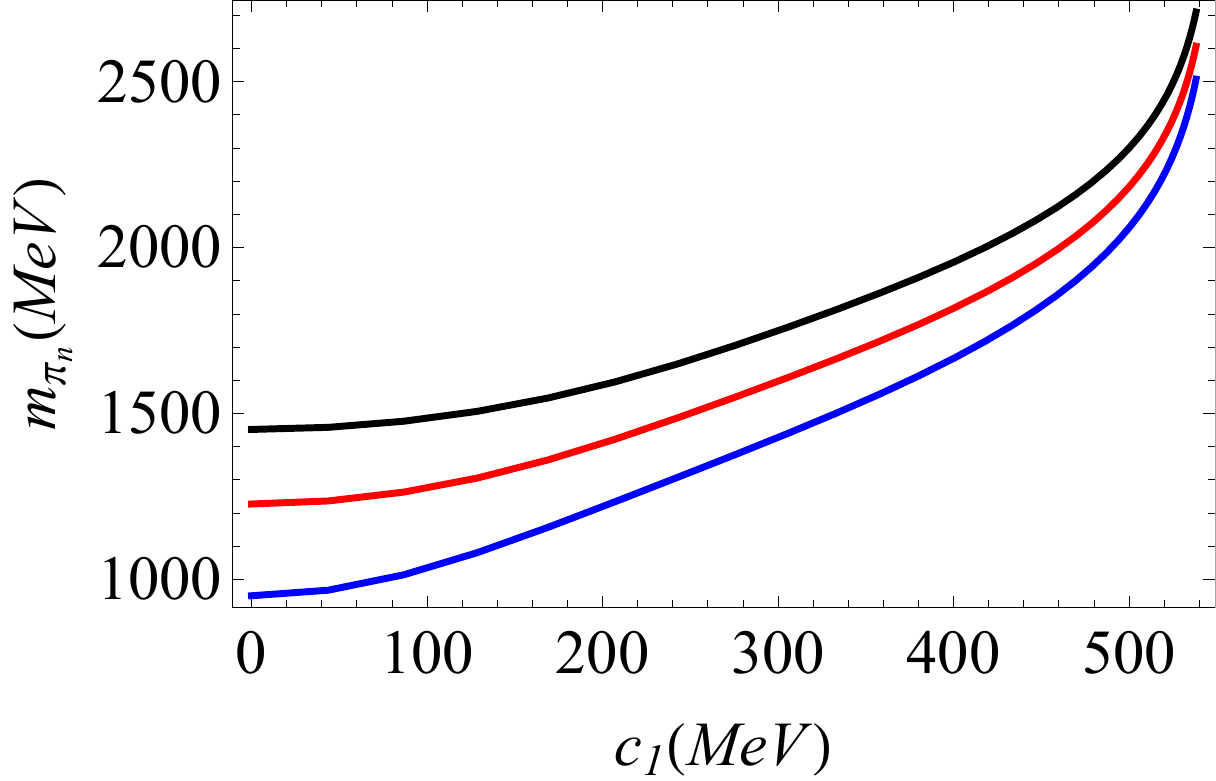}
\caption{
The mass as a function of $C_0$ (left) and $c_1$ (right). The source parameter $c_1$ is related to the quark mass by $c_1 = m_q \zeta$ with $\zeta=\sqrt{N_c}/(2\pi)$. Solid lines represent the results for 
$\lambda=-2$, while dashed lines for $\lambda=0$. The 
results were obtained setting $\phi_{\infty}=(388\,\text{MeV})^2$.
}
\label{Fig:PSMassFig}
\end{figure}

In Fig.~\ref{Fig:PSMassFig} we show the evolution of the pseudoscalar meson masses with the parameter $C_0$ (left panel) and $c_1$ (right panel), compared against the results of the linear soft wall model plotted with dashed lines. Note that the masses increase with the quark mass parameter $c_1$, as expected in QCD. However, since there is an unphysical cutoff for the quark mass parameter $c_1$ the pseudoscalar meson masses grow very fast and tend to diverge when $c_1$ reaches the cutoff. 
\noindent
\noindent

\subsection{Meson spectrum ($\lambda>0$)}

As described in subsection \ref{Subsec:VM}, the spectrum of vector mesons is insensitive to the nonlinear potential so it is identical to the linear case ($\lambda=0$). Below we describe the spectrum of scalar, axial-vector and pseudoscalar mesons for the case of positive $\lambda$, which corresponds to a Mexican hat potential in \eqref{Eq:Higgspot}.

\subsubsection{Spectrum of the scalar sector}
\label{Sec:ScalarSectorLneg}

\begin{figure}[ht]
\centering
\includegraphics[width=7cm]{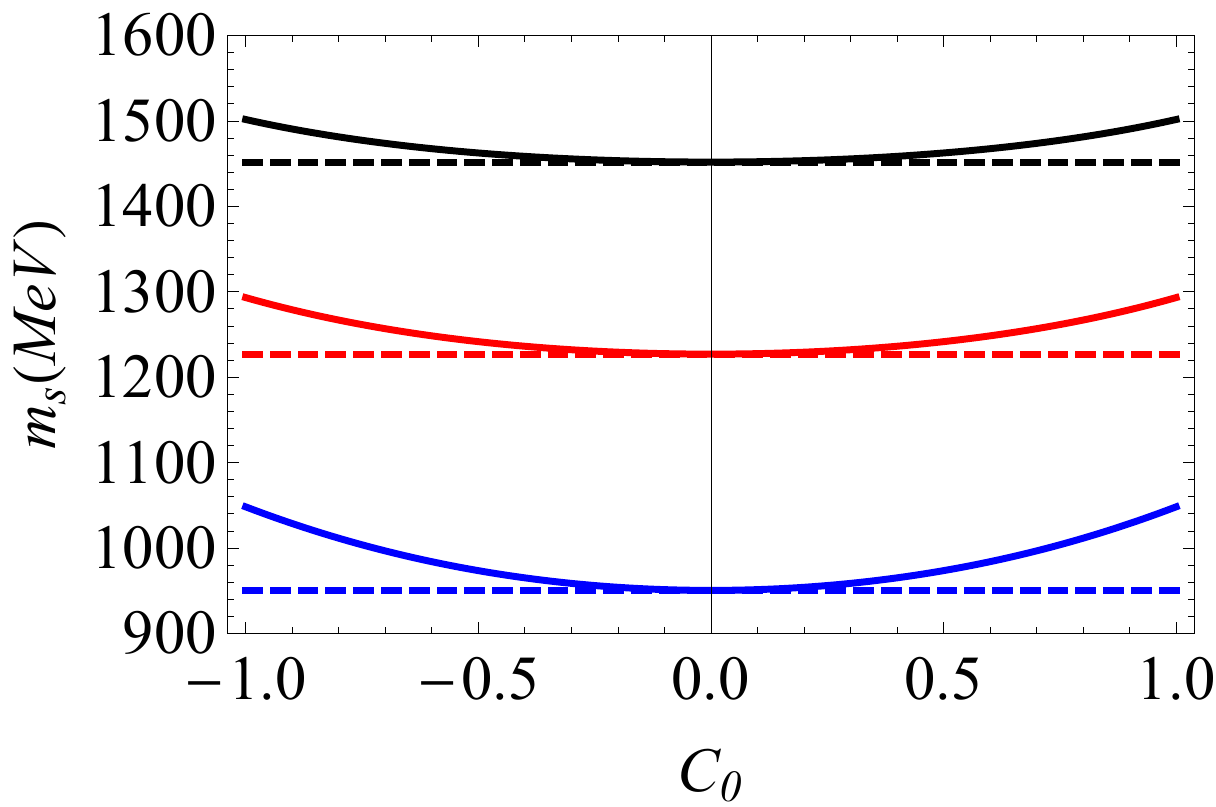}
\hfill
\includegraphics[width=7cm]{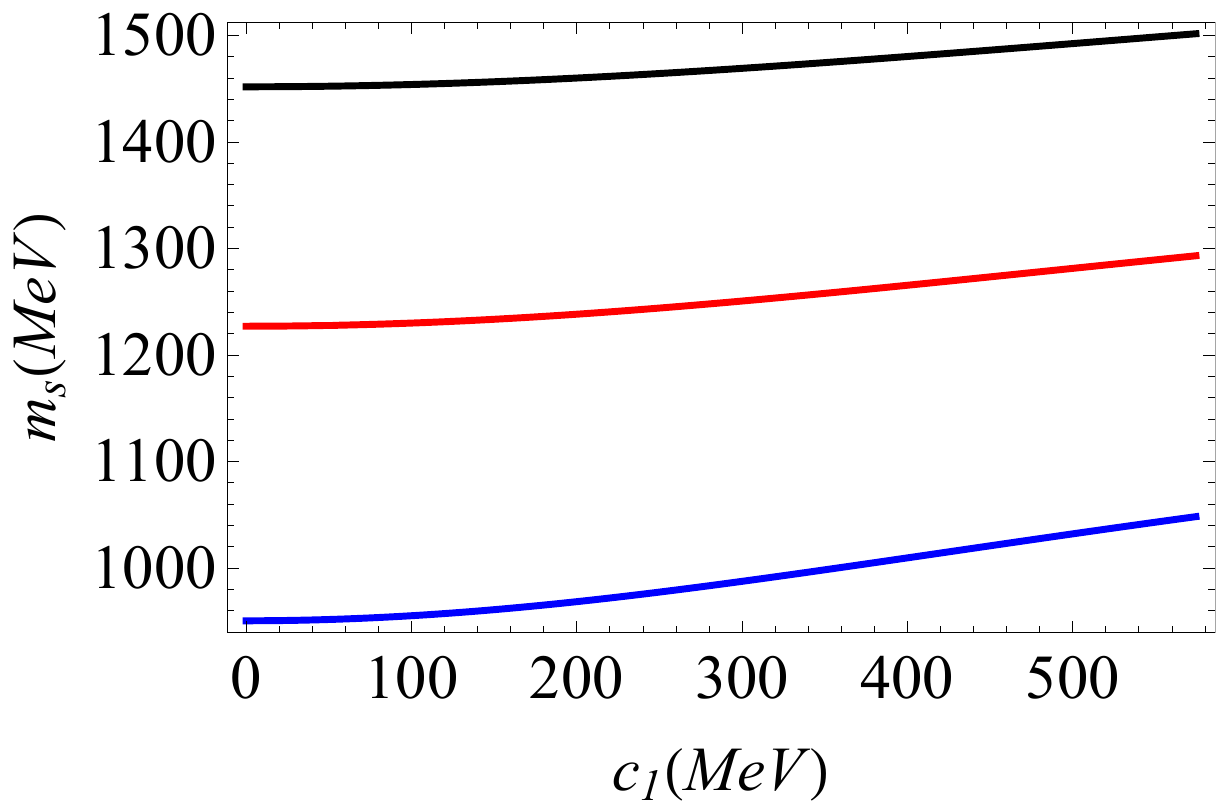}
\caption{
The masses of the scalar mesons as a function of $C_0$ (left panel) and $c_1$ (right panel). The source parameter $c_1$ is related to the quark mass by $c_1 = m_q \zeta$ with $\zeta=\sqrt{N_c}/(2\pi)$. Solid lines represent the result for $\lambda=2$, while dashed lines for $\lambda=0$. The results were obtained setting $\phi_{\infty}=(388\,\text{MeV})^2$.
}
\label{Fig:ScalarMassC1Lneg}
\end{figure}

The differential equation of the scalar sector, written in a Schrödinger form, was given in \eqref{Eq:SchroScalarEq}. The effect of going from negative to positive $\lambda$ was displayed in Fig.~\ref{Fig:PotScalarIGKK}. We concluded that  states become heavier for $\lambda>0$ compared to the cases $\lambda=0$ and $\lambda<0$. As described in the previous subsection, we consider  $f_0(980)$ state as the first scalar state and use that value  to fix the parameter $C_0$, for given $\lambda$. Our results for that parameter choice are displayed in Table~\ref{Taba:01Lneg}, labeled as NLSW, compared against the linear soft wall, the holographic model  of \cite{Gherghetta:2009ac} and experimental data. For $\lambda=7$ we obtain $C_0=0.3$ and find $c_1=142.4$(MeV), which implies a large value for the quark mass. 
In Fig.~\ref{Fig:ScalarMassC1Lneg} we show the evolution of the scalar meson masses as functions of the parameter $C_0$ (left panel) and $c_1$ (right panel). We observe that the masses increase  with $C_0$ and $c_1$. The lower bound around $950$ MeV for the  first scalar meson implies that we never reach the state $f_0(550)$, whose status is controversial anyway \cite{Tanabashi:2018oca}.
Meson masses increasing with the quark mass is a behaviour expected in QCD so we conclude that the model with $\lambda>0$ provides the best scenario. It is worth mentioning that  the masses of scalar mesons do not vanish in the chiral limit.
\noindent

\begin{table}[tbp]
\centering
\begin{tabular}{l |c|c|c|l}
\hline 
\hline
 $n$ & NLSW &  SW \cite{Karch:2006pv} &
 GKK \cite{Gherghetta:2009ac}&
$f_0$ experimental \cite{Tanabashi:2018oca} \\
\hline 
 $1$ & 980  & 950  & 799   & $980\pm10$  \\
 $2$ & 1246 & 1227 & 1184  & $1350\pm 150$  \\
 $3$ & 1466 & 1452 & 1466  & $1505\pm 6$ \\
 $4$ & 1657 & 1646 & 1699  & $1724\pm 7$  \\
 $5$ & 1829 & 1820 & 1903  & $1992\pm 16$  \\
 $6$ & 1986 & 1978 & 2087  & $2103\pm 8$ \\
 $7$ & 2132 & 2125 & 2257  & $2314\pm 25$ \\
 $8$ & 2268 & 2262 & 2414  &  \\
\hline\hline
\end{tabular}
\caption{
The masses of the scalar mesons (in MeV) obtained in the nonlinear soft wall model with $\lambda>0$, compared against the linear soft wall model \cite{Karch:2006pv}, the holographic model of  \cite{Gherghetta:2009ac} and experimental data \cite{Tanabashi:2018oca}. The value of the parameters are $\lambda=7$, $C_0=0.3$ and $\phi_{\infty}=(388\,\text{MeV})^2$.
}
\label{Taba:01Lneg}
\end{table}

\subsubsection{Spectrum of the axial-vector sector}

\begin{figure}[ht]
\centering
\includegraphics[width=7cm]{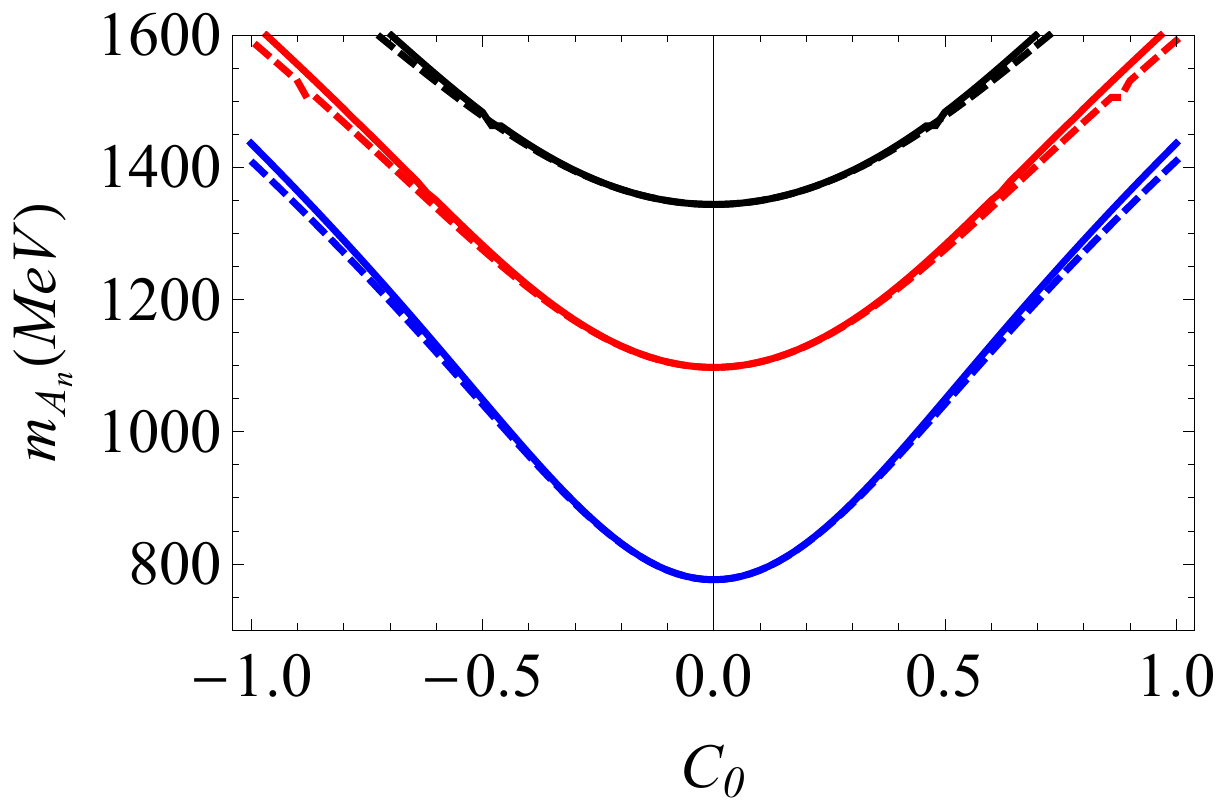}
\hfill
\includegraphics[width=7cm]{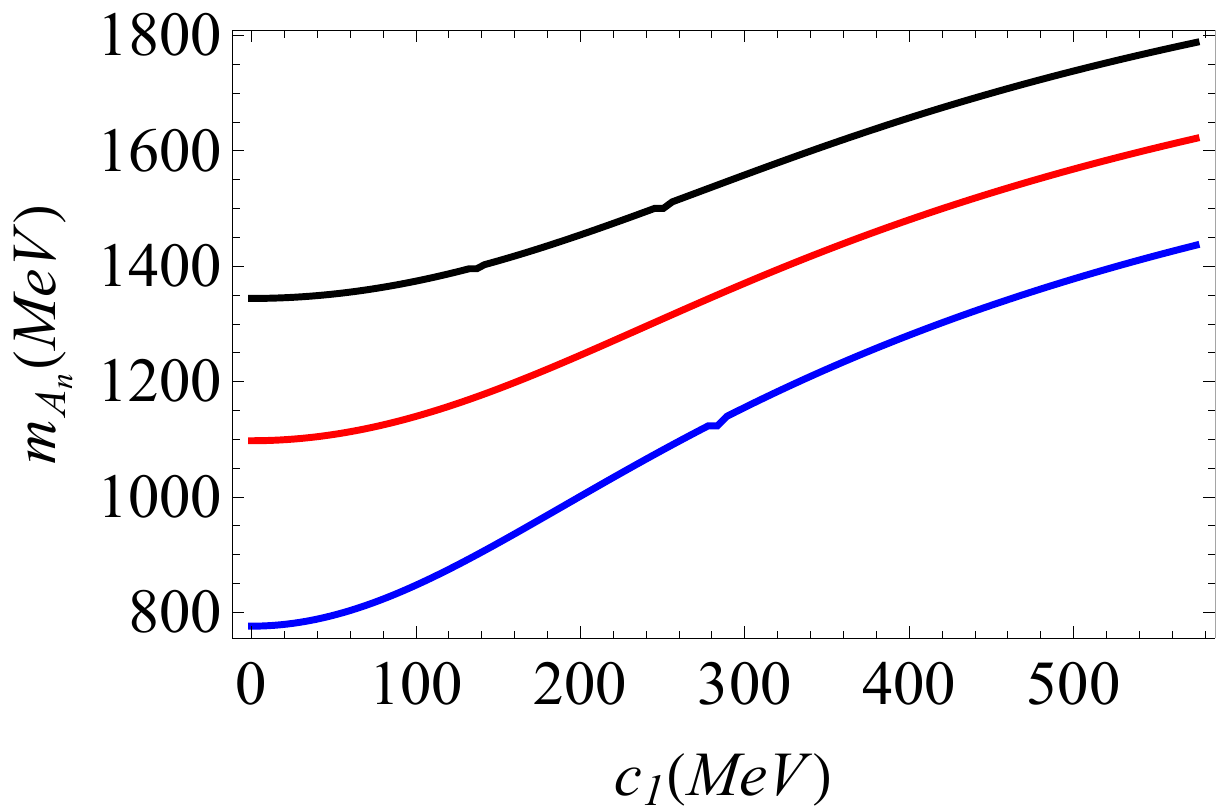}
\caption{
The masses of the axial-vestor mesons as a function of $C_0$ (left panel) and $c_1$ (right panel). The source parameter $c_1$ is related to the quark mass by $c_1 = m_q \zeta$ with $\zeta=\sqrt{N_c}/(2\pi)$. Solid lines represent the result for $\lambda=2$, while dashed lines for $\lambda=0$. The results were obtained setting $\phi_{\infty}=(388\,\text{MeV})^2$.
}
\label{Fig:AVMassFigLneg}
\end{figure}

The Schrödinger equation for the axial-vector sector was given in Eq. \eqref{Eq:AVSchroEq} and this time we consider the case $\lambda>0$. Solving that equation we find the axial-vector meson masses. In Fig.~\ref{Fig:AVMassFigLneg} we show the evolution of the masses as functions of the parameters $C_0$ (left panel) and $c_1$ (right panel). The meson masses  increase with the IR parameter $C_0$ and the UV quark mass parameter $c_1$. Meson masses increasing with the quark mass is a behaviour expected in QCD. In the chiral limit $c_1 \to 0$ the masses of axial-vector mesons and vector mesons become degenerate. 
\noindent
\noindent

In Table \ref{Taba:02} we present our results for the parameter choice $\lambda=7$, $C_0=0.3$, fixed in the scalar sector. The results, labeled  as NLSW,  are compared against the linear soft wall model \cite{Karch:2006pv}, the holographic model of  \cite{Gherghetta:2009ac} and experimental data \cite{Tanabashi:2018oca}. 
We observe that our results are close to the results obtained in the linear soft wall model. This is explained by the smallness  of the tachyon field for the parameters chosen to calculate the spectrum. For a small tachyon field the last term in the potential \eqref{Eq:AVPotential} is relatively small and the potential gets closer to the case of linear soft wall model. 

We  point out that  the spectrum of the vector and axial-vector mesons presented in Tables \ref{Taba:VectorSF} and \ref{Taba:02} is not degenerate (the parameter $c_1$ is outside the chiral limit). In our model, the mass difference between the vector and axial-vector mesons at relatively large values of $c_1$ is purely a consequence of explicit chiral symmetry breaking, in contrast with QCD where we expect another contribution coming from spontaneous chiral symmetry breaking. 

\begin{table}[ht]
\centering
\begin{tabular}{l |c|c|c|l}
\hline 
\hline
 $n$ & NLSW & SW \cite{Karch:2006pv}&
 GKK \cite{Gherghetta:2009ac}&
$a_1$ experimental \cite{Tanabashi:2018oca} \\
\hline 
 $1$ & 897  & 891  & 1185   & $1230\pm 40$  \\
 $2$ & 1172 & 1168 & 1591  & $1647\pm 22$  \\
 $3$ & 1398 & 1395 & 1900  & $1930^{+30}_{-70}$ \\
 $4$ & 1594 & 1592 & 2101  & $2096\pm 122$  \\
 $5$ & 1770 & 1768 & 2279  & $2270^{+55}_{-40}$  \\
 $6$ & 1930 & 1928 &   &  \\
 $7$ & 2079 & 2077 &   &  \\
\hline\hline
\end{tabular}
\caption{
The masses of the axial-vector mesons (in MeV) obtained in the 
nonlinear soft wall model, compared against the linear soft wall model \cite{Karch:2006pv}, the holographic model of  \cite{Gherghetta:2009ac} and experimental data \cite{Tanabashi:2018oca}. The value of the parameters used are 
$\lambda=7$, $C_0=0.3$ and $\phi_{\infty}=(388\,\text{MeV})^2$.
}
\label{Taba:02}
\end{table}

\subsubsection{Spectrum of the pseudoscalar sector}
\label{subsec:pionposlambda}

\begin{figure}[ht]
\centering
\includegraphics[width=7cm]{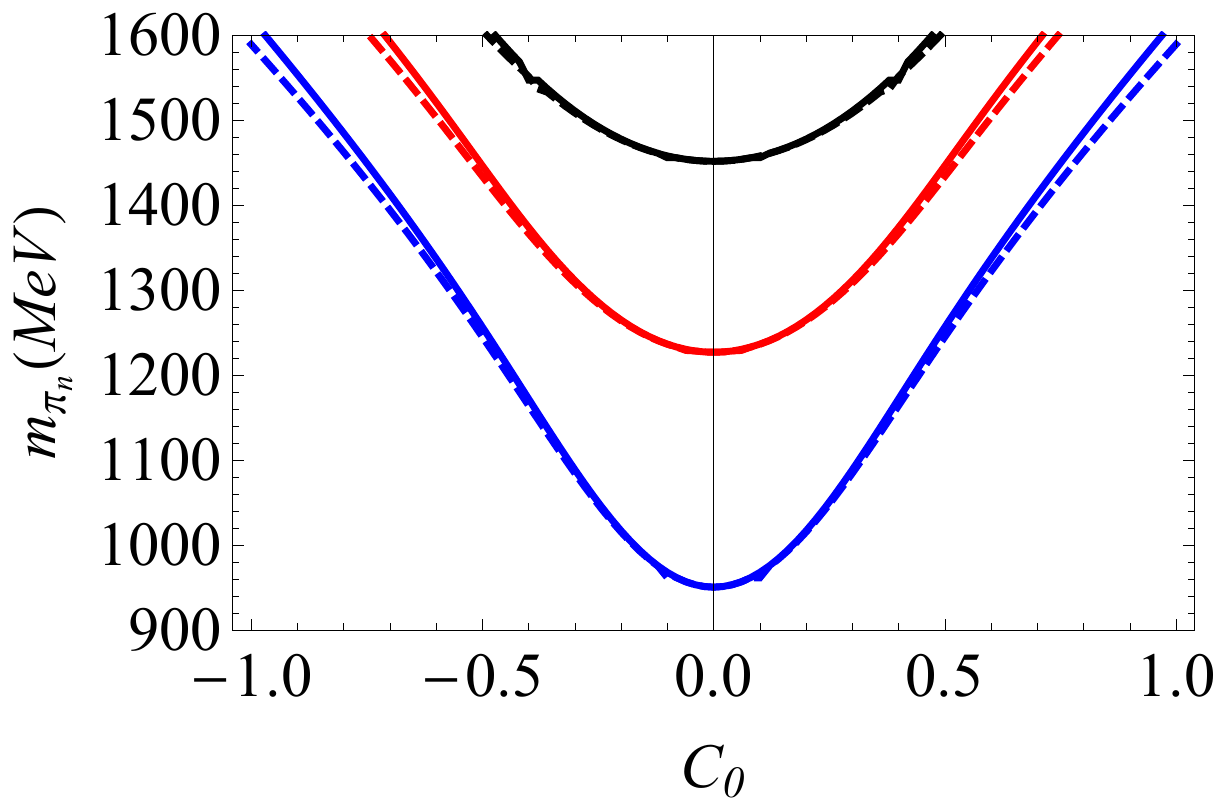}
\hfill
\includegraphics[width=7cm]{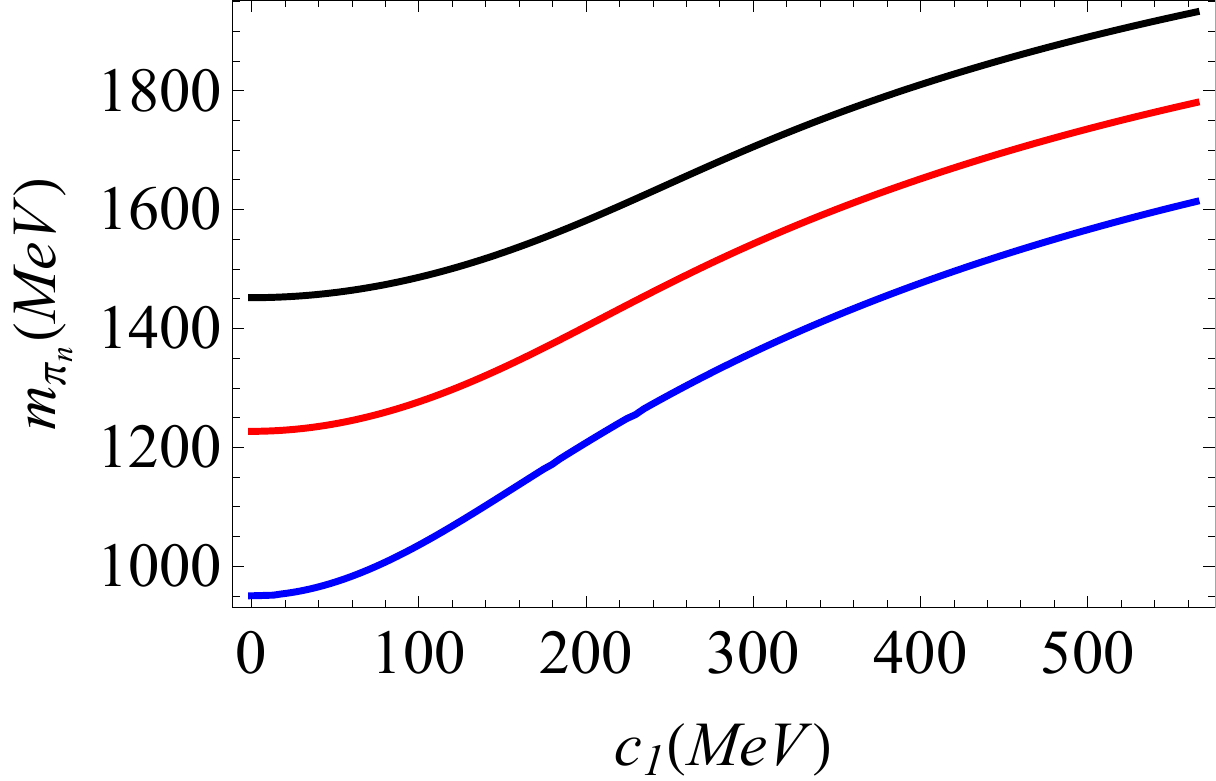}
\caption{
The masses of the pseudoscalar mesons as a function of $C_0$ (left panel) and $c_1$ (right panel). The source parameter $c_1$ is related to the quark mass by $c_1 = m_q \zeta$ with $\zeta=\sqrt{N_c}/(2\pi)$. Solid lines represent the result for $\lambda=2$, while dashed lines for $\lambda=0$. The results were obtained setting $\phi_{\infty}=(388\,\text{MeV})^2$.
}
\label{Fig:PionMassC1Lneg}
\end{figure}

The pseudoscalar sector is described by the coupled equations \eqref{Eq:EqPion1}-\eqref{Eq:EqPion2}. The system was decoupled and written in a Schrödinger form in \eqref{Eq:SchroPion} and this time we consider $\lambda>0$. Fig.~\ref{Fig:PionMassC1Lneg} shows our numerical results for the pseudoscalar meson masses as functions of the parameters $C_0$ (left panel) and $c_1$ (right panel). Again, we observe that the masses increase with the quark mass parameter $c_1$. Note that in the chiral limit $c_1\to 0$ the mass of the first pseudoscalar state has a finite value. This result can be interpreted as the absence of pseudo NG bosons in the spectrum and supports the background analysis leading to a vanishing chiral condensate in the chiral limit.

As described previously, the potential well in the Schrödinger potential of the pseudoscalar sector, shown on the left panel of  Fig. \ref{Fig:PionPotential}, is not deep enough  to support the presence of a very light state. Therefore, the first state behaves as a pion resonance instead of a true pseudo NG boson.  This is contrast with the Schrödinger potential of \cite{Gherghetta:2009ac,Kelley:2010mu}, shown on the right panel of Fig. \ref{Fig:PionPotential}, which is deep enough  to allow for a pseudo NG boson. We will show in  section \ref{Sec:DecayConstantLneg} that all the  decay constants of the pseudoscalar mesons in our model go to zero in the chiral limit  $c_1 \to 0$, characterising them as resonances. 

What new feature does  the holographic model in ~\cite{Gherghetta:2009ac} have to allow for a pseudo NG mode \cite{Kelley:2010mu}? It is worth stressing that the main difference lies in the behavior of the dilaton field. In this model we consider a monotonically increasing positive dilaton whereas the dilaton profile in  ~\cite{Gherghetta:2009ac} is negative in the UV and positive in the IR  with a minimum lying in the intermediate region, see Fig.~\ref{Fig:DilGKK} of Appendix \ref{Sec:GKK}. This difference in the dilaton profile has an important consequence in the tachyon solution near the boundary. In  the chiral limit, the tachyon solution in our model vanishes whilst the tachyon profile in \cite{Gherghetta:2009ac,Kelley:2010mu} is still nonzero. These two important differences have an effect on the Schrödinger potential 
that allows for a pseudo NG mode in ~\cite{Gherghetta:2009ac,Kelley:2010mu}. We remark, however, that we have followed a more conservative approach of taking a positive monotonically increasing dilaton to avoid instabilities in the meson spectrum. As described in appendix \ref{Sec:GKK}, the dilaton profile in \cite{Gherghetta:2009ac} leads to instabilities in the scalar sector. It would be interesting if the model of \cite{Gherghetta:2009ac} allowed for an extension that avoids those instabilities.

We also note from Fig.~\ref{Fig:ScalarMassC1Lneg} and  Fig.~\ref{Fig:PionMassC1Lneg} that in the chiral limit $c_1 \to 0$ the masses of the pseudoscalar and scalar mesons reach the same values. We therefore conclude that the scalar and pseudoscalar mesons become degenerate in the chiral limit.  As explained previously, the vector and axial-vector mesons also become degenerate in the chiral limit. 

\noindent
\noindent

The spectrum obtained using the parameters $\lambda=7$ and $C_0=0.3$, fixed in the scalar sector, is displayed in Table~\ref{Taba:03Lneg}. Our results, labeled as NLSW, are compared against the linear soft wall model \cite{Karch:2006pv}, the holographic model of \cite{Gherghetta:2009ac,Kelley:2010mu}  and experimental data \cite{Tanabashi:2018oca}. We also point out that the masses of scalar and pseudoscalar mesons shown in Tables \ref{Taba:01Lneg} and \ref{Taba:03Lneg} are not degenerate (the parameter $c_1$ is outside the chiral limit).
\begin{table}[ht]
\centering
\begin{tabular}{l |c|c|c|l}
\hline 
\hline
 $n$ & NLSW & SW \cite{Karch:2006pv}&
 KBK \cite{Kelley:2010mu}&
$\pi$ experimental \cite{Tanabashi:2018oca} \\
\hline 
 $1$ & -  & -  & 144    & $140$  \\
 $2$ & 1100 & 951 & 1557 & $1300\pm 100$  \\
 $3$ & 1321 & 1227 & 1887 & $1816\pm 14$ \\
 $4$ & 1518 & 1452 & 2090 & $2070$  \\
 $5$ & 1697 & 1646 & 2270 & $2360$  \\
 $6$ & 1861 & 1820 & 2434 &  \\
 $7$ & 2013 & 1980 & 2586 &  \\
\hline\hline
\end{tabular}
\caption{
The masses of the pseudoscalar mesons (in MeV) obtained in the 
nonlinear soft wall model with $\lambda>0$, compared against the linear soft wall model \cite{Karch:2006pv}, the holographic model of \cite{Kelley:2010mu}  and experimental data \cite{Tanabashi:2018oca}. The value of the parameters are 
$\lambda=7$ and $C_0=0.3$.
}
\label{Taba:03Lneg}
\end{table}

\begin{figure}[ht]
\centering
\includegraphics[width=7cm]{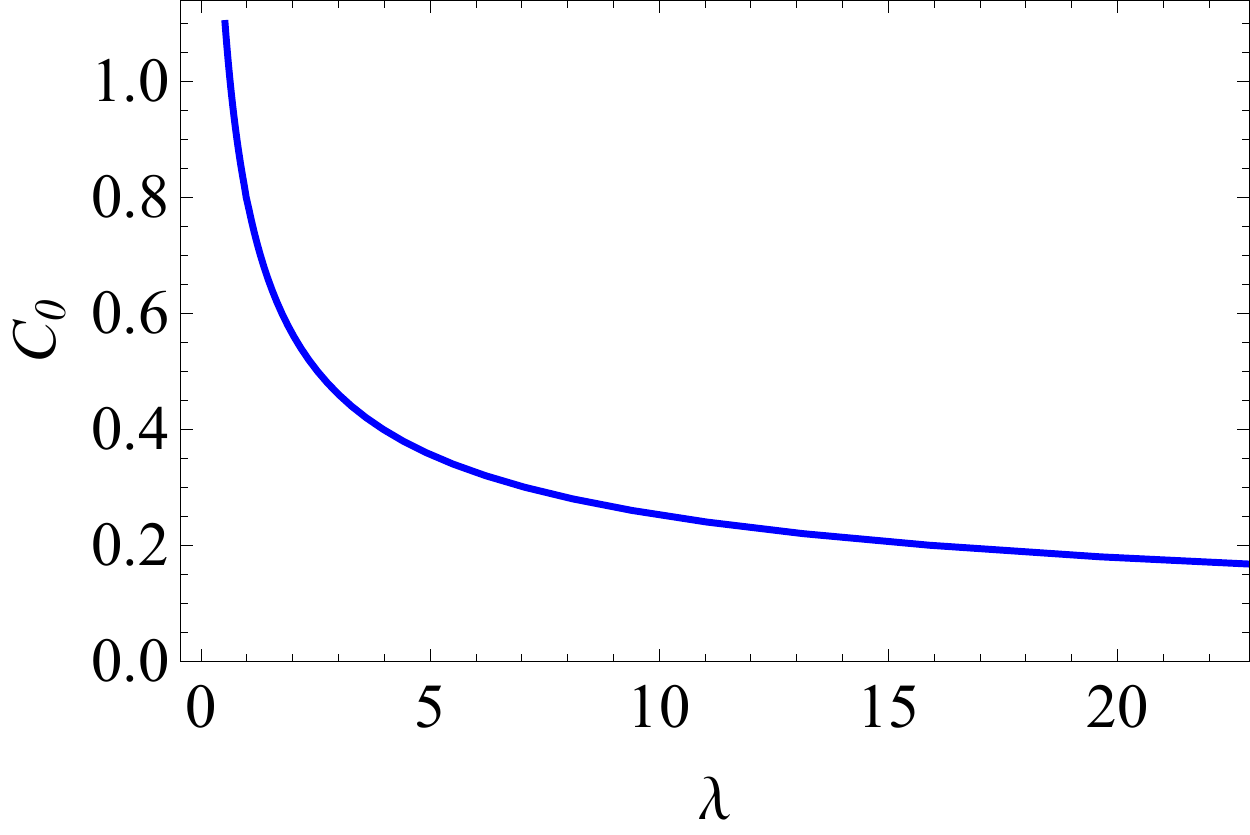}
\hfill
\includegraphics[width=7cm]{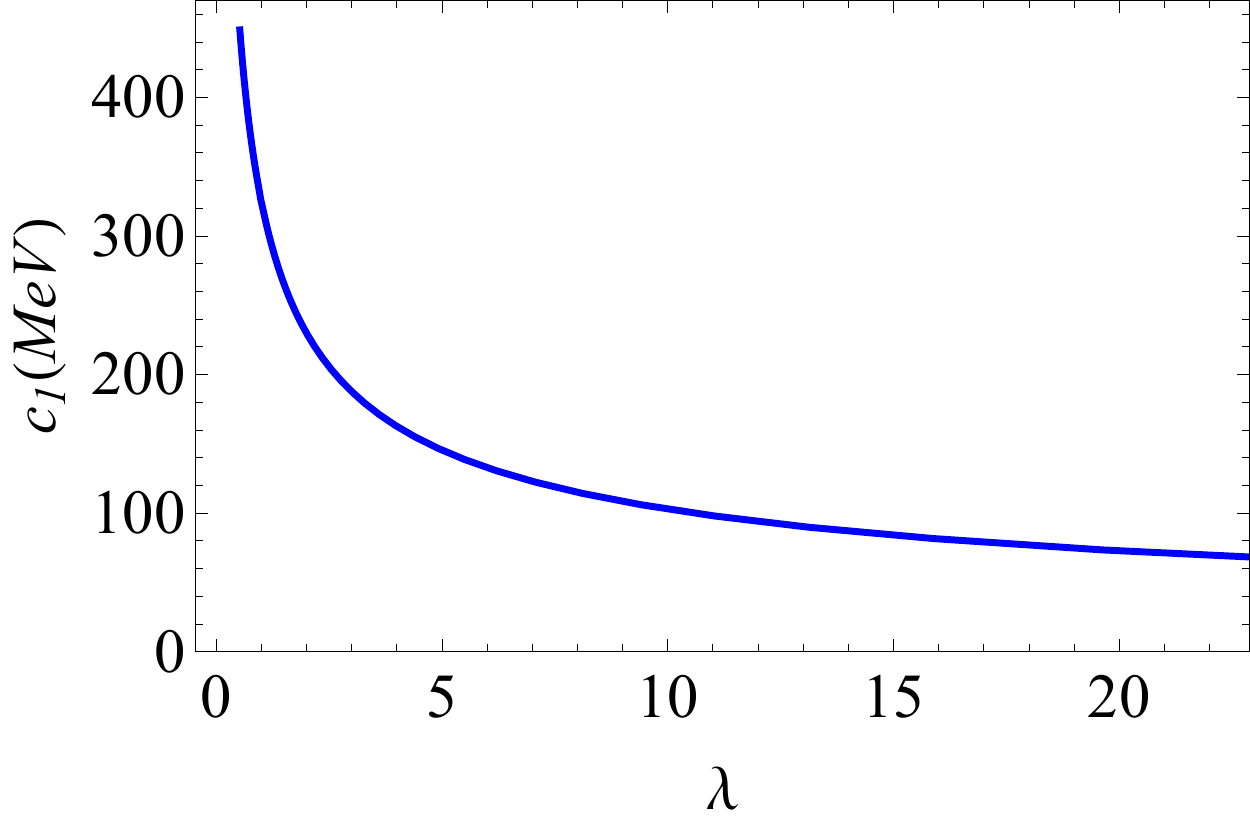}
\caption{
The evolution of $C_0$ as a function of $\lambda$ (left panel) and $c_1$ as a function of $\lambda$ (right panel), when matching the first scalar state to the meson $f_0(980)$. The source parameter $c_1$ is related to the quark mass by $c_1 = m_q \zeta$ with $\zeta=\sqrt{N_c}/(2\pi)$.  We remind the reader that $\phi_{\infty}=(388\,\text{MeV})^2$, fixed in the vector sector.}
\label{Fig:C0C1Lneg}
\end{figure}

We finish this subsection showing in Fig.~\ref{Fig:C0C1Lneg} the evolution of the parameters $C_0$  (left panel) and $c_1$ (right panel) when varying $\lambda$, matching the first scalar state to the meson $f_0(980)$. In this figure we observe why it is not possible to get small values for $c_1$ when $\lambda$ is small. To get a small quark mass, for example, $m_q=8 \text{MeV}$, we would need $c_1=2.21 \text{MeV}$, and therefore the value of $\lambda$ would be very large, i.e., $\lambda \approx 3\times 10^{4}$, with the corresponding value for $C_0 \approx 4.7\times 10^{-3}$. Those results would have dramatic consequences in the spectrum, because the IR parameter $C_0$ would be so small that the contribution to  the Schrödinger equations would be negligible and the spectrum of scalar and pseudoscalar mesons as well as  vector and axial-vector mesons would be degenerate.
\noindent
\noindent

\subsection{Decay constants}
\label{Sec:DecayConstantLneg}

\begin{figure}[ht]
\centering
\includegraphics[width=7cm]{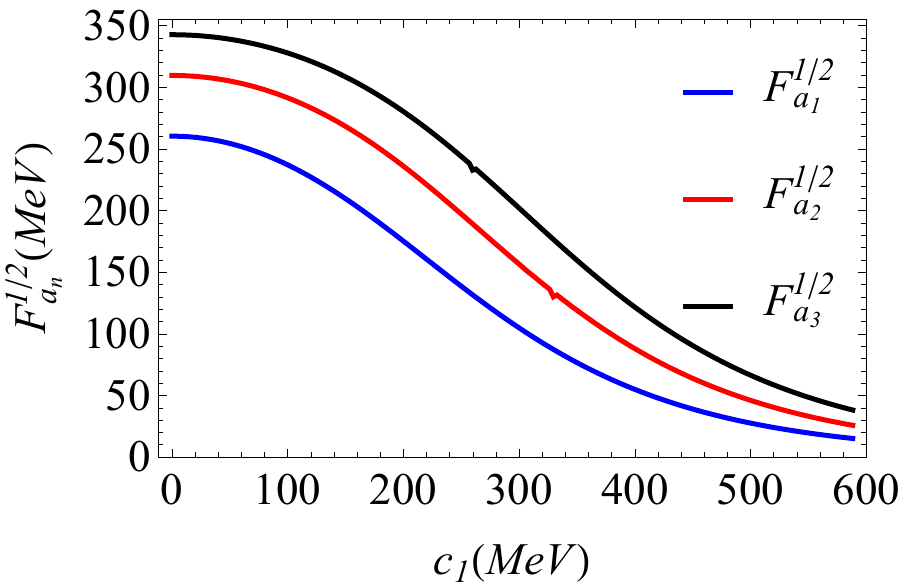}
\hfill
\includegraphics[width=7cm]{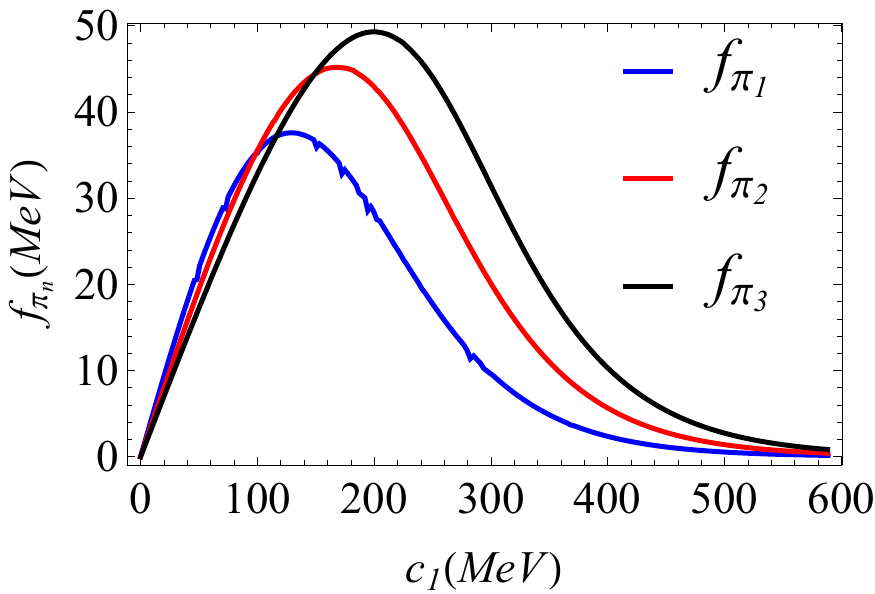}
\caption{
The decay constants of the axial-vector mesons (left panel) and pseudoscalar mesons (right panel) as a function of $c_1$ obtained in the NLSW model for $\phi_{\infty}=(388\,\text{MeV})^2$ and $\lambda=2$. The source parameter $c_1$ is related to the quark mass by $c_1 = m_q \zeta$ with $\zeta=\sqrt{N_c}/(2\pi)$.
}
\label{Fig:AVDecayL}
\end{figure}

In this subsection we calculate the decay constants of the vector, axial-vector and pseudoscalar mesons. In holographic QCD models the meson decay constants are related to the normalisation constants for the field perturbations, see e.g. \cite{Ballon-Bayona:2017bwk}. The normalisation condition for the vector field is given by 
\noindent
\begin{equation}
\int dz\, e^{A_s-\Phi}v_m(z)\,v_n(z)=\delta_{mn}, \label{Eq:Normvec}
\end{equation}
\noindent
where $A_s = - \ln z$ is the AdS warp factor, $\Phi(z) = \phi_{\infty} z^2$ the dilaton and $v_n$ is the vector meson mode related to the wave function by $v_n(z)=\,e^{-B_{V}}\psi_{v_{n}}(z)$. In turn, the wave function $\psi_{v_n}$ satisfies the Schrödinger equation \eqref{Eq:SchrodingerVector}. As described in appendix \ref{Sec:4daction}, the decay constants are related to 4d conserved currents and they are defined through the relations in Eq. \eqref{Eq:DecayConstDic}
\noindent
\begin{equation}
F_{v_n}=-\lim_{\epsilon\to 0}\frac{e^{A_s-\Phi}}{g_5}\,\partial_z v_n\bigg{|}_{z=\epsilon}=\frac{2}{g_5}N_{v_n}.
\end{equation}
\noindent
The normalisation constant $N_{v_n}$ appears as the UV coefficient near the boundary of a vector mode, i.e.  $v_n(z)= N_{v_n} z^2 + \dots $ satisfying the normalisation condition \eqref{Eq:Normvec}. Thus,  the decay constants are proportional to the normalisation constants of the field perturbations. We follow the same procedure for the axial-vector sector, where the normalisation condition is given by
\noindent
\begin{equation}
\int dz\, e^{A_s-\Phi}a_m(z)\,a_n(z)=\delta_{mn}, \label{Eq:NormAxial}
\end{equation}
\noindent
where $a_m(z)=\,e^{-B_{A}}\psi_{a_{n}}(z)$ is the axial vector model satisfying the UV behaviour $a_m(z)=N_{a_n} z^2 + \dots $ and the normalisation condition \eqref{Eq:NormAxial}. $N_{a_n}$ is the normalisation constant and $\psi_{a_n}$ is the solution of the Schrödinger equation \eqref{Eq:AVSchroEq}. Thus, the decay constants are given by
\noindent
\begin{equation}
F_{a_n}=-\lim_{\epsilon\to 0}\frac{e^{A_s-\Phi}}{g_5}\,\partial_z a_n\bigg{|}_{z=\epsilon}=\frac{2}{g_5}N_{a_n}.
\end{equation}
\noindent
The problem of finding decay constants have reduced to the problem of calculating normalisation constants of the field perturbations (or equivalently  wave functions). In the left panel of Fig.~\ref{Fig:AVDecayL} we display the results for the decay constants of axial-vector mesons as functions of the quark mass parameter $c_1$ .The decay constants do not vary significantly in the region of small $c_1$ but they decrease fast at large $c_1$; approximately as $F^{1/2}_{a_1}\sim 1/c_1^{b}$, with $b=2.6$, for the first state. In Table \ref{Taba:DecayConstantDSW} we display our numerical results for the ground state $(n=1)$ decay constants for the parameter choice $\lambda=7$ and $C_0=0.3$. 
\begin{table}[ht]
\centering
\begin{tabular}{l |c|c|l}
\hline 
\hline
 & NLSW $(\lambda>0)$ & SW \cite{Karch:2006pv} &  experimental \cite{Tanabashi:2018oca} \\
\hline 
 $F_{\rho}^{1/2}$ & 260.12 & 261 & $346.2\pm 1.4$  \\
 $F_{a_1}^{1/2}$  & 215.37 & 261 & $433\pm 13$ \\
\hline\hline
\end{tabular}
\caption{
The decay constants (in MeV) obtained in the nonlinear soft wall model, compared against the result obtained in the linear soft wall model ~\cite{Karch:2006pv} and experimental results of PDG \cite{Tanabashi:2018oca}. The results for $\lambda>0$ were obtained setting $\lambda=7$ and $C_0=0.3$.
}
\label{Taba:DecayConstantDSW}
\end{table}

For the pseudoscalar sector the normalisation condition is given by \cite{Ballon-Bayona:2017bwk}
\noindent
\begin{equation}
\int dz\, e^{A_s-\Phi}\beta(z)\left(\partial_z\pi_m\right)\left(\partial_z\pi_n\right)=m_{\pi_n}^2\delta_{mn}.
\end{equation}
\noindent
There are different approaches to calculate the decay constants of  pseudoscalar mesons in the literature. First of all, we consider the prescription used in  hard wall model \cite{Erlich:2005qh} (see also \cite{Kwee:2007dd,Kwee:2007nq}). The authors considered the massless case, $m_{\pi}^2=0$. In that limit the pion decay constant can be extracted from the axial current correlator  by the relation 
\noindent
\begin{equation}\label{Eq:DecayErlich}
f^2_{\pi}=-\lim_{\epsilon\to 0}\frac{e^{A_s-\Phi}}{g_{5}^{2}}\partial_z\,A(0,\epsilon),
\end{equation}
\noindent
where $A(0,\epsilon)$ is the non-normalisable solution for the axial-vector field dual to the 4d axial-vector current. We point out that the prescription is valid only in the case $m_{\pi}^2=0$ and does not allow the investigation of pion resonances. 

A prescription for calculating decay constants for the pion and their resonances was developed in \cite{Ballon-Bayona:2014oma}. Details on the derivation are given in appendix \ref{Sec:4daction}. The holographic dictionary for the decay constant maybe written in the form  
\noindent
\begin{equation}\label{Eq:DecayPi1}
f_{\pi_n}=-\lim_{\epsilon\to 0}\frac{e^{A_s-\Phi}}{g_5}\partial_z\varphi_n(z)\bigg{|}_{z=\epsilon},
\end{equation}
\noindent
where $\varphi_n(z)$ is the normalised wave function satisfying the normalisation condition
\noindent
\begin{equation}
\int dz\,\frac{e^{A_s-\Phi}}{\beta(z)}\left(\partial_z\varphi_m\right)\left(\partial_z\varphi_n\right)=\delta_{mn}.
\end{equation}
\noindent
In terms of $\partial_z\pi_n$ the decay constant is obtained by plugging Eq.~\eqref{Eq:EqPion2} into Eq.~\eqref{Eq:DecayPi1} 
\noindent
\begin{equation}\label{Eq:DecayPi2}
f_{\pi_n}=-\lim_{\epsilon\to 0}\frac{e^{A_s-\Phi}}{g_5\,m_{\pi_n}^{2}}\beta(z)\partial_z\pi_n(z)\bigg{|}_{z=\epsilon}.
\end{equation}
\noindent
Hence, the normalisation condition takes the form 
\noindent
\begin{equation}
\int dz\,e^{A_s-\Phi}\beta(z)\left(\partial_z\pi_m\right)\left(\partial_z\pi_n\right)=m_{\pi_n}^{2}\delta_{mn}.
\end{equation}
\noindent
Finally, in terms of the Schrödinger wave function defined in  Eq.~\eqref{Eq:SchroPion}, the decay constant takes the form 
\noindent
\begin{equation}
f_{\pi_n}=-\lim_{\epsilon\to 0}\frac{e^{(A_s-\Phi)/2}}{g_5\,m_{\pi_n}^{2}}\beta^{1/2}(z)\psi_{\pi_n}(z)\bigg{|}_{z=\epsilon},
\end{equation}
\noindent
where the normalisation condition is given by
\noindent
\begin{equation}
\int dz\,\psi_{\pi_m}(z)\,\psi_{\pi_n}(z)=m_{\pi_n}^{2}\delta_{mn}.
\end{equation}
\noindent
At the end of the day, the decay constant depends only on the normalisation constant. Thus, the procedure above allows us to calculate  decay constants of the fundamental state and its resonances. In our calculations for the pseudoscalar mesons we have used the three formulae described above to show the consistency of our numerical results. In the right panel of Fig.~\ref{Fig:AVDecayL} we plot the decay constants as a function of the parameter $c_1$ of the first three pseudoscalar mesons. In the chiral limit the decay constants of all  pseudoscalar states go to zero linearly, i.e.  $f_{\pi_n} \sim \, c_1$, which agrees with the observed result for pion resonances in the hard wall model, cf. ~\cite{Ballon-Bayona:2017bwk} and in QCD cf. \cite{Holl:2004fr} (see also ~\cite{Krassnigg:2006ps}). On the other hand, as the quark mass increases the pseudoscalar meson decay constants display a non-monotonic behaviour and, in particular, go to zero in the heavy quark limit $c_1 \to \infty$.  It is worth mentioning that in the perturbative QCD approach for heavy quarks \cite{Manohar:2000dt} one expects the behaviour $f\sim 1/\sqrt{M}$ for meson decay constants, see also ~\cite{Maris:2005tt}, where $M$ is the mass of the heavy mesons. In our case, the numerical data displayed in the right panel of Fig.~\ref{Fig:AVDecayL} indicates the approximate behaviour $f_{\pi_1}\sim 1/c_1^{a}$, with $a=4.4$ for the first pseudoscalar state.
\noindent
\noindent

Finally, we note in the right panel of Fig.~\ref{Fig:AVDecayL} a crossing of the different curves. The hierarchy between the decay constants change when going from the regime of small quark mass   to the regime of heavy quark mass. Numerical results are displayed in Table \ref{Taba:DecayConstantDSWPionL}, for $\lambda=7$ and specific values of $C_0$. We see that the  decay constants increase  with the radial number for $C_0=0.4$, corresponding to $c_1=203.68\text{MeV}$, whereas for the case $C_0=0.1$ , corresponding to $c_1=44.17 \text{MeV}$, the decay constants decrease with the radial number. In the regime of small quark mass, the hierarchy $f_{\pi_1} > f_{\pi_2} > f_{\pi_3} $ was found in 
~\cite{Ballon-Bayona:2014oma} for pion resonances in the hard wall model. We have found the same hierarchy in the (nonlinear) soft wall model in the regime of small quark mass and, interestingly, we have found an inversion of that hierarchy in the regime of heavy quark mass.
\begin{table}[ht]
\centering
\begin{tabular}{l |c|c|l}
\hline 
\hline
 & $f_{\pi_1}$ & $f_{\pi_2}$ & $f_{\pi_3}$ \\
\hline 
 NLSW $(C_0=0.4)$  & 27.20 & 40.84 & 47.36\\
 \hline 
 NLSW $(C_0=0.1)$  & 19.63 & 17.25 & 15.25\\
\hline\hline
\end{tabular}
\caption{
The decay constants (in MeV) obtained in the nonlinear soft wall model. The results for $\lambda=7$ were obtained setting $C_0=0.4$ $(c_1=203.68\text{MeV})$,  $C_0=0.1$ $(c_1=44.17\text{MeV})$ and $\phi_{\infty}=(388\text{MeV})^2$.
}
\label{Taba:DecayConstantDSWPionL}
\end{table}

\section{Nonlinear soft wall models with running mass}
\label{Sec:NLSWRM}

Recent works in holographic soft wall models have considered the possibility of a tachyon squared mass $m_X^2$ depending on the radial coordinate $z$ \cite{Vega:2010ne,Fang:2016nfj}. The motivation for a tachyon running mass was to gain a nontrivial IR contribution to the tachyon differential equation and therefore find a richer dynamics. In holographic QCD a 5d running mass for the tachyon would correspond to the anomalous dimension for the 4d quark mass operator \cite{Jarvinen:2011qe,Fang:2016nfj}. 

For the tachyon running mass we take the ansatz
\noindent
\begin{equation}
m_{X}^{2}(z)=-3-\phi_{c}\,z^2 \, ,
\label{Eq:runningmass}
\end{equation}
\noindent
with $\phi_C>0$. The tachyon differential equation now becomes 
\begin{equation}
\Big [ z^2 \partial_z^2 - (3 + 2 \phi_{\infty} z^2)  z \partial_z  + 3  + \phi_c z^2 \Big ] v - \frac{\lambda}{2} v^3 = 0 \,. 
\label{Eq:TachyonRM}
\end{equation}

\subsection{Asymptotic analysis}

In the UV we consider again the Frobenius ansatz
\noindent
\begin{equation}
v(z)=c_1 z + d_3 z^3 \ln z + c_3 z^3 + d_5 z^5 \ln z + c_5 z^5 + \dots
\end{equation}
Plugging this ansatz into \eqref{Eq:TachyonRM} we find the UV coefficients
\begin{align}
d_3 &=  \frac14 {c_1} \left (c_1^2 \lambda + 4 \phi_{\infty} - 2 \phi_c \right )
\, , \, \, 
d_5 = \frac{1}{64} c_1 \left (-c_1^2 \lambda - 4 \phi_{\infty} + 2 \phi_c \right ) \left ( - c_1^2 \lambda - 12 \phi_{\infty} + 2 \phi_c \right )  \, , \nonumber \\
c_5 &=   \frac{1}{256}   \Big [ - 9 c_1^5 \lambda^2  - ( - 24 c_1^3 \phi_c + 56 c_1^3 \phi_{\infty} +48 c_1^2 c_3 ) \lambda \nonumber \\
&- 12 c_1 \phi_c^2 + 64 c_1 \phi_c \phi_{\infty} - 80 c_1 \phi_{\infty}^2 - 32 c_3 \phi_c - 192 c_3 \phi_{\infty} \Big ]   \, , \, \dots 
\end{align}
In the special case $c_1^2 \lambda + 4 \phi_{\infty} - 2 \phi_c=0$ we have $d_3 = d_5 =0$ and  $c_5 = \frac14 \phi_c c_3$ so we do not expect logarithmic terms but there maybe another solution besides the linear solution $v(z) = c_1 z$ due to a nonzero $c_3$.

In the IR we work with the variable $y=1/z$. Eq.~\eqref{Eq:TachyonRM} becomes 
\noindent
\begin{equation}
\Big [ (y\partial_y)^2 + 2(2+\phi_{\infty}\,y^{-2})(y\partial_y)+(3+\phi_{c}\,y^{-2}) \Big ]v
-\frac{\lambda}{2}v^3=0 \, . \label{Eq:TachyonRMv2}
\end{equation}
We take again the power ansatz
\begin{equation}
v^{IR}(y) = C_0 y^{\alpha} \,. 
\end{equation}
Plugging this ansatz into \eqref{Eq:TachyonRMv2} we obtain the polynomial equation
\begin{equation}
C_0 \, y^{\alpha} ( \alpha^2 + 4 \alpha +3 ) + C_0 \, y^{\alpha-2} (2 \alpha \phi_{\infty}+ \phi_c) - \frac{\lambda}{2} C_0^3 \, y^{3 \alpha} = 0 \,.  
\label{Eq:IRPolynRM}
\end{equation}
Again we distinguish 3 cases: $\alpha>-1$, $\alpha=-1$ and $\alpha<-1$. The latter case is trivial because it leads to $C_0=0$.

In the case $\alpha>-1$ the second term dominates and we find 
\begin{equation}
\alpha = - \frac{\phi_c}{2 \phi_{\infty}} \quad , \quad 
\phi_c < 2 \phi_{\infty}   \,.     
\end{equation}
This is a natural deformation of the regular solution found in the previous section. For $0 >\alpha >-1$ the solution is actually divergent and admits the expansion
\begin{equation}
v^{IR}(y)=  y^{\alpha} \left (C_0 + C_2 \, y^\beta  + \dots \right ) \quad , \quad \beta = 2 - \frac{\phi_c}{\phi_{\infty}} \, , 
\label{Eq:IRsolRM}   
\end{equation}
with 
\begin{equation}
C_2 =  - \frac{C_0^3 \lambda}{4 (\phi_c - 2 \phi_{\infty})} \,.
 \label{IRcoeffsRM}
\end{equation}
We will focus on the case $\phi_{c}=\phi_{\infty}$ where, in terms of $z$, the tachyon solution reads 
\noindent
\begin{equation}
v=C_0 \sqrt{z} \left(1+\frac{C_0^2\,\lambda}{4\phi_{\infty}} z^{-1}
+\frac{3C_0^4\lambda^2-10\phi_{\infty}}{32\phi_{\infty}^2} z^{-2}
+\cdots \right) \, .\label{Eq:IRsolRMv2}
\end{equation}
\noindent
This is a divergent solution depending on only one parameter, $C_0$.
The IR leading behaviour $v\propto \sqrt{z}$ was considered as an IR constraint in a previous approach ~\cite{Sui:2009xe} 

In the special case $\alpha=-1$,  the first term in \eqref{Eq:IRPolynRM} vanishes whereas the second and third terms lead to the condition $-C_0^2 \lambda +2 \phi_c- 4 \phi_{\infty}=0$. This linear solution, i.e. $v(z) = C_0 \, z$ is valid only for $(\lambda<0 , \phi_c < 2 \phi_{\infty})$ or  ($\lambda>0 , \phi_c > 2 \phi_{\infty}$).

\medskip 

{\bf Running mass with $\lambda=0$}

\medskip 

We finish this subsection pointing out that the divergent solution  \eqref{Eq:IRsolRM} survives in the case $\lambda=0$. This corresponds to the linear soft wall model with running mass.  In the case $\phi_c=\phi_{\infty}$, this solution has the following UV and IR behaviour
\begin{eqnarray}
v^{UV}(z) &= c_1 z +
\frac{c_1\,\phi_{\infty}}{2}z^3\,\ln{z} + c_3 z^3 + '\cdots \, \, , \\
v^{IR}(z) &=  C_0 \sqrt{z} \left(1-\frac{5}{16\phi_{\infty}} z^{-2}
+  \cdots \right) \, .
\end{eqnarray}

\subsection{Numerical solution}

The numerical results for the nonlinear soft wall model in the presence of a tachyon running mass are qualitatively similar to the case without the running mass. The main effect of the running mass will be extending the range for the IR parameter $C_0$. In particular, for the case $\lambda>0$ the upper bound $ C_0^2 \lambda <6$ found in the previous section is not present anymore.

\begin{figure}[ht]
\centering
\includegraphics[width=7cm]{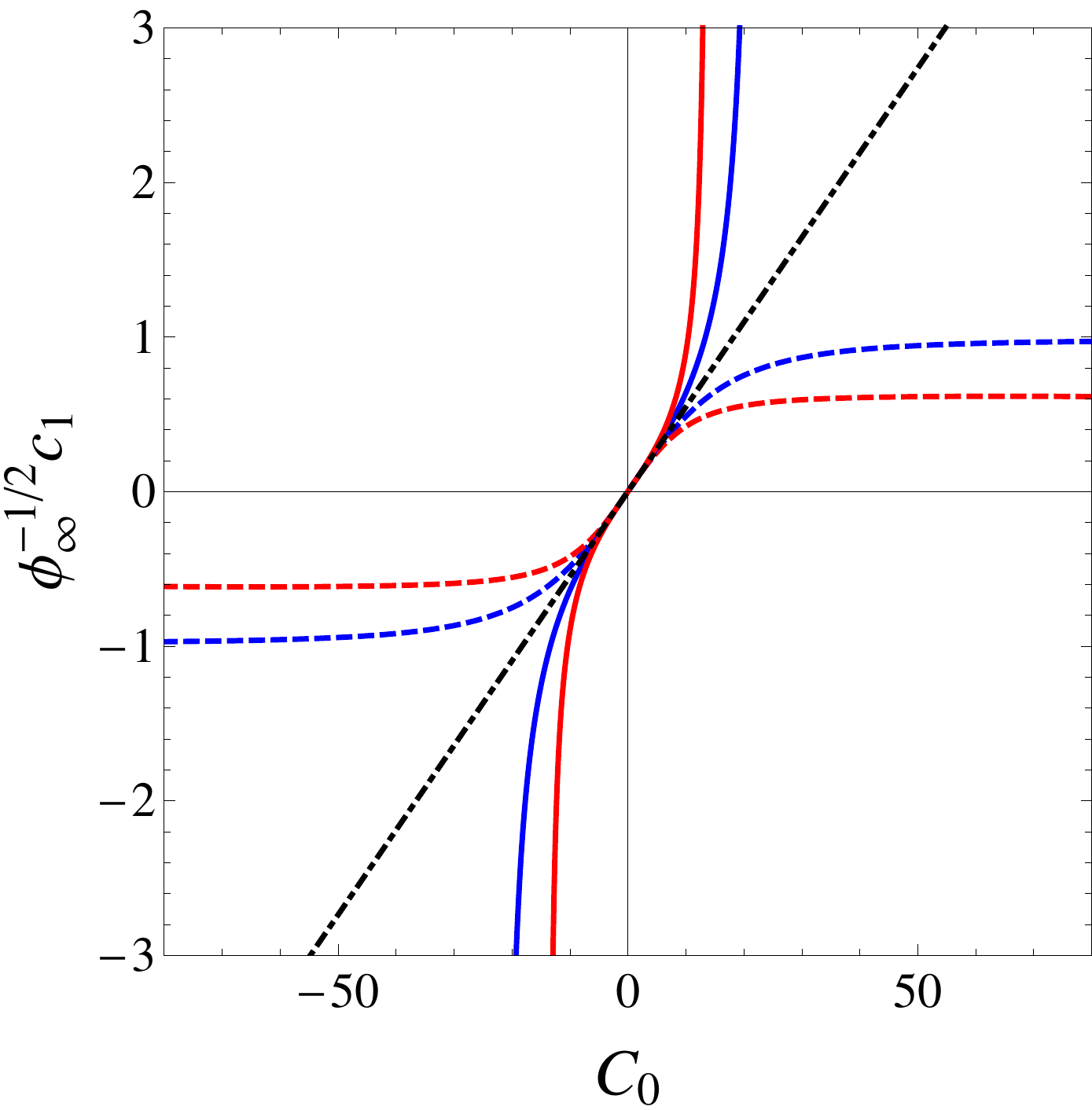}
\hfill
\includegraphics[width=7cm]{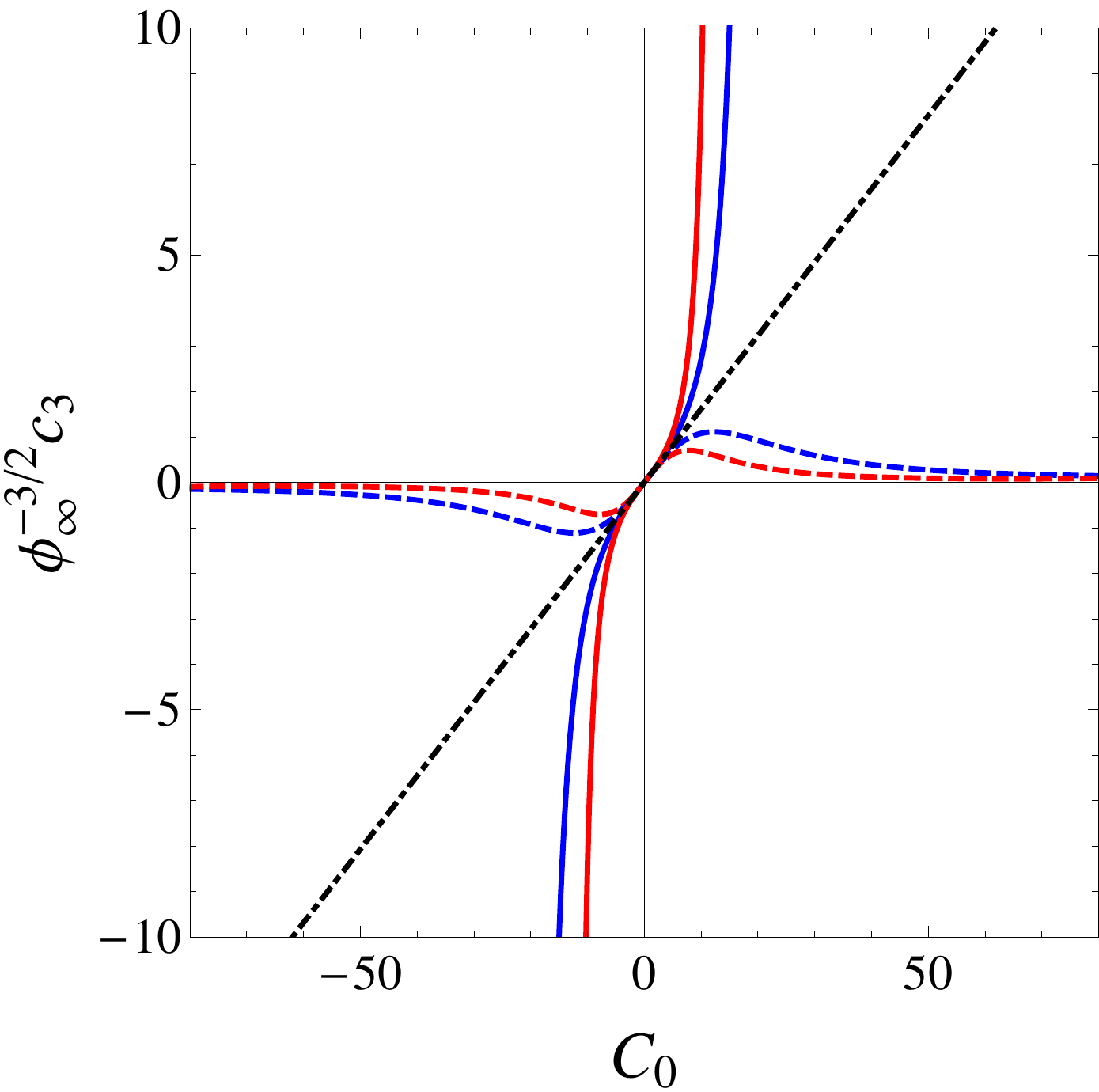}
\caption{Numerical results for $c_1$ (left panel) and $c_3$ (right panel) as functions of $C_0$ in units of $\sqrt{\phi_{\infty}}$. The blue and red solid lines (dashed lines) correspond to $\lambda=2$ and $\lambda=5$ ($\lambda=-2$ and $\lambda=-5$). The black dotdashed line represents the case $\lambda=0$ (linear soft wall model with running mass). }
\label{Fig:Plotc1c3vsC0run}
\end{figure}

\begin{figure}[ht]
\centering
\includegraphics[width=7cm]{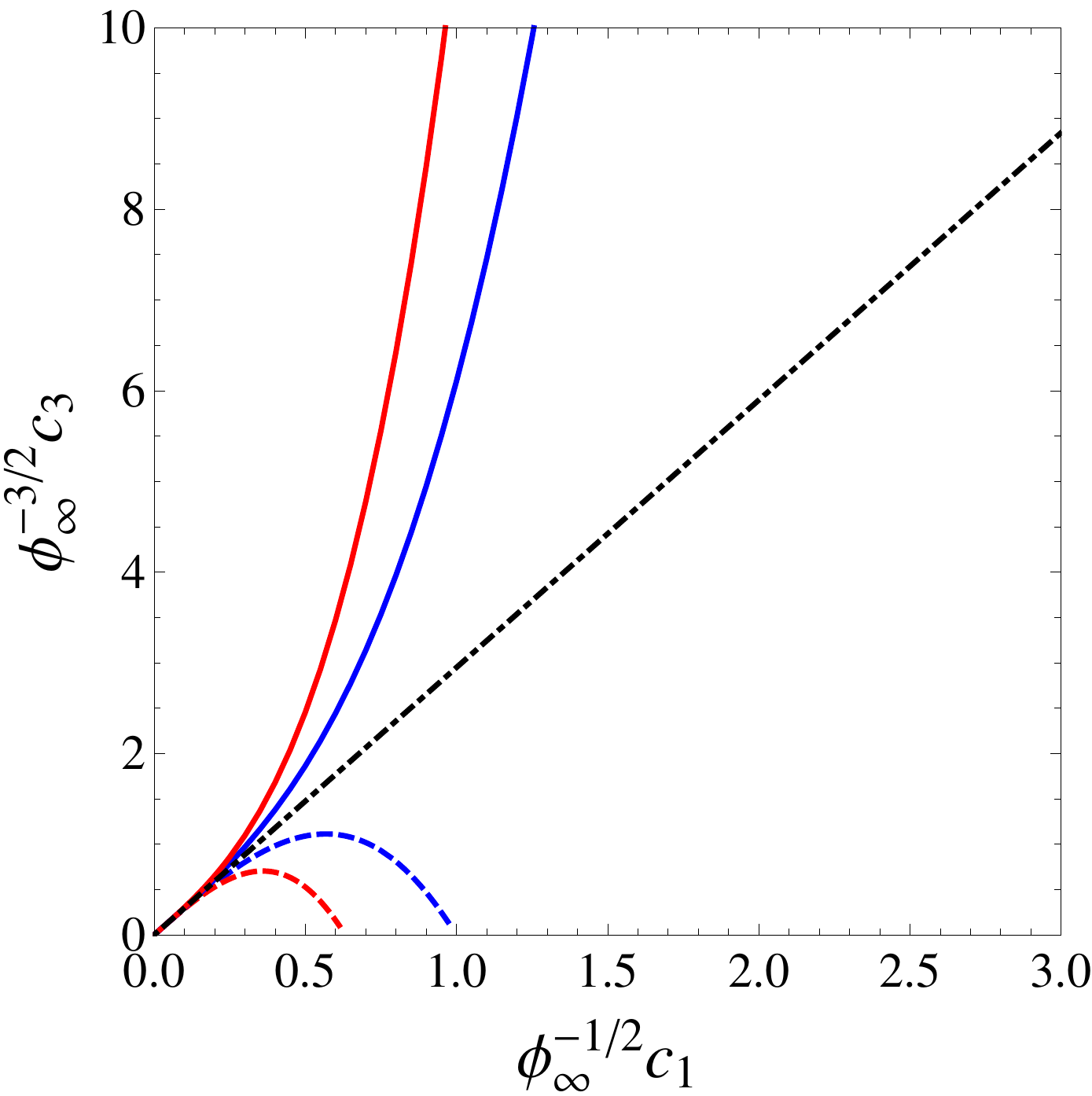}
\caption{The VEV parameter $c_3$ as a function of the source parameter $c_1$, in units of $\sqrt{\phi_{\infty}}$,  for different values of $\lambda$. The blue and red solid lines (dashed lines) correspond to $\lambda=2$ and $\lambda=5$ ($\lambda=-2$ and $\lambda=-5$). The black dotdashed line represents the case $\lambda=0$ (linear soft wall model with running mass) }
\label{Fig:Plotc3vsc1run}
\end{figure}

We present numerical results for the case $\phi_{c}= \phi_{\infty}$  corresponding to the IR behaviour \eqref{Eq:IRsolRMv2}. Fig.~\ref{Fig:Plotc1c3vsC0run}  displays the UV parameters $c_1$ and $c_3$ as functions of the IR parameter $C_0$ for different values of $\lambda$. The case $\lambda=0$, represented by  black dotdashed lines, corresponds to the (linear) soft wall model with a running mass for the tachyon. 
Fig.~\ref{Fig:Plotc3vsc1run} displays $c_3$ as a function of $c_1$ which can be interpreted in terms of the 4d chiral condensate as a function of the quark mass, as described in subsection \ref{subsec:Hologrenorm}. 
\noindent
\noindent
\noindent
\noindent
For the case $\lambda<0$ we note a decrease in the range of $c_1$ and $c_3$, when compared to the case without running mass described in the previous section. For the case $\lambda>0$, despite having a bigger $C_0$ range, we do not notice a significant difference in $c_3$ vs $c_1$ when compared to the case without the running mass. Again, we conclude that the case $\lambda>0$ provides the more realistic scenario for chiral symmetry breaking. 

One of the motivations of considering a 5d running mass for the tachyon was to gain a nontrivial dynamics in the IR depending on the parameter $\phi_c$ in \eqref{Eq:runningmass}. However, we have found that the numerical results were very similar, despite having this time a tachyon solution divergent in the IR. This may be related to the fact that the divergence is of the form $z^{\alpha}$ with $0<\alpha<1$.  As in the case without running mass, in the chiral limit $c_1 \to 0$ all the parameters go to zero. This is a negative result for nonlinear soft wall models because they not describe spontaneous symmetry breaking in the chiral limit, in constrast with  QCD.

\subsection{Meson Spectrum \texorpdfstring{($\lambda<0$)}{} }

As in the case without running mass, the vector meson is insensitive to the tachyon dynamics and therefore we focus on the scalar, axial-vector and pseudoscalar sectors. 

\subsubsection{Spectrum of the scalar sector}

\begin{figure}[ht]
\centering
\includegraphics[width=7cm]{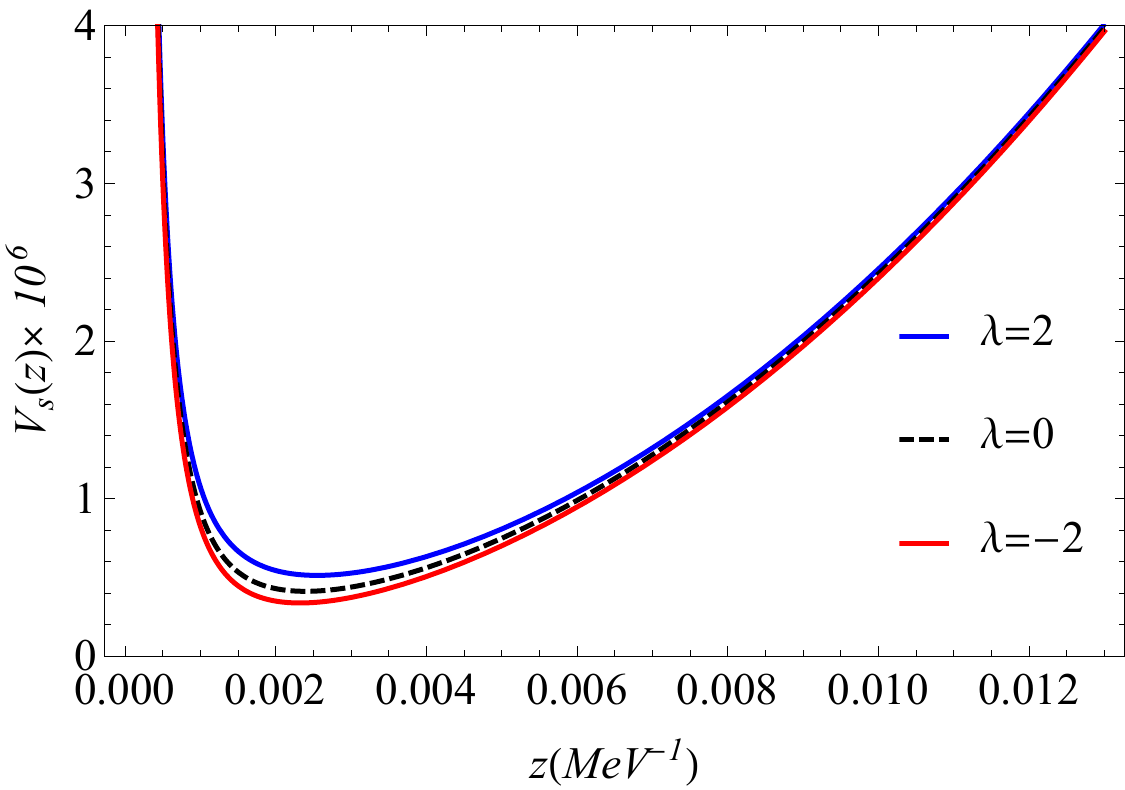}
\caption{The potential of the Schrödinger equation in the scalar sector for $\phi_{c}=\phi_{\infty}=(388\,\text{MeV})^2$ and three different values of the parameter $\lambda$.}
\label{Fig:PotScalarRM}
\end{figure}

The equation of the scalar sector is again \eqref{Eq:ScalarEq} but this time considering a running mass term $m_X^2(z)=-3-\phi_c\,z^2$. We focus on the case $\phi_{\infty}=\phi_c$, where the tachyon solution diverges in the IR region as $v\sim \sqrt{z}$. We expect a different behavior of the potential in the Schrödinger equation, i.e.,  \eqref{Eq:SchroScalarEq},  and consequently a different spectrum. A plot of the potential is displayed in Fig.~\ref{Fig:PotScalarRM} for different values of $\lambda$; we observe the main difference between the models with  negative, zero or positive $\lambda$. In the case $\lambda<0$  it is possible to find a very light state, just like in the model without running mass, cf. subection \ref{Sec:ScalarL}. However, as in that case, we follow a more conservative approach and consider $f_0(980)$ as the first scalar meson.
\noindent
\noindent

\begin{figure}[ht]
\centering
\includegraphics[width=7cm]{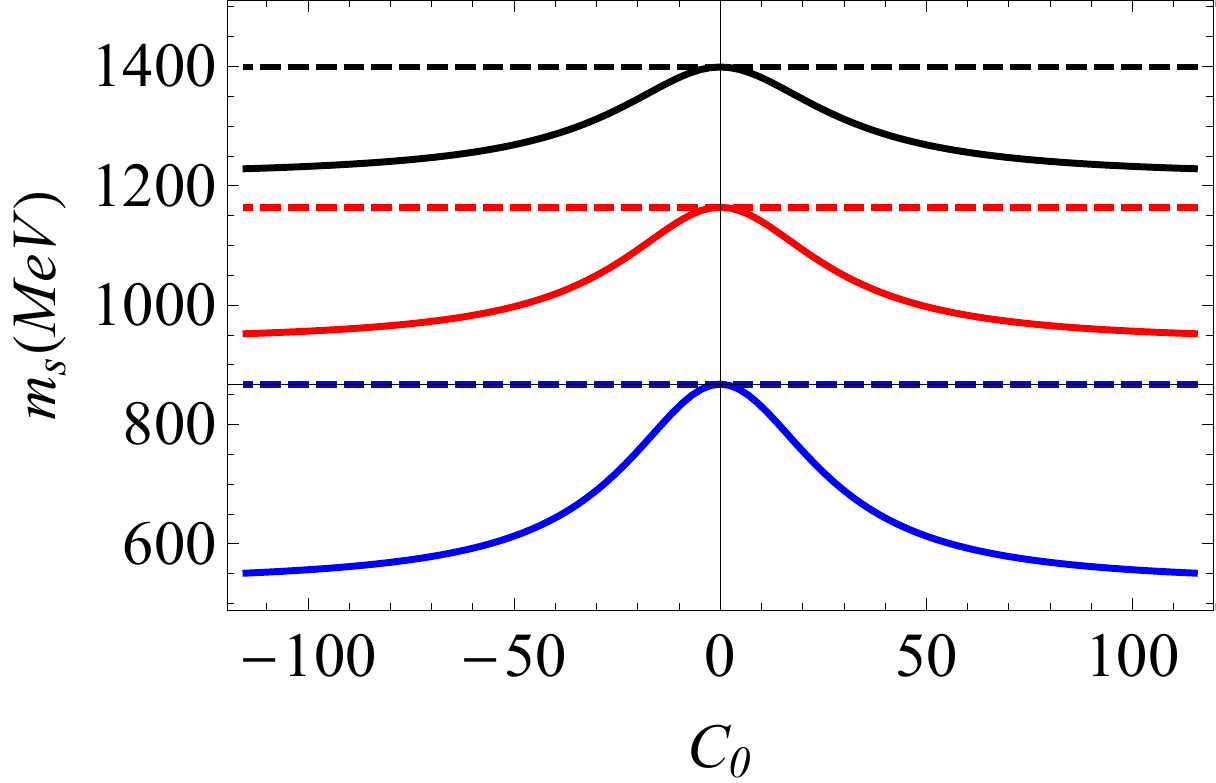}
\hfill
\includegraphics[width=7cm]{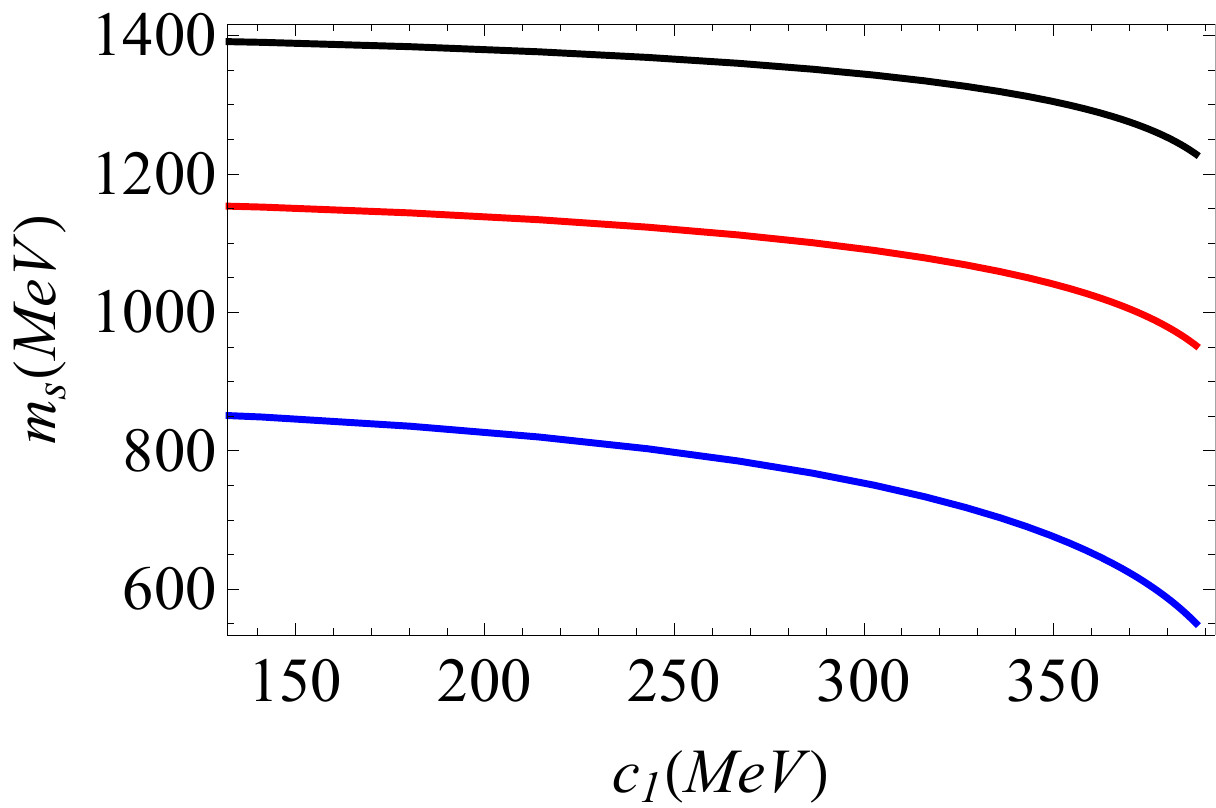}
\caption{
The mass of the scalar mesons as a function of $C_0$ (left panel) and $c_1$ (right panel). The source parameter $c_1$ is related to the quark mass by $c_1 = m_q \zeta$ with $\zeta=\sqrt{N_c}/(2\pi)$. Solid lines represent the result for $\lambda=-2$, while dashed lines for $\lambda=0$. The results were obtained setting $\phi_{\infty}=\phi_c=(388\,\text{MeV})^2$.
}
\label{Fig:ScalarMassDWRM}
\end{figure}

In Fig.~\ref{Fig:ScalarMassDWRM} we show the mass of the scalar mesons as a function of the IR parameter $C_0$ (left panel). From this figure we observe that the scalar meson masses decreases with $C_0$ (solid lines), suggesting the possibility of a light state in the spectrum. Right panel of Fig.~\ref{Fig:ScalarMassDWRM} shows the evolution of the scalar meson masses with the quark mass, i.e., $c_1\propto m_q$. All the masses decrease as the quark mass increases, which is not expected in QCD. This  pathology of the case $\lambda<0$ had arised previously in the model without running mass and maybe related to the absence of a minimum in the Higgs potential \eqref{Eq:Higgspot} when $\lambda<0$. 
\noindent
\noindent

\subsubsection{Spectrum of the axial-vector sector}

\begin{figure}[ht]
\centering
\includegraphics[width=7cm]{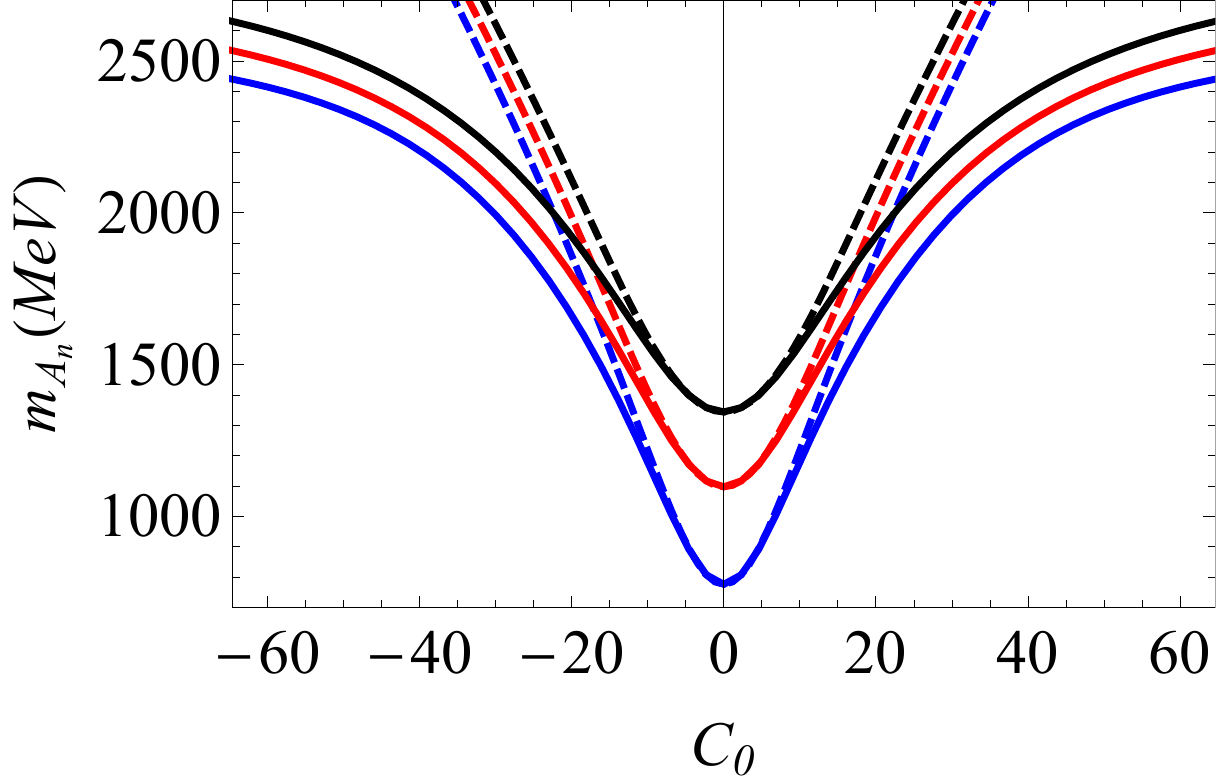}
\hfill
\includegraphics[width=7cm]{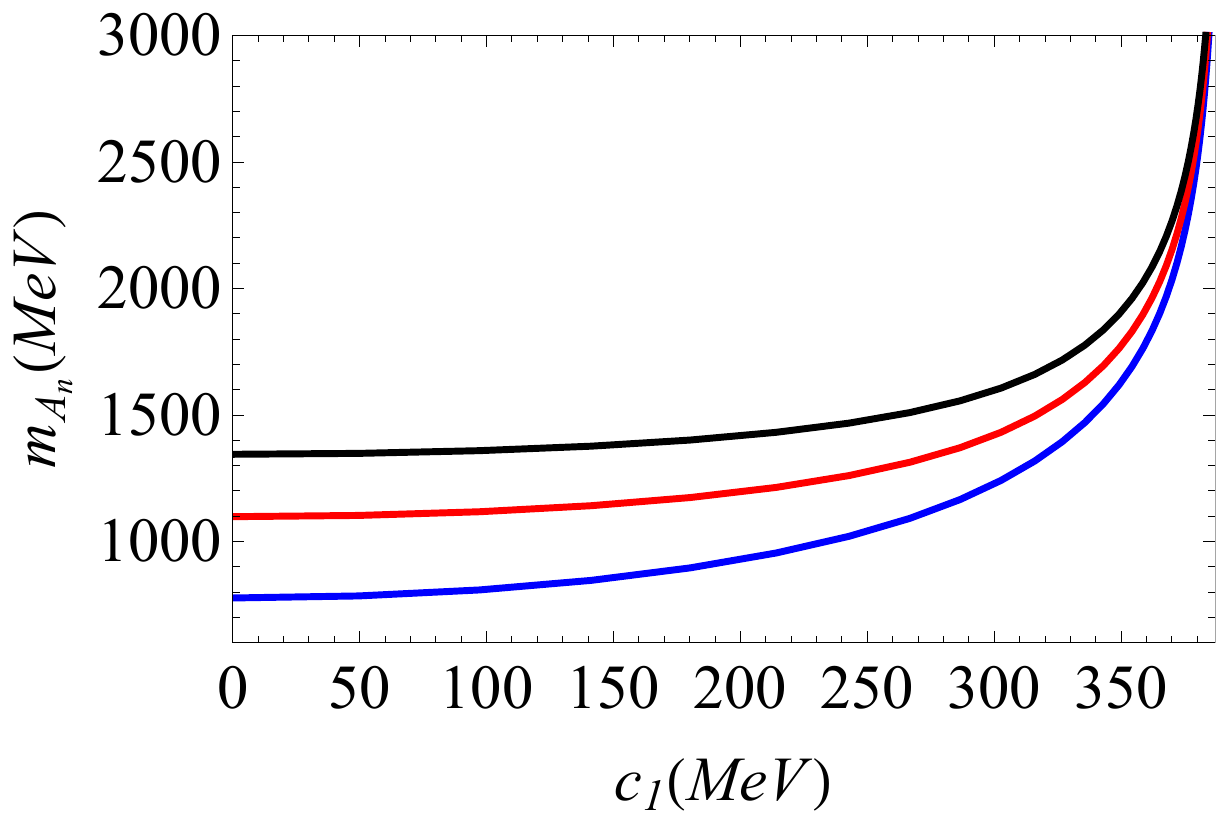}
\caption{
Masses of axial-vector mesons as functions of $C_0$ (left panel) and $c_1$ (right panel). The source parameter $c_1$ is related to the quark mass by $c_1 = m_q \zeta$ with $\zeta=\sqrt{N_c}/(2\pi)$. Solid lines represent the results for 
$\lambda=-2$, while dashed lines for $\lambda=0$. The 
results were obtained setting $\phi_{\infty}=\phi_c=(388\,\text{MeV})^2$.
}
\label{Fig:AVMassFigDSWRM}
\end{figure}

The Schrödinger equation describing the axial-vector sector is the same as Eq. \eqref{Eq:AVSchroEq}; this time with a running mass term $m_X^2(z)=-3-\phi_c\,z^2$. We display results of the evolution of the mass (for the first three states) as a function of the parameter $C_0$ in the left panel of Fig. \ref{Fig:AVMassFigDSWRM} with solid lines, while dashed lines represent the results for $\lambda=0$. From this figure we see that the masses increase as the parameter $C_0$ increases. The right panel of the same figure display the axial-vector meson masses as functions of the quark mass parameter $c_1$. The masses of the axial-vector mesons initially increase slowly with the quark mass but then all grow  rapidly becoming divergent as the quark mass parameter reaches its upper bound. As described previously, an upper bound for the quark mass is not expected in QCD. In our model this upper bound arises as a saturation effect due to a negative quartic coupling $\lambda$ for the Higgs potential \eqref{Eq:Higgspot}.
\noindent
\noindent

\subsubsection{Spectrum of the pseudoscalar sector}

\begin{figure}[ht]
\centering
\includegraphics[width=7cm]{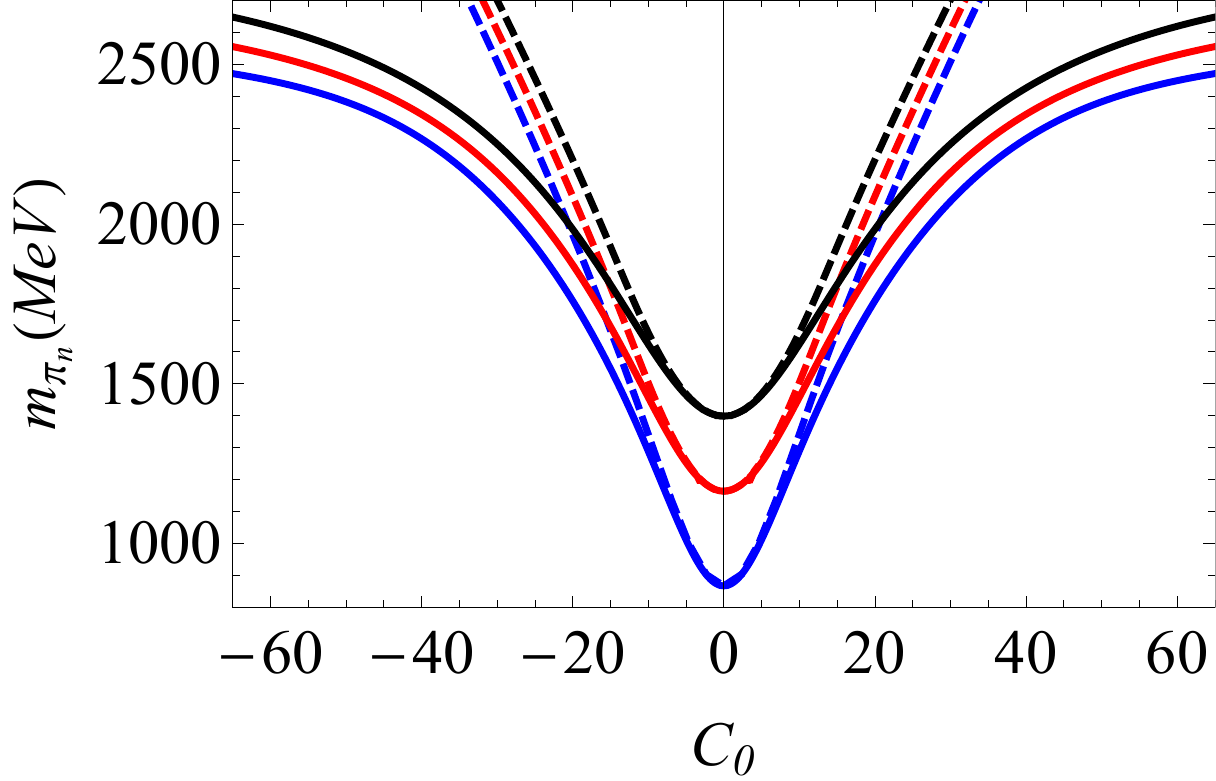}
\hfill
\includegraphics[width=7cm]{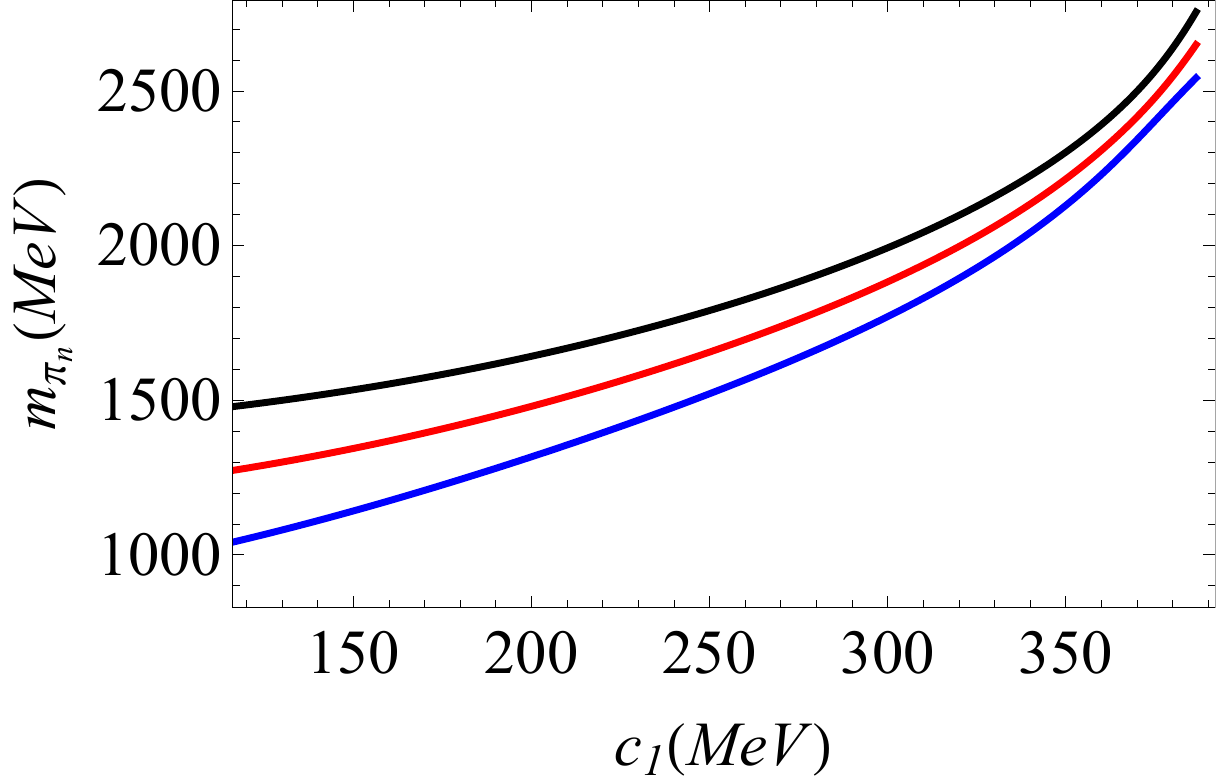}
\caption{
The mass of the pseudoscalar mesons as a function of $C_0$ (left panel) and $c_1$ (right panel). The source parameter $c_1$ is related to the quark mass by $c_1 = m_q \zeta$ with $\zeta=\sqrt{N_c}/(2\pi)$. Solid lines represent the result for $\lambda=-2$, while dashed lines for $\lambda=0$. The results were obtained setting $\phi_{\infty}=\phi_c=(388\,\text{MeV})^2$.
}
\label{Fig:PiMassFigDSWRM}
\end{figure}

The coupled equations of the pseudoscalar mesons are the same as Eqs.~\eqref{Eq:EqPion1} and \eqref{Eq:EqPion2}. Combining those equations we reduced them into a Schrödinger form in Eq.~\eqref{Eq:SchroPion}. Using the same numerical procedure applied in the case without running mass, we are able to find the masses of pseudoscalar masses as functions of the parameters $C_0$ and $c_1$, We display the numerical results in Fig.~\ref{Fig:PiMassFigDSWRM}. The left panel of this figure shows the variation of the masses as  functions of $C_0$ with solid lines, while the results for $\lambda=0$ are represented with dashed lines; we see that the masses of pseudoscalar mesons always increase with $C_0$. We also point out that the mass increases faster close to $C_0=0$ and slowly for large values of $C_0$. The right panel of the same figure displays the evolution of the masses as function of the quark mass parameter $c_!$. We observe that the masses of pseudoscalar mesons increase slowly in the region of small quark mass and faster in the intermediate and large quark mass region. Again, it seems that the masses diverge when the quark mass parameter $c_1$ reaches its upper bound. 
\noindent
\noindent

\subsection{Meson Spectrum \texorpdfstring{($\lambda>0$)}{} }

\subsubsection{Spectrum of the scalar sector}

\begin{figure}[ht]
\centering
\includegraphics[width=7cm]{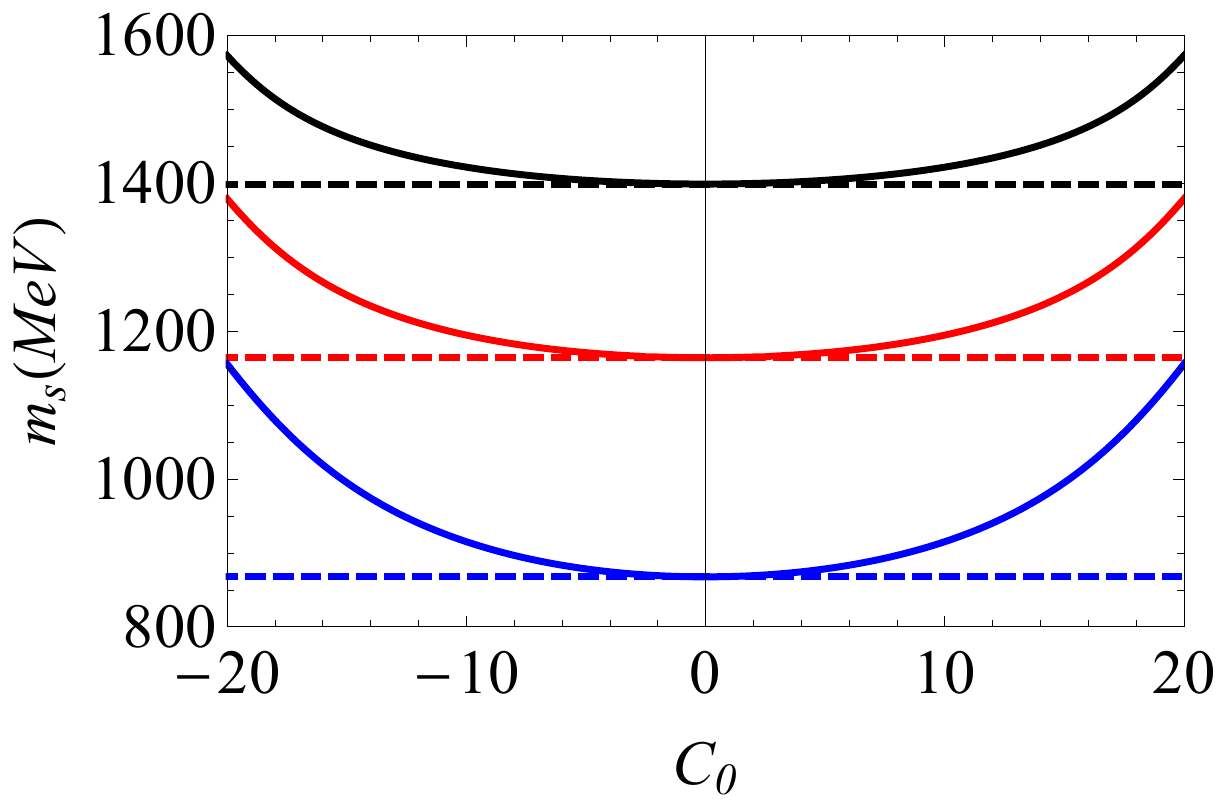}
\hfill
\includegraphics[width=7cm]{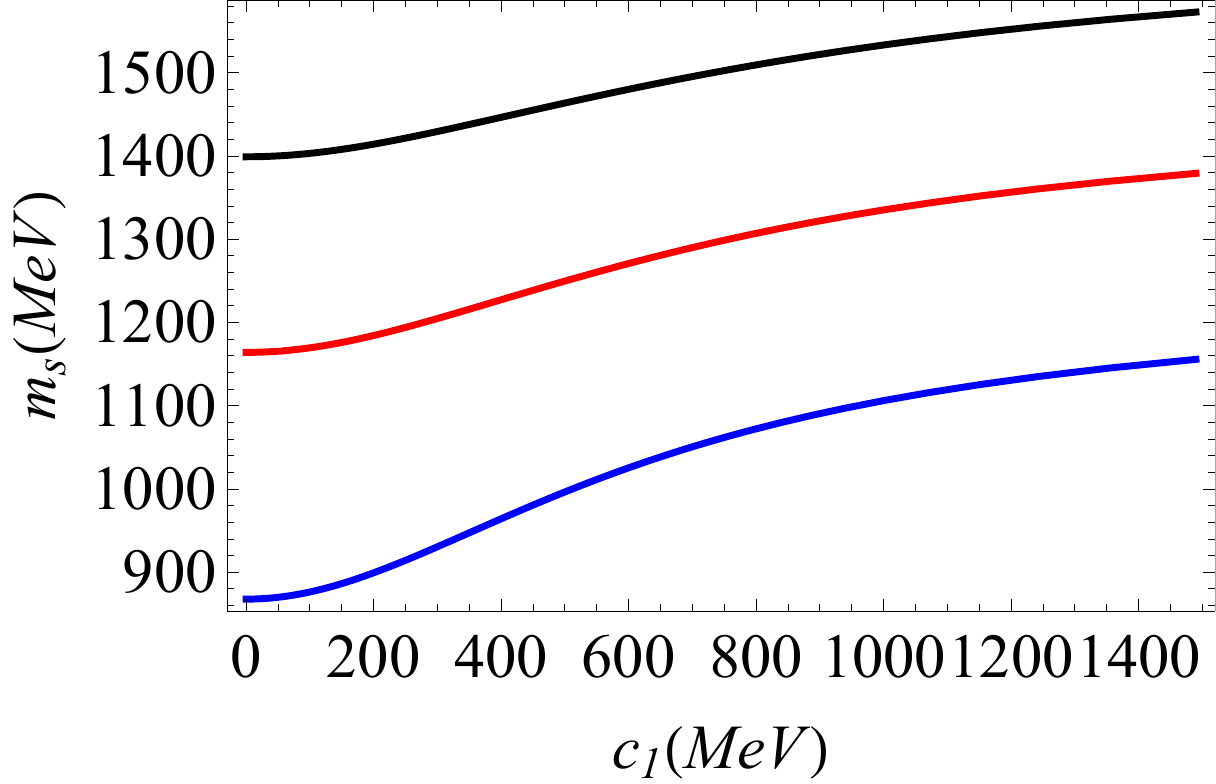}
\caption{
Masses of scalar mesons as functions of $C_0$ (left panel) and $c_1$ (right panel). The source parameter $c_1$ is related to the quark mass by $c_1 = m_q \zeta$ with $\zeta=\sqrt{N_c}/(2\pi)$. Solid lines represent the result for $\lambda=2$, while dashed lines for $\lambda=0$. The results were obtained setting $\phi_{\infty}=\phi_c=(388\,\text{MeV})^2$.
}
\label{Fig:ScalarMassC1RMLneg}
\end{figure}

The potential of the Schrödinger equation in the scalar sector was displayed in Fig.~\ref{Fig:PotScalarRM} for $\lambda=2$. Our numerical results of the masses as functions of the parameter $C_0$ ($c_1$) are displayed in the left panel (right panel) of Fig.~\ref{Fig:ScalarMassC1RMLneg}. We see that the scalar meson masses increase with both parameters. This allows us to use the same strategy, implemented in Sec.~\ref{Sec:ScalarSectorLneg}, to fix the parameters, i.e., using the state $f_0(980)$. Having fixed the parameters we calculate the spectrum, displayed in Table \ref{Taba:01RMLneg} as NLSW-RM. Although we are able to find a spectrum compatible with experimental results, this does not guarantee a physical value for the quark mass. For example, the parameter choice  $\lambda=7$ and $C_0=7.6$ corresponds to a very large value for the quark mass parameter $c_1$,  even larger than the value obtained in the case without running mass.
\noindent
\noindent

\begin{table}[ht]
\centering
\begin{tabular}{l |c|c|c|l}
\hline 
\hline
 $n$ & NLSW-RM & FLZ A \cite{Fang:2016nfj} 
     & GKK \cite{Gherghetta:2009ac} 
     & $f_0$ experimental \cite{Tanabashi:2018oca} \\
\hline 
 $1$ & 980  & 586  & 799   & $980\pm 10$  \\
 $2$ & 1238 & 1346 & 1184  & $1350\pm 150$  \\
 $3$ & 1455 &      & 1466  & $1505\pm 6$ \\
 $4$ & 1645 & 1743 & 1699  & $1724\pm 7$  \\
 $5$ & 1816 & 2232 & 1903  & $1992\pm 16$  \\
 $6$ & 1973 & 2420 & 2087  & $2103\pm 8$ \\
 $7$ & 2118 &      & 2257  & $2314\pm 25$ \\
 $8$ & 2255 &      & 2414  &   \\
\hline\hline
\end{tabular}
\caption{
The masses os the scalar mesons (in MeV) obtained in the nonlinear soft wall model with running mass, compared against the results of Refs.~\cite{Gherghetta:2009ac,Fang:2016nfj} and experimental results of PDG \cite{Tanabashi:2018oca}. The value of the parameters used are $\lambda=7$ and $C_0=7.6$.
}
\label{Taba:01RMLneg}
\end{table}

\subsubsection{Spectrum of the axial-vector sector}

\begin{figure}[ht]
\centering
\includegraphics[width=7cm]{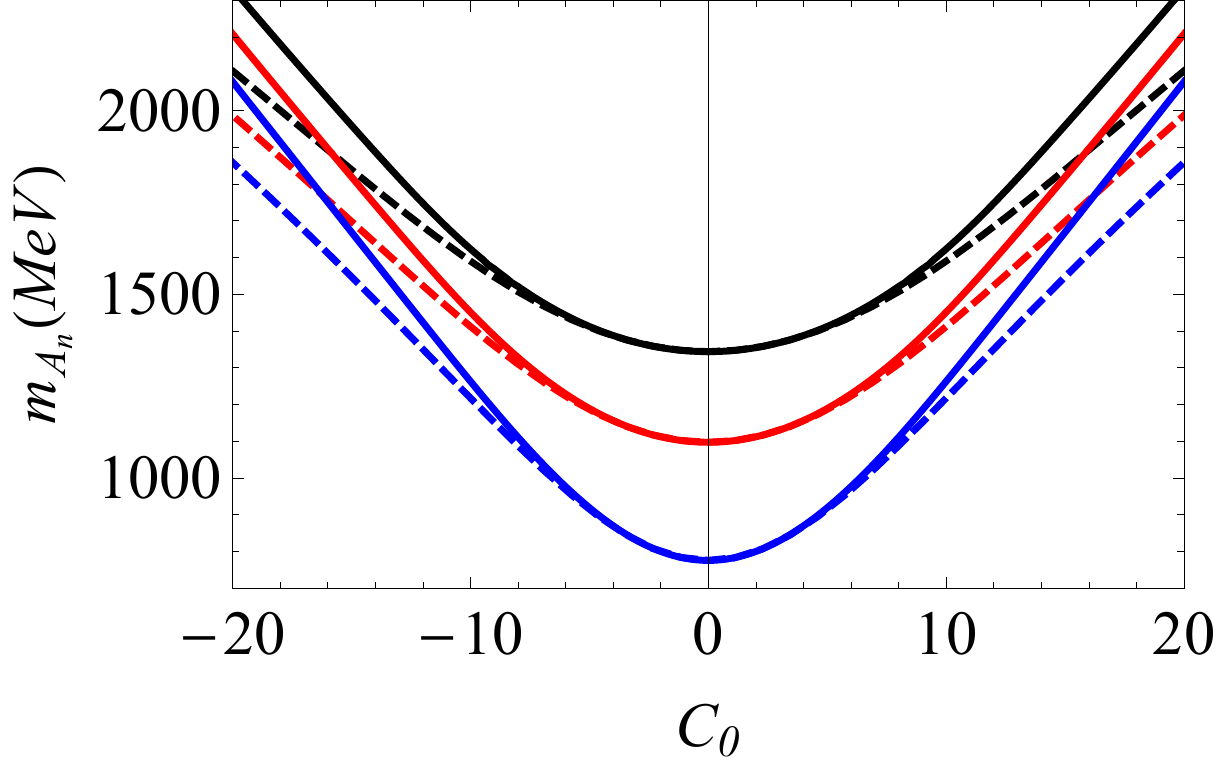}
\hfill
\includegraphics[width=7cm]{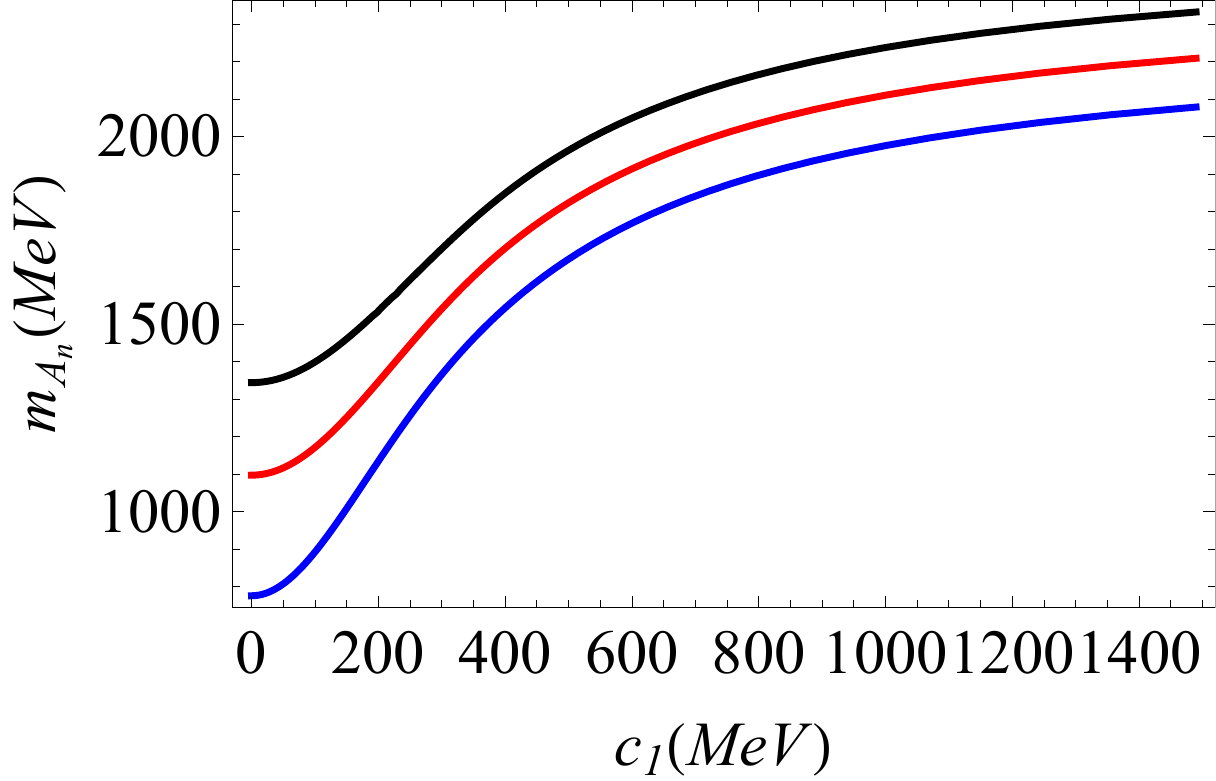}
\caption{
The mass of the axial-vector mesons as a function of $C_0$ (left panel) and $c_1$ (right panel). The source parameter $c_1$ is related to the quark mass by $c_1 = m_q \zeta$ with $\zeta=\sqrt{N_c}/(2\pi)$. Solid lines represent the results for $\lambda=2$, while dashed lines for $\lambda=0$. The 
results were obtained setting $\phi_{\infty}=\phi_c=(388\,\text{MeV})^2$.
}
\label{Fig:AMassC1RMLneg}
\end{figure}

The axial-vector sector is described by the same equations of subsection \ref{Sec:AV}; this time with running mass term $m_X^2(z)=-3-\phi_c\,z^2$. The evolution of the masses as functions of the parameter $C_0$ i($c_1$) is displayed in the left panel (right panel) of Fig.~\ref{Fig:AMassC1RMLneg}. We see that the masses of  axial-vector mesons increase slowly with $c_1$ and seem to reach asymptotic finite values in the limit $c_1 \to \infty$.
\noindent
\noindent
In Table \ref{Taba:02RMLneg} we display the results for the parameter choice $(\lambda=7, C_0=7.6)$, fixed previously in the scalar sector. 
\begin{table}[ht]
\centering
\begin{tabular}{l |c|c|c|l}
\hline 
\hline
 $n$ & NLSW-RM & FLZ A \cite{Fang:2016nfj} &
 GKK \cite{Gherghetta:2009ac}&
$a_1$ experimental \cite{Tanabashi:2018oca} \\
\hline 
 $1$ & 1147  & 1121  & 1185   & $1230\pm 40$  \\
 $2$ & 1359 & 1608 & 1591  & $1647\pm 22$  \\
 $3$ & 1547 & 1922 & 1900  & $1930^{+30}_{-70}$ \\
 $4$ & 1718 & 2156 & 2101  & $2096\pm 122$  \\
 $5$ & 1876 & 2352 & 2279  & $2270^{+55}_{-40}$  \\
 $6$ & 2023 & 2526 &   &  \\
 $7$ & 2161 &  &   &  \\
\hline\hline
\end{tabular}
\caption{The masses of the axial-vector mesons (in MeV) obtained in the nonlinear soft wall model with running mass, compared against the results of Refs.~\cite{Fang:2016nfj,Gherghetta:2009ac} and experimental results of RPP \cite{Tanabashi:2018oca}. The value of the parameters used are $\lambda=7$, $C_0=7.6$ and $\phi_{\infty}=\phi_c=(388\,\text{MeV})^2$.
}
\label{Taba:02RMLneg}
\end{table}

It is worth mentioning that the spectrum of axial-vector mesons is very different from the spectrum of the vector mesons, cf. Table~\ref{Taba:VectorSF}, which means that both sectors are not degenerate. This non-degeneracy of the spectrum is enhanced by the tachyon runnning mass because the tachyon field becomes divergent in the IR. This is also a signal that chiral symmetry is never restored in the axial-vector excited states.

\subsubsection{Spectrum of the pseudoscalar sector}

\begin{figure}[ht]
\centering
\includegraphics[width=7cm]{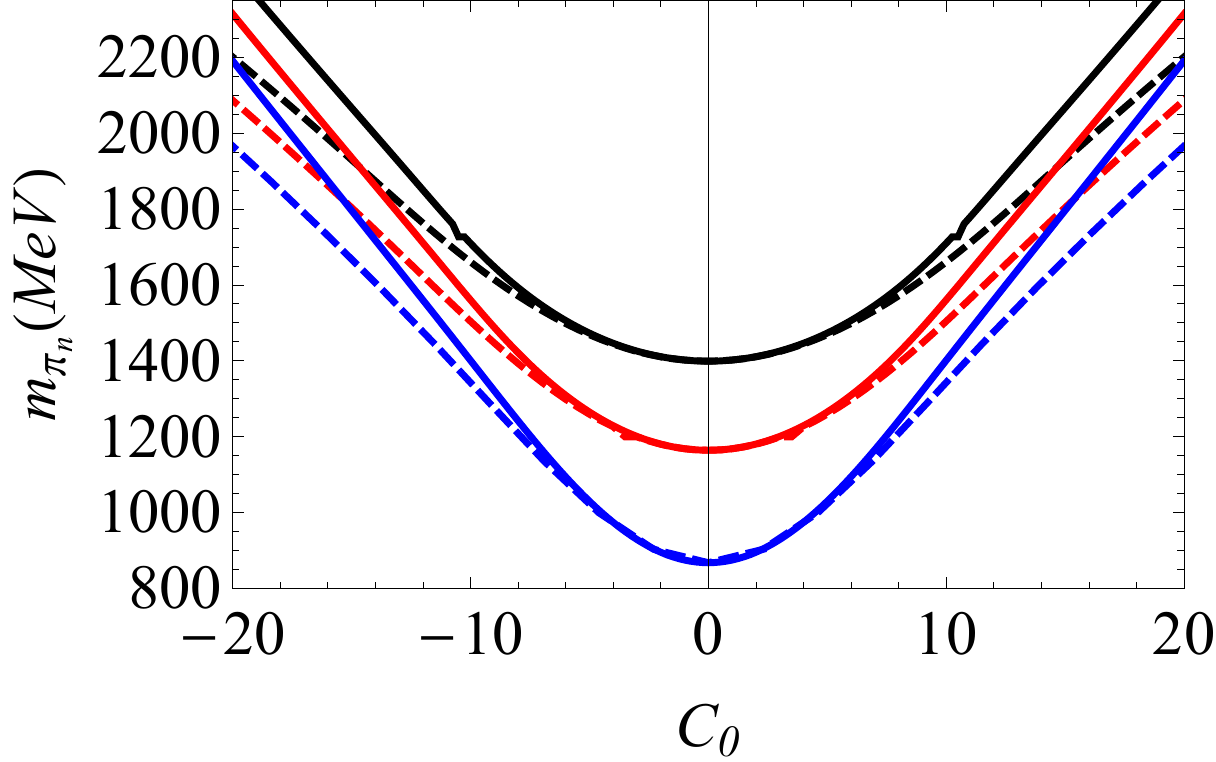}
\hfill
\includegraphics[width=7cm]{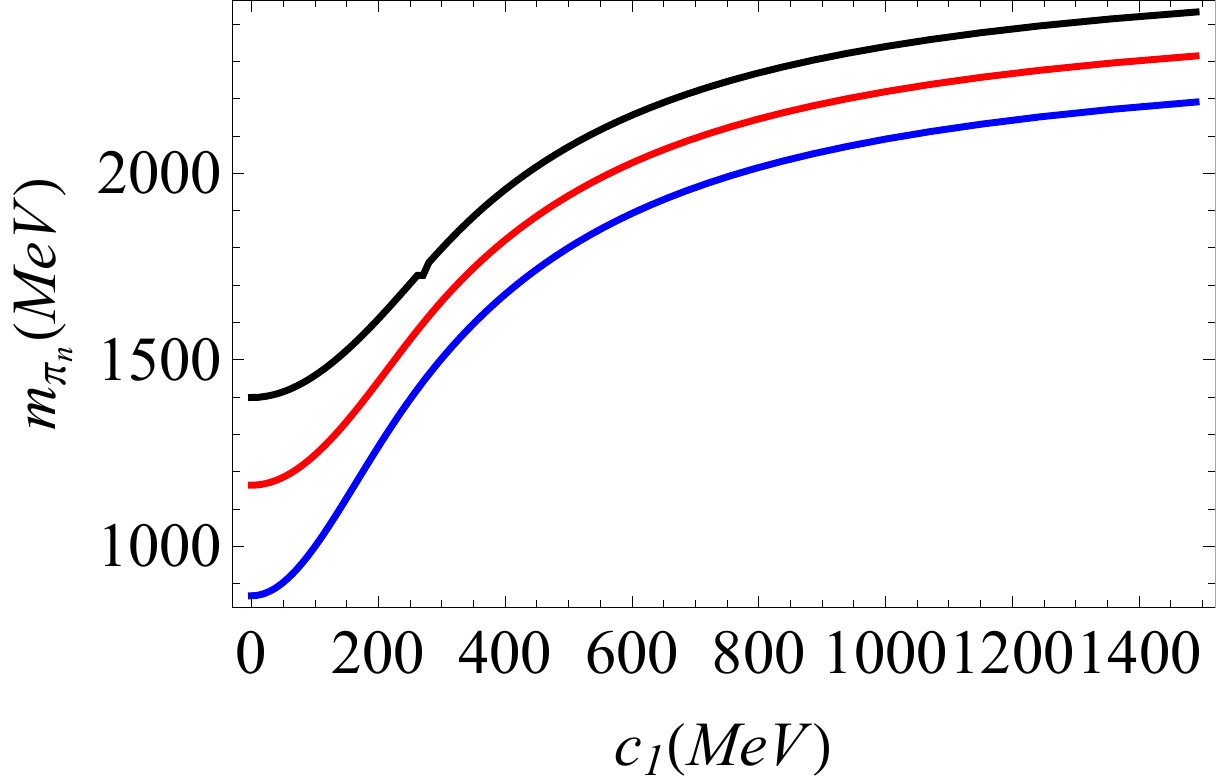}
\caption{
The mass of the pseudoscalar mesons as a function of $C_0$ (left panel) and $c_1$ (right panel). The source parameter $c_1$ is related to the quark mass by $c_1 = m_q \zeta$ with $\zeta=\sqrt{N_c}/(2\pi)$. Solid lines represent the result for $\lambda=2$, while dashed lines for $\lambda=0$. The results were obtained setting $\phi_{\infty}=\phi_c=(388\,\text{MeV})^2$.
}
\label{Fig:PionMassC1RMLneg}
\end{figure}

The pseudoscalar mesons are described by the same equations of subsection~\ref{Sec:Pion};  this time with running mass term $m_X^2(z)=-3-\phi_c\,z^2$. The evolution of the pseudoscalar meson masses as functions of the parameter $C_0$ ($c_1$) is displayed in the left panel (right panel) of Fig.~\ref{Fig:PionMassC1RMLneg}.  As in the axial-vector sector, the masses in the pseudoscalar sector increase slowly with $c_1$ and seem to reach asymptotic finite values in the limit $c_1 \to \infty$. Again, in the chiral limit $c_1 \to 0$ the mass of the lightest state does not vanish. This means that we do not have a pseudo NG boson in the spectrum and the pseudoscalar mesons behave just as pion resonances.

\noindent
\noindent

 Our numerical results for the parameter choice $\lambda=7$ and $C_0=7.6$ are displayed in Table~\ref{Taba:03RMLneg} as NLSW-RM. Since a pseudo NG boson is missing in the spectrum, the model at hand allows us to calculate only the spectrum of pion resonances. We also compare our results against the results available in Refs.~\cite{Fang:2016nfj,Kelley:2010mu} and  experimental data \cite{Tanabashi:2018oca}.
\begin{table}[ht]
\centering
\begin{tabular}{l |c|c|c|l}
\hline 
\hline
 $n$ & NLSW-RM & FLZ A \cite{Fang:2016nfj}&
 KBK \cite{Kelley:2010mu}&
$\pi$ experimental \cite{Tanabashi:2018oca} \\
\hline 
 $1$ & -  & 139.6  & 144    & $140$  \\
 $2$ & 1301 & 1269  & 1557 & $1300\pm 100$  \\
 $3$ & 1479 & 1753  & 1887 & $1816\pm 14$ \\
 $4$ & 1642 & 2051  & 2090 & $2070$  \\
 $5$ & 1796 & 2277  & 2270 & $2360$  \\
 $6$ & 1942 & 2467  & 2434 &  \\
 $7$ & 2081 &       & 2586 &  \\
\hline\hline
\end{tabular}
\caption{The masses of the pseudoscalar mesons (in MeV) obtained in the 
nonlinear soft wall model with running mass, 
compared against the results 
of Refs.~\cite{Fang:2016nfj,Kelley:2010mu} and experimental results of 
RPP \cite{Tanabashi:2018oca}. The value of the parameters used are 
$\lambda=7$, $C_0=7.6$ and $\phi_{\infty}=\phi_c=(388\,\text{MeV})^2$.
}
\label{Taba:03RMLneg}
\end{table}

\subsection{Decay constants}

\begin{figure}[ht]
\centering
\includegraphics[width=7cm]{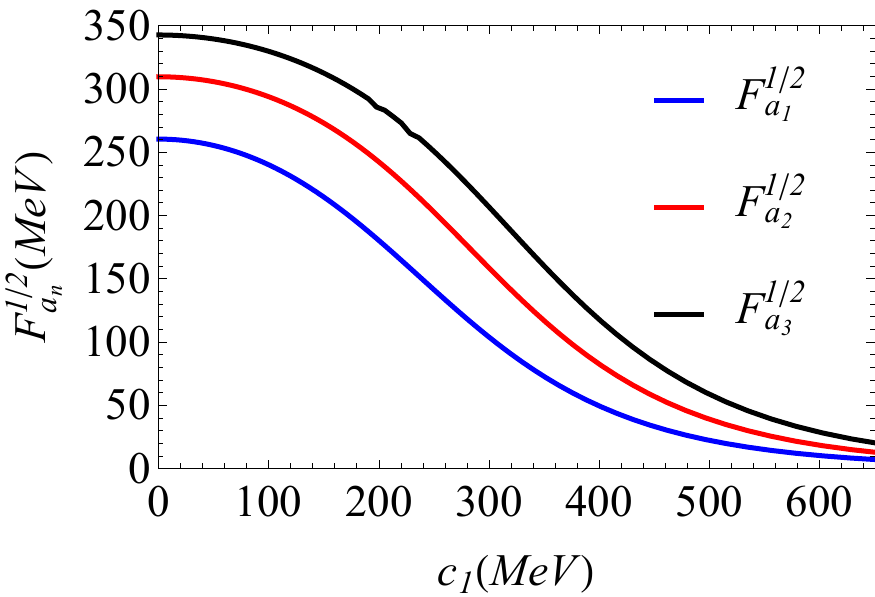}
\hfill
\includegraphics[width=7cm]{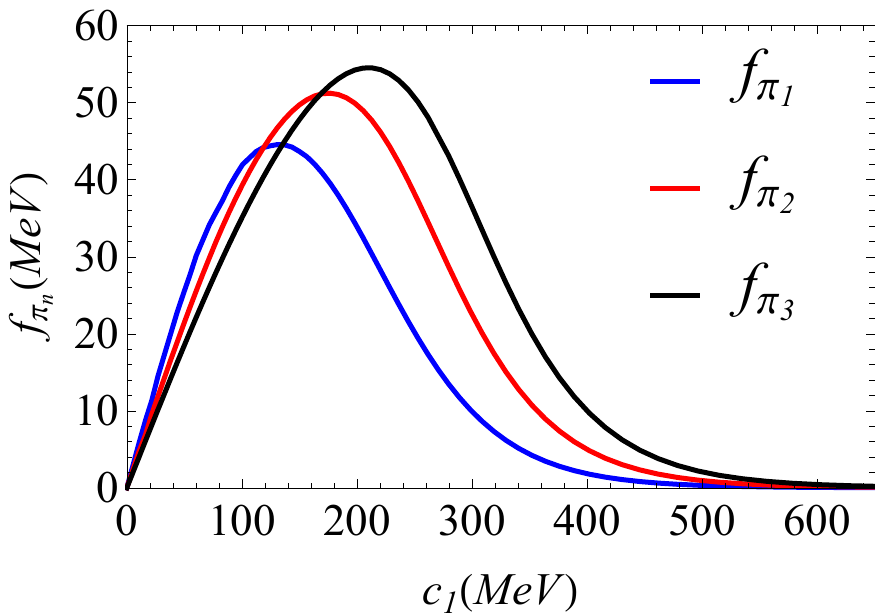}
\caption{
The decay constants of the axial-vector mesons (left panel) and pseudoscalar mesons (right panel) as a function of $c_1$ obtained in the NLSW model with running mass for $\phi_{\infty}=\phi_c=(388\,\text{MeV})^2$ and $\lambda=2$. The source parameter $c_1$ is related to the quark mass by $c_1 = m_q \zeta$ with $\zeta=\sqrt{N_c}/(2\pi)$.
}
\label{Fig:PionDecayRMLneg}
\end{figure}

In this subsection we calculate the decay constants of the vector, axial and pseudoscalar mesons in the nonlinear soft wall model with running mass. As explained in subsection~\ref{Sec:DecayConstantLneg}, the decay constants are related to the normalisation condition on the field perturbations. Again, we investigate the evolution of the decay constants of axial-vector and pseudoscalar mesons as a function of the quark mass parameter $c_1$. In the left panel of Fig.~\ref{Fig:PionDecayRMLneg} we display the evolution of the decay constants of the first three axial-vector mesons; the results are qualitatively similar to the case without running mass, cf. left panel of Fig.~\ref{Fig:AVDecayL}. In turn, the evolution of the decay constants of the first three pseudoscalar mesons are displayed in the right panel of Fig.~\ref{Fig:PionDecayRMLneg}.  Again, the results are qualitatively similar to the case without running mass, displayed in the right panel of Fig.~\ref{Fig:AVDecayL}. In particular, we find again an inversion of hierarchy for the pseudoscalar meson decay constants. 
\noindent
\noindent
Considering the specific parameter choice $\lambda=7$, $C_0=7.6$ that fixes the mass of the first scalar state, we may calculate the decay constants of the first state of the vector and axial-vector mesons. These results are displayed in Table~\ref{Taba:DecayConstantDSWRM} as NLSW-RM.

\begin{table}[ht]
\centering
\begin{tabular}{l |c|c|l}
\hline 
\hline
 & NLSW-RM $(\lambda>0)$ & FLZ A \cite{Fang:2016nfj} & Exp. \cite{Tanabashi:2018oca} \\
\hline 
 $F_{\rho}^{1/2}$ &  260.12 & 296 & $346.2\pm 1.4$  \\
 $F_{a_1}^{1/2}$  & 152.78 &  389 & $433\pm 13$ \\
\hline\hline
\end{tabular}
\caption{
The decay constants (in MeV) obtained in the nonlinear soft wall model with running mass, compared against the results of Refs.~\cite{Fang:2016nfj} and experimental results of PDG \cite{Tanabashi:2018oca}. The results were obtained setting $\lambda=7$, $C_0=7.6$ and $\phi_{\infty}=\phi_c=(388\text{MeV})^2$.
}
\label{Taba:DecayConstantDSWRM}
\end{table}
 In Table~\ref{Taba:DecayConstantDSWRMPionL} we display the decay constants for the first three pseudoscalar mesons in the case $\lambda=7$ and specific values for $C_0$. These results show explicitly the inversion of hierarchy for the pseudoscalar decay constants when going from the regime of small quark mass to the regime of heavy quark mass, see the end of subsection \ref{Sec:DecayConstantLneg}.

\begin{table}[ht]
\centering
\begin{tabular}{l |c|c|l}
\hline 
\hline
 & $f_{\pi_1}$ & $f_{\pi_2}$ & $f_{\pi_3}$ \\
\hline 
 NLSW-RM  $(C_0=7.6)$    & 21.81 & 37.58 & 48.11\\
 \hline 
 NLSW-RM  $(C_0=0.2)$    & 2.45 & 1.93 & 1.63\\
\hline\hline
\end{tabular}
\caption{
The decay constants of the first three pseudoscalar mesons (in MeV) obtained in the nonlinear soft wall model with running mass. The results were obtained setting $\lambda=7$, $C_0=7.6$ $(c_1=237.2\text{MeV})$, $C_0=0.2$ $(c_1=4.25\text{MeV})$ and $\phi_{\infty}=\phi_c=(388\text{MeV})^2$.
}
\label{Taba:DecayConstantDSWRMPionL}
\end{table}

\section{Discussion and conclusions}
\label{Sec:Conclusions}

In this work we have investigated a nonlinear realisation  for chiral symmetry breaking in soft wall models based on a Higgs potential. 
Soft wall models allow for a more realistic description of the meson spectrum because a positive quadratic dilaton in the IR guarantees linear Regge trajectories.  Solving the nonlinear differential equation of the tachyon we found that the tachyon solution in the IR depends on only one parameter $C_0$, Integrating numerically the tachyon differential equation we found non-trivial relations between the UV parameters $c_1$ and $c_3$ and the IR parameter $C_0$. Moreover, implementing the procedure of holographic renormalisation we obtained a dictionary for the 4d chiral condensate in terms of the VEV parameter $c_3$. This allowed us to find the evolution of the 4d chiral condensate with the quark mass.  For the case $\lambda>0$, corresponding to a Mexican hat Higgs potential, we found that the chiral condensates grows nonlinearly with the quark mass, as expected in QCD. We found, however, that a nonlinear Higgs potential for the tachyon is not sufficient to provide spontaneous symmetry breaking in the chiral limit because the chiral condensate vanishes as the quark mass goes to zero. 

We have calculated the spectrum of the scalar, vector, axial-vector and pseudoscalar mesons. The spectrum of vector mesons decoupled from the other sectors and their solutions were the same as in the linear soft wall model. We concluded that the case $\lambda>0$, corresponding to a Mexican hat Higgs potential, provides the most realistic scenario for the meson spectrum. The spectrum of the scalar mesons presented a pathology in the case $\lambda<0$, because the masses decreased with the quark mass. In the case $\lambda>0$ we found that it was possible to match the scalar state $f_0(980)$ state by fixing appropriately  the parameters $\lambda$ and $C_0$. The analysis of the spectrum showed us that nonlinear soft wall models led to a non-degeneracy between vector and axial-vector mesons and also between scalar and pseudoscalar mesons. However, we found that it was not possible to fit the meson spectrum with a small quark mass parameter, cf. end of  subsection \ref{subsec:pionposlambda}. We calculated the decay constants of vector, axial-vector and pseudoscalar mesons.  The vector meson decay constants were insensitive to the tachyon dynamics and hence the quark mass. The axial-vector meson decay constants decreased with the quark mass whereas the pseudoscalar meson decay constants presented a non-monotonic behaviour consistent with pion resonances. We found, in particular that all the pseudoscalar decay constants vanish in the chiral limit, indicating the absence of pseudo NG bosons.  A separate comment deserves the decreasing behavior of the pseudoscalar decay constants in the regime of heavy quark mass, which agrees qualitatively with the perturbative QCD prediction,  cf.  end of subsection \ref{Sec:DecayConstantLneg}.

As an attempt to allow for  spontaneous symmetry breaking in the chiral limit,  we also investigated nonlinear soft wall models with a tachyonic running mass. The consequence of the running mass was to increase the interval of $C_0$ in relation to the model without running mass. However, we realised that the results were qualitatively similar to the case without running mass. Again, we found that $\lambda>0$ provides the most realistic scenario for the meson spectrum and decay constants and we were able to match again the $f_0(980)$ state in the scalar sector by fixing appropriately the parameters. Again, we found that fixing the parameter to describe the meson spectrum leads to a very large value for the quark mass.  We found  that the non-degeneracy between vector and axial-vector mesons was enhanced by the running mass and the same was true for the scalar and pseudoscalar mesons.

Let us discuss briefly some other approaches to the problem of chiral symmetry in  holographic QCD. The model proposed in ~\cite{Iatrakis:2010jb}  (see also ~\cite{Iatrakis:2010zf}), which implemented the ideas of ~\cite{Casero:2007ae} claim that the tachyon must blow up in the IR when the background has confinement properties. This statement may be related to the Coleman-Witten theorem \cite{Coleman:1980mx}. As in our case, the IR tachyon solution in \cite{Iatrakis:2010jb} depends only on one parameter. In contrast to our case, the model in \cite{Iatrakis:2010jb} leads to spontaneous chiral symmetry breaking in the chiral limit. Running mass models were considered in \cite{Vega:2010ne},\cite {Cui:2013xva} and \cite{Fang:2016nfj} as an attempt to allow for spontaneous symmetry breaking in the chiral limit. We have found, however, that in a consistent description of nonlinear soft wall models based on a Higgs potential, the tachyon running mass does not solve the lack of spontaneous symmetry breaking in the chiral limit. An interesting approach to the problem of chiral symmetry breaking is considering a negative profile for the dilaton \cite{Chelabi:2015gpc}. A negative dilaton profile has some issues, see for instance ~\cite{Karch:2010eg} for a discussion. For example, it would violate the null energy condition in the gravitational background \cite{Kiritsis:2006ua}. Nevertheless, the authors of ~\cite{Chelabi:2015gpc} proposed that the profile of the dilaton in the UV may be negative, while in the IR must be positive in order to guarantee confinement and Regge-like behaviour. They claim that in this way is possible to describe spontaneous chiral symmetry breaking in the chiral limit. 

In conclusion, nonlinear soft wall models based on a Higgs potential, with and without a tachyon running mass, and a positive quadratic dilaton do not provide spontaneous chiral symmetry breaking in the chiral limit. Consequently, there are no pseudo NG bosons in the spectrum of pseudoscalar mesons. This conclusion is supported by the study of masses and  decay constants in the region of small quark mass. We found, however, in the case $\lambda>0$ a very reasonable behaviour for the chiral condensate in the regime of large quark mass, similar to the behaviour expected in QCD. For the case $\lambda>0$ we also found a reasonable behaviour for all the meson masses as growing functions of the quark mass. Interestingly, we found that the decay constants of axial-vector mesons and pseudoscalar mesons decrease in the regime of heavy quark mass. This behaviour is also expected in QCD and encourages us to continue the investigation of nonlinear soft wall models. Finally, a natural extension of this work would be the investigation of backreacted Einstein-Dilaton backgrounds, where the confinement criterion is satisfied. This would allow for a consistent description of confinement and chiral symmetry breaking in a minimal holographic setup.

\section*{Acknowledgments}
 The authors would like to thank Carlisson Miller for stimulating discussions during the early stages of this project. The authors would also like to acknowledge very useful conversations with Saulo Diles,  Diego Rodrigues, Jonathan Shock and Dimitrios Zoakos during the development of this work. A.B-B would like to thank the organizers of WONPAQCD 2019 for providing a  stimulating environment.  The work of A.B-B is partially funded by Conselho Nacional de Desenvolvimento Cientifico e Tecnologico (CNPq), grants No. 306528/2018-5 and No. 434523/2018-6. L.~A.~H.~M. has financial support from Coordena\c{c}\~ao de Aperfei\c{c}oamento do Pessoal de N\'ivel Superior - Programa Nacional de P\'os-Doutorado (PNPD/CAPES, Brazil).

\appendix

\section{Equations of motion and decay constants}
\label{Sec:4daction}

In this Appendix we write details about the derivation of the equations of motion used to calculate the meson spectrum as well as the holographic dictionary for meson decay constants. We will expand the action \eqref{HiggsAction} up to second order on the fields and take the 5d metric as in \eqref{Eq:hQCDmetric}.

 We follow ~\cite{Ballon-Bayona:2017bwk} (see also \cite{Abidin:2009aj}). For simplicity, we take $N_f=2$ and assume isospin symmetry ($m_u=m_d$).
First we  decompose the bifundamental field $X$ in the form
\noindent
\begin{equation}
X=e^{2 i\,\pi^{a}\,T^{a}}\,\left( \frac12 v(z) +S \right),
\label{Eq:ScalarFlucts}
\end{equation}
\noindent
where $\pi^{a}(x^{\mu},z)$ is the pseudoscalar field, $T^a$ are the generators of $SU(2)$ and $S(x^{\mu},z)$ the scalar fluctuation related to the scalar mesons. We also rewrite the fluctuation for the gauge fields as
\noindent
\begin{equation}
A_m^{(L/R)}=V_m \pm A_m  \, , 
\label{Eq:VecFlucts}
\end{equation}
\noindent
where  $V^m=V^{m}_{a}T^a$ and $A^m=A^{m}_{a}T^a$ are the vector and axial fields, respectively. 
Plugging \eqref{Eq:ScalarFlucts} and \eqref{Eq:VecFlucts} into the action \eqref{HiggsAction} and expanding on the fields $\pi^a$, $V^m_a$ and $A^m_a$ up to second order we obtain $S=S_0+S_2 + \dots$ with $S_0$ the effective 1d action for the background $v(z)$ and 
\begin{equation}
\begin{split}
S_2=&-\int dx^5\sqrt{-g}\,e^{-\Phi}\Big [2 (\partial_{m}S )^2+2m_X^2S^2 + 3\lambda v^2(z) S^{2}\\
& +\frac{1}{4g_5^2}v^{mn}_{a}v^{a}_{mn} +\frac{1}{4g_5^2}a^{mn}_{a}a^{a}_{mn} 
+ \frac12 v^2(z) (\partial_m\pi_a-A_{m,a})^2 \, , \Big ]
\end{split}
\end{equation}
\noindent
the action describing the kinetic terms for the 5d field fluctuations $\pi^a$, $V^m_a$ and $A^m_a$.  We have defined the Abelian tensors
\noindent
\begin{equation}
v_{mn}^a=\partial_{m}V_n^a-\partial_{n}V_m^a \, ,\quad a_{mn}^a=\partial_{m}A_n^a-\partial_{n}A_m^a \label{Eq:Abelian}
\end{equation}
\noindent
To obtain the Euler-Lagrange equations we write the action as $S_2=\int dx^5 \mathcal{L}_2$. The variation takes the form
\noindent
\begin{equation}\label{Eq:QuadraticAction}
\begin{split}
\delta S_2\,&=\int d^5 x \Big [\left(\frac{\partial \mathcal{L}_2}{\partial S}-\partial_{m}P^{m}_{S}\right)\delta S+\left(\frac{\partial \mathcal{L}_2}{\partial V^{a}_{l}}-\partial_{m}P^{ml}_{V,a}\right)\delta V^{a}_{l}
+\left(\frac{\partial \mathcal{L}_2}{\partial A^{a}_{l}}-\partial_{m}P^{ml}_{A,a}\right)\delta A^{a}_{l}\\
&+\left(\frac{\partial \mathcal{L}_2}{\partial \pi^{a}}-\partial_{m}P^{m}_{\pi,a}\right)\delta \pi^{a} \Big ]+\int dx^5\partial_{m}\left(P^{m}_{S}\delta S+P^{ml}_{V,a}\delta V^{a}_{l}+P^{ml}_{A,a}\delta A^{a}_{l}+P^{m}_{\pi,a}\delta \pi^{a}\right),
\end{split}
\end{equation}
\noindent
where $P^{m}_{S}$, $P^{ml}_{V,a}$, $P^{ml}_{A,a}$ and $P^{m}_{\pi,a}$ are the conjugate momenta associated with the scalar, vector, axial-vector and pseudoscalar fields, respectively. Explicitly, the conjugate momenta are given by
\noindent
\begin{equation}
\begin{split}
P^{m}_{S}&=\frac{\partial \mathcal{L}_2}{\partial (\partial_m S)}=-4e^{-\Phi}\sqrt{-g}\partial^{m}S, \quad 
P^{ml}_{V,a}=\frac{\partial \mathcal{L}_2}{\partial (\partial_m V^{a}_{l})}=-\frac{e^{-\Phi}}{g^{2}_{5}}\sqrt{-g}\,v^{ml}_{a},\\
P^{ml}_{A,a}&=\frac{\partial \mathcal{L}_2}{\partial (\partial_m A^{a}_{l})}=-\frac{e^{-\Phi}}{g^{2}_{5}}\sqrt{-g}\,a^{ml}_{a},\\
P^{m}_{\pi,a}&=\frac{\partial \mathcal{L}_2}{\partial (\partial_m \pi^{a})}=- e^{-\Phi}\sqrt{-g} \, v^2(z) \left(\partial^{m}\pi^a-A^{m,a} \right) \, .
\end{split}
\end{equation}
\noindent
In turn, the derivatives of the Lagrangian are:
\noindent
\begin{equation}
\begin{split}
\frac{\partial \mathcal{L}_2}{\partial S}&=-e^{-\Phi}\sqrt{-g} [ 4 m_{X}^{2}-6v^{2}(z) ] S \, , \quad 
\frac{\partial \mathcal{L}_2}{\partial V_{l}^{a}}=-0 \, ,\\
\frac{\partial \mathcal{L}_2}{\partial A_{l}^{a}}&=e^{-\Phi} \sqrt{-g} \, v^2(z) (\partial^{l}\pi^a-A^{l,a} ) \, , \quad 
\frac{\partial \mathcal{L}_2}{\partial \pi^{a}}=0 \, .
\end{split}
\end{equation}
\noindent
From these results we find that the equation of motion of the vector sector takes the form
\noindent
\begin{equation}
\frac{e^{\Phi}}{\sqrt{-g}}\partial_{m}\left(e^{-\Phi}\sqrt{-g} \, v^{ml}_{a}\right) =0 \,. 
\end{equation}
\noindent
The Abelian field strength was defined in \eqref{Eq:Abelian}. Working in the axial gauge $V_z^a=0$, we get
\noindent
\begin{equation}
e^{-A_s+\Phi}\partial_{z}\left(e^{A_s-\Phi} \partial_z V_{\nu,a}\right)+\square V_{\nu,a}=0.
\label{Eq:VectorMesonsEq}
\end{equation}
\noindent
The equation of motion of the scalar sector takes the form
\noindent
\begin{equation}
\frac{e^{\Phi}}{\sqrt{-g}}\partial_{m}\left(e^{-\Phi}\sqrt{-g}\,g^{mn}\partial_{n}S\right)-\left(m^{2}_{X}+\frac{3}{2}\lambda v^2\right)S=0,
\end{equation}
\noindent
which may be written in the form
\noindent
\begin{equation}\label{Eq:ScalarMesonsEq}
e^{-3A_s+\Phi}\partial_z\left(e^{3A_s-\Phi}\partial_z S\right)+\square S-e^{2A_s}\left(m^{2}_{X}+\frac{3}{2}\lambda v^2\right)S=0,
\end{equation}
\noindent
The remaining equations of motion are
\begin{equation}\label{Eq:AxialScalarEqs}
\begin{split}
\frac{e^{\Phi}}{\sqrt{-g}} \, \partial_m\left(e^{-\Phi}\sqrt{-g} a^{ml}_a\right) + v^2(z) g^{2}_{5}\left(\partial^{l}\pi^a-A^{l,a} \right)&=0,\\
\partial_m \Big [ e^{-\Phi}\sqrt{-g} \, v^2(z) g_5^{2} \left(\partial^{m}\pi^a-A^{m,a} \right) \Big ] &=0.
\end{split}
\end{equation}
\noindent
We work in the axial gauge, $A_z^a=0$, and decompose the gauge field as $A^{\mu}_{b}=A^{\mu}_{b\perp}+\partial^{\mu} \varphi_b$, where $A^{l}_{b\perp}$ is the transverse part and $\partial^{\mu}\varphi_b$ the longitudinal part. The tranverse part leads to the equation of motion for the axial-vector sector
\noindent
\begin{equation}\label{Eq:AxialMesonsEq}
e^{A_s-\Phi}\partial_z\left(e^{A_s-\Phi}\partial_z A^{\mu,a}_{\perp}\right)+\square A^{\mu,a}_{\perp}-v^2(z)g_{5}^{2}e^{2A_s}A^{\mu,a}_{\perp}=0,
\end{equation}
\noindent
 For the pseudoscalar sector we find the coupled equations 
\noindent
\begin{eqnarray}
e^{A_s-\Phi}\partial_z\left(e^{A_s-\Phi}\partial_z \varphi^a\right)+v^2(z)g_{5}^{2}e^{2A_s}\left(\pi^a-\varphi^a\right)=0 \,, \label{Eq:PionMesonsEq1} \\
-\partial_z \square\varphi^a+v^2(z)g_{5}^{2}e^{2A_s}\partial_z\pi^a=0 \,. \label{Eq:PionMesonsEq2}
\end{eqnarray}
The dictionary for the decay constants is obtained from the holographic currents. The latter arise in the surface term of \eqref{Eq:QuadraticAction}, which may be written as
\noindent
\begin{equation}\label{Eq:CurrentsAction}
\delta S_2^{{\text Bdy}}=-\int dx^4\left(\langle J_s\rangle(\delta S)+\langle J^{\mu}_{V,a}\rangle(\delta V^{a}_{\mu})+\langle J^{\mu}_{A,a}\rangle(\delta A^{a}_{\mu})+\langle J_{\pi,a}\rangle(\delta \pi^{a})\right)_{z=\epsilon}
\end{equation}
\noindent
The VEV of 4d operators appearing in \eqref{Eq:CurrentsAction} are defined by
\noindent
\begin{equation}\label{Eq:Currents}
\begin{split}
\langle J_s\rangle&=P^z_s=-4e^{-\Phi}\sqrt{-g}\,\partial^z S, \quad
\langle J^{\mu}_{V,a}\rangle=P^{z\mu}_{V,a}=-\frac{e^{-\Phi}}{g_5^2}\sqrt{-g}\,v^{z\mu},\\
\langle J^{\mu}_{A,a}\rangle&=P^{z\mu}_{A,a}=-\frac{e^{-\Phi}}{g_5^2}\sqrt{-g}\, a^{z\mu},\quad 
\langle J_{\pi,a}\rangle=P^{z}_{\pi,a}=-e^{-\Phi}\sqrt{-g} \,  v^2(z)\left(\partial^z \pi^a-A^{z,a} \right) \, .
\end{split}
\end{equation}
\noindent
We identify $\langle J^{\mu}_{V,a}\rangle$ and $\langle J^{\mu}_{A,a}\rangle$
as the holographic vector and axial currents leading to the meson decay constants. The next step is to decompose the fields into their Kaluza-Klein modes as follows. 
\noindent
\begin{equation}\label{Eq:KKDecomposition}
\begin{split}
S(x,z)&=\sum_{n=0}^{\infty}s_n(z) \hat S_n(x),\quad
V^{\mu}_a(x,z)=g_5\sum_{n=0}^{\infty}v_{a,n}(z) \hat V^{\mu}_{a,n}(x),\\
A^{\mu,\perp}_a(x,z)&=g_5\sum_{n=0}^{\infty}a_{a,n}(z) \hat{A}^{\mu}_{a,n}(x),\quad 
\pi_{a}(x,z)=g_5\sum_{n=0}^{\infty}\pi_{a,n}(z) \hat \pi_{a,n}(x),\\
\varphi_{a}(x,z)&=g_5\sum_{n=0}^{\infty}\varphi_{a,n}(z) \hat \pi_{a,n}(x).
\end{split}
\end{equation}
\noindent
Plugging \eqref{Eq:KKDecomposition} into the vector and axial-vector currents \eqref{Eq:Currents} we obtain the expansions
\noindent
\begin{equation}
\begin{split}
\langle J^{\mu}_{V,a}\rangle&=\sum_{n}\left(-\frac{e^{A_s-\Phi}}{g_5}\partial_z v_{a,n}(z)\right)\hat V^{\mu}_{a,n}(x),\\
\langle J^{\mu}_{A,a}\rangle&=\sum_{n}\left(-\frac{e^{A_s-\Phi}}{g_5}\partial_z a_{a,n}(z)\right)\hat A^{\mu}_{a,n}(x)+\sum_{n}\left(-\frac{e^{A_s-\Phi}}{g_5}\partial_z \varphi_{a,n}(z)\right)\partial^{\mu}\hat \pi_{a,n}(x),
\end{split} \label{Eq:CurrentExp}
\end{equation}
\noindent
On the other hand, the meson decay constant are defined by the following relations 
\noindent
\begin{equation}
\begin{split}
\langle 0|J^{\mu}_{V,a}(x)|V^{b,m}(p,\lambda)\rangle
&=F_{v^{a,m}}e^{-ip\cdot x}\epsilon^{\mu}(p,\lambda)\delta^{ab},\\
\langle 0|J^{\mu}_{A,a}(x)|A^{b,m}(p,\lambda)\rangle
&=F_{a^{a,m}}e^{-ip\cdot x}\epsilon^{\mu}(p,\lambda)\delta^{ab},\\
\langle 0|J^{\mu}_{A,a}(x)|\pi^{b,m}(p)\rangle
&=f_{\pi^{a,m}}e^{-ip\cdot x} i  p^\mu \delta^{ab},
\end{split} \label{Eq:DecayConstDef}
\end{equation}
\noindent
The quantities $F_{v^{a,m}}$, $F_{a^{a,m}}$, $f_{\pi^{a,m}}$ are the decay constant of the vector, axial-vector and pseudoscalar mesons. Comparing \eqref{Eq:CurrentExp} and \eqref{Eq:DecayConstDef} we arrive at the holographic dictionary for meson decay constants
\noindent
\begin{equation}
\begin{split}
F_{v_{a,n}}&=-\frac{e^{A_s-\Phi}}{g_5}\partial_z v_{a,n}(z) \, , \quad 
F_{a_{a,n}}=-\frac{e^{A_s-\Phi}}{g_5}\partial_z a_{a,n}(z) \, \\
f_{\pi_{a,n}}&=-\frac{e^{A_s-\Phi}}{g_5}\partial_z \varphi_{a,n}(z) \,.
\end{split} \label{Eq:DecayConstDic}
\end{equation}

\section{The GKK model: A review}
\label{Sec:GKK}

In this section we summarize the model investigated in ~\cite{Gherghetta:2009ac}, known as the Gherghetta-Kapusta-Kelley (GKK) model (see also \cite{Kelley:2010mu}). This model was motivated by the original soft wall model \cite{Karch:2006pv}, which considers a quadratic dilaton from the UV to the IR. In turn, the GKK model proposes to reconstruct the dilaton profile by rewriting the tachyon differential equation as
\noindent
\begin{equation}\label{Eq:GKKDilaton}
\partial_z\Phi(z)=\frac{\partial^2_z v}{\partial_z v}
-e^{2A_{s}(z)}\left(m_{X}^2
-\frac{\kappa}{2}\,v^2\right)\,\frac{v}{\partial_z v}
+3\partial_z A_{s}(z).
\end{equation}
\noindent

\begin{figure}[ht]
\centering
\includegraphics[width=7cm]{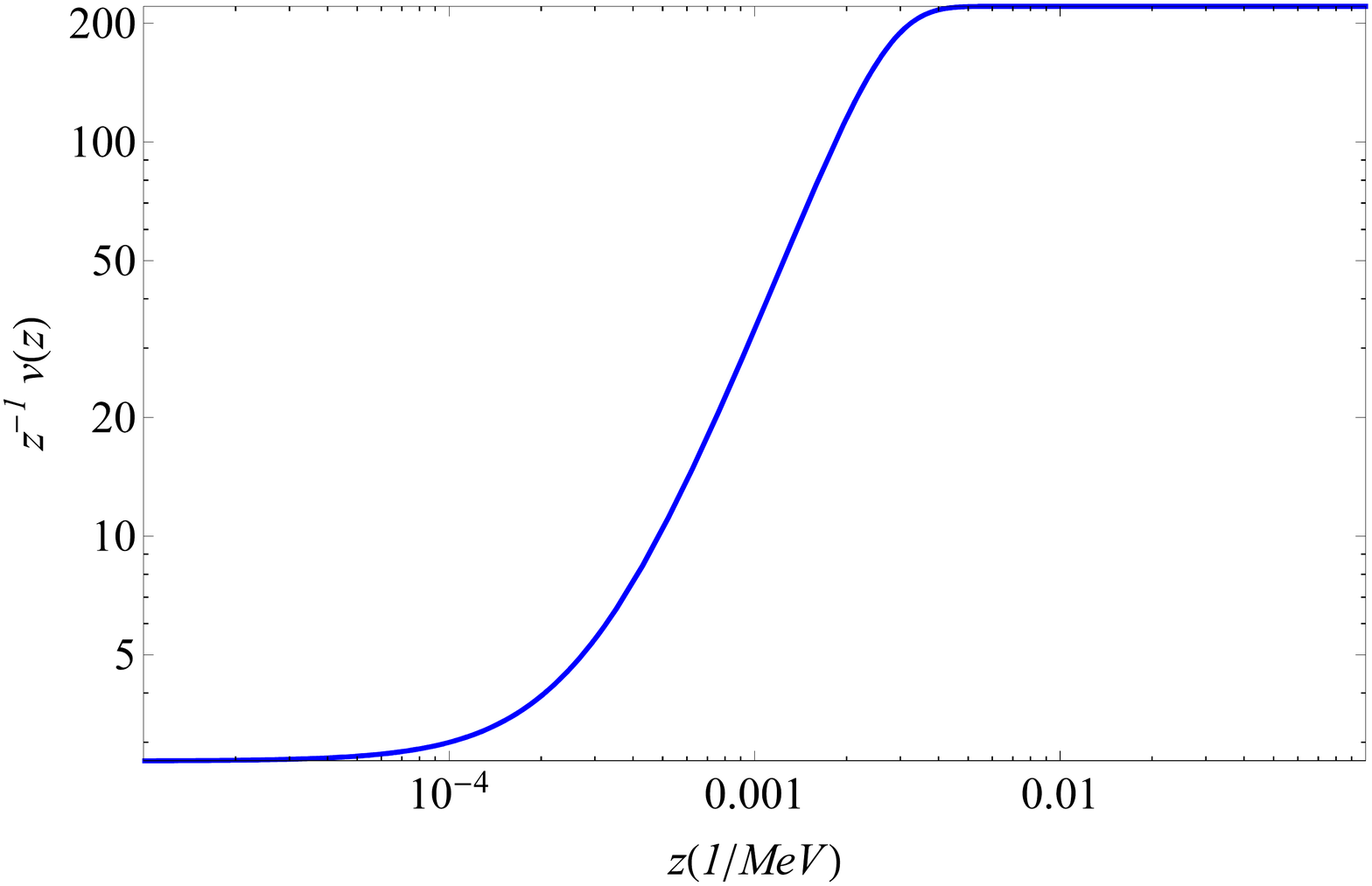}
\caption{Profile of the tachyon, to get this figure we 
fix the parameters: $\kappa=15,$ $m_q=9.75$Mev, 
$\Sigma=(204.5\text{MeV})^3$ and 
$\phi_{\infty}=0.1831\text{GeV}^2$.}
\label{Fig:TachGKK}
\end{figure}

\begin{figure}[ht]
\centering
\includegraphics[width=7cm]{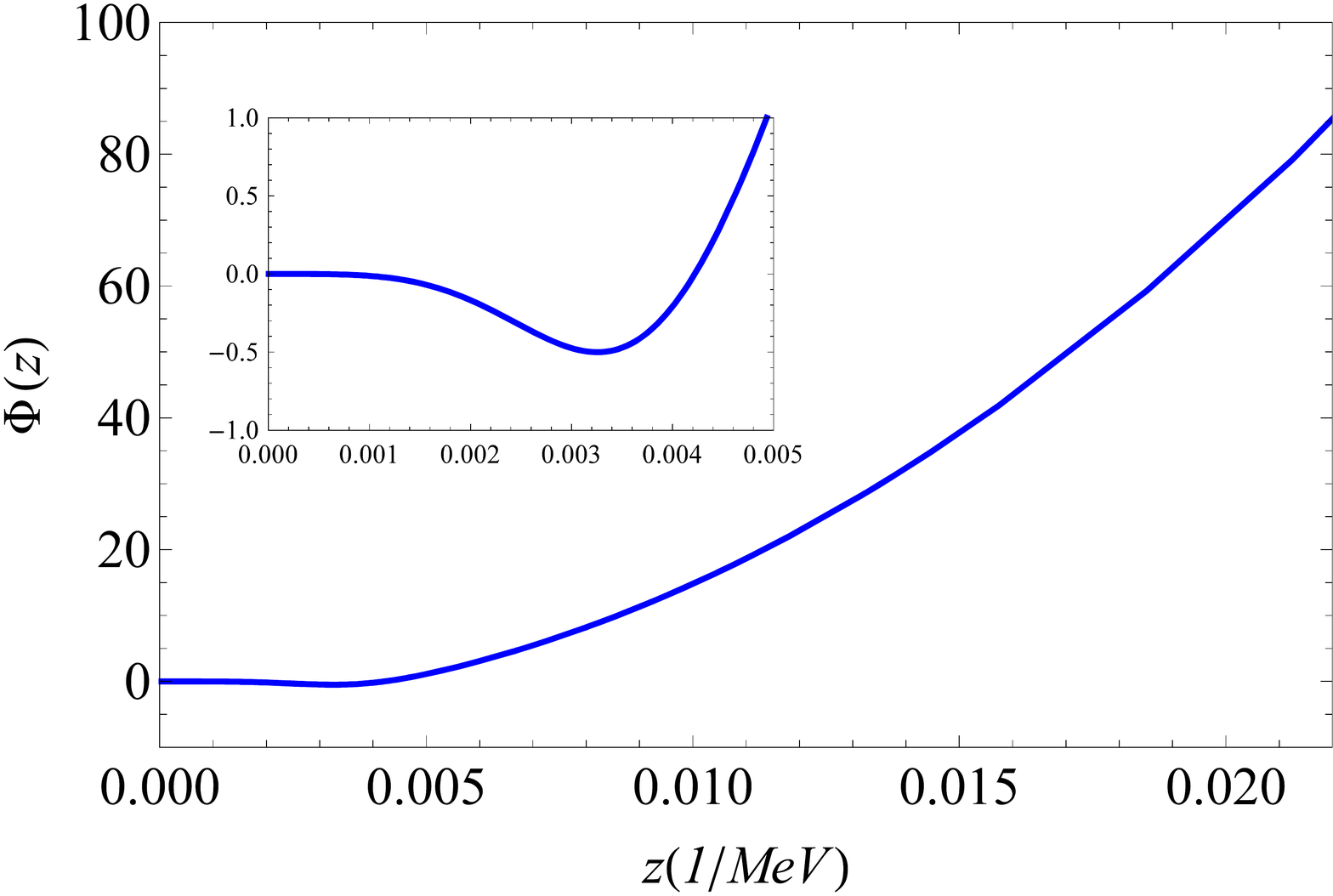}
\hfill
\includegraphics[width=7cm]{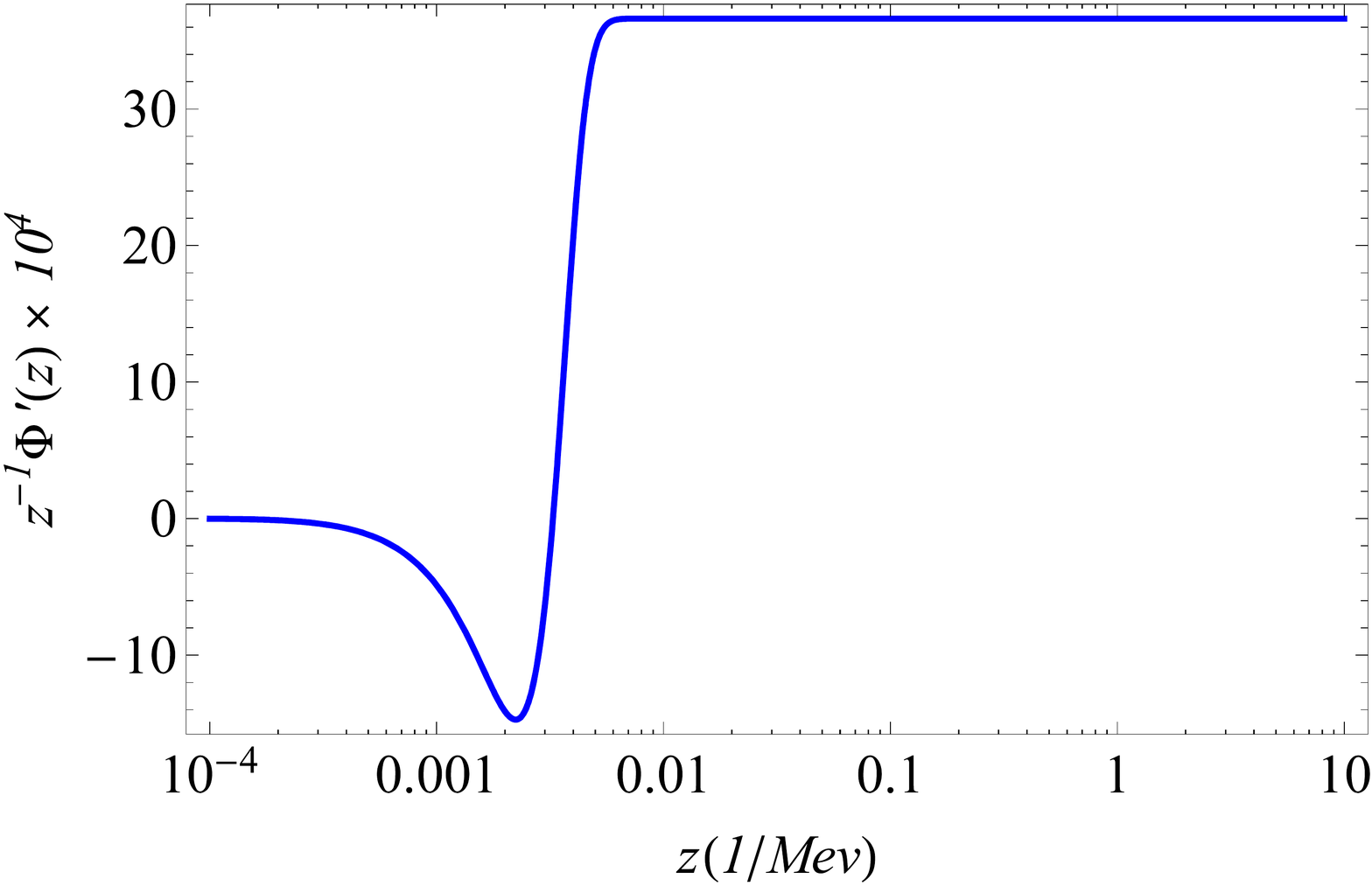}
\caption{Left: profile of the dilaton field. 
Right: profile of the derivative of the dilaton. Both results 
were obtained setting $\kappa=15,$ $m_q=9.75$Mev, 
$\Sigma=(404.5\text{MeV})^3$ and 
$\phi_{\infty}=0.1831\text{GeV}^2$.}
\label{Fig:DilGKK}
\end{figure}

We observe that this equation depends on the tachyon field. thus, to solve this equation we must know the tachyon. On the other hand, the asymptotic expansion of the tachyon field close to the boundary takes the form $v= c_1 z+c_3z^3$. In turn, in the IR the tachyon is linearly divergent, $v\sim z$. Additional constraints were imposed by the phenomenology, see ~\cite{Gherghetta:2009ac} for details. Thus, the following interpolation function recovers the asymptotic behavior in the UV and IR
\noindent
\begin{equation}\label{Eq:GKKTachyon}
v(z)=z\left(A+B\tanh{\left(Cz^2\right)}\right),
\end{equation}
where the parameters of the model are defined as 
\noindent
\begin{equation}\label{Eq:GKKParameters}
A=\frac{\sqrt{N_c}\,m_q}{2\pi},\quad 
B=2\sqrt{\frac{\phi_{\infty}}{\kappa}}-\frac{\sqrt{N_c}\,m_q}{2\pi},\quad 
C=\frac{2\pi\Sigma}{\sqrt{N_c}\,B}.
\end{equation}
\noindent
Close to the UV \eqref{Eq:GKKTachyon} takes the form
\noindent
\begin{equation}
v(z)=A\,z+BC\,z^3+\cdots,
\end{equation}
\noindent
where $A\propto m_q$ and $BC\propto \Sigma$. In the way the parameters were defined in Eq.~\eqref{Eq:GKKParameters}, we can see that in the chiral limit, i.e., $m_q\to 0$, $\Sigma\neq 0$, which means the spontaneous chiral symmetry breaking in nonzero. In the IR the tachyon reduces to 
\noindent
\begin{equation}
v(z)=(A+B)z=2\sqrt{\frac{\phi_{\infty}}{\kappa}}z.
\end{equation}
\noindent
A plot of the tachyon field is shown in Fig.~\ref{Fig:TachGKK}. In turn, the dilaton field and its derivative are displayed in the left and right panels of Fig.~\ref{Fig:DilGKK}, respectively. as we can see, we notice that the dilaton field becomes negative in a small region close to the boundary. 
However, regarding the analysis developed in ~\cite{Kiritsis:2006ua}, a negative dilaton violates the null energy condition in the gravitational side, where the dilaton must rise monotonically so that $\Phi'(z)>0$.\footnote{It is worth mentioning that the analysis done by Kiritsis and Nitti in ~\cite{Kiritsis:2006ua} states that the dilaton field should be monotonically increasing. However, as the soft wall model in not backreacted maybe the null energy condition can be ``relaxed'' in some kind at least locally.} Hence, maybe this pathological behavior will have consequences in the spectrum.

\subsection{Scalar sector}

\begin{figure}[ht]
\centering
\includegraphics[width=7cm]{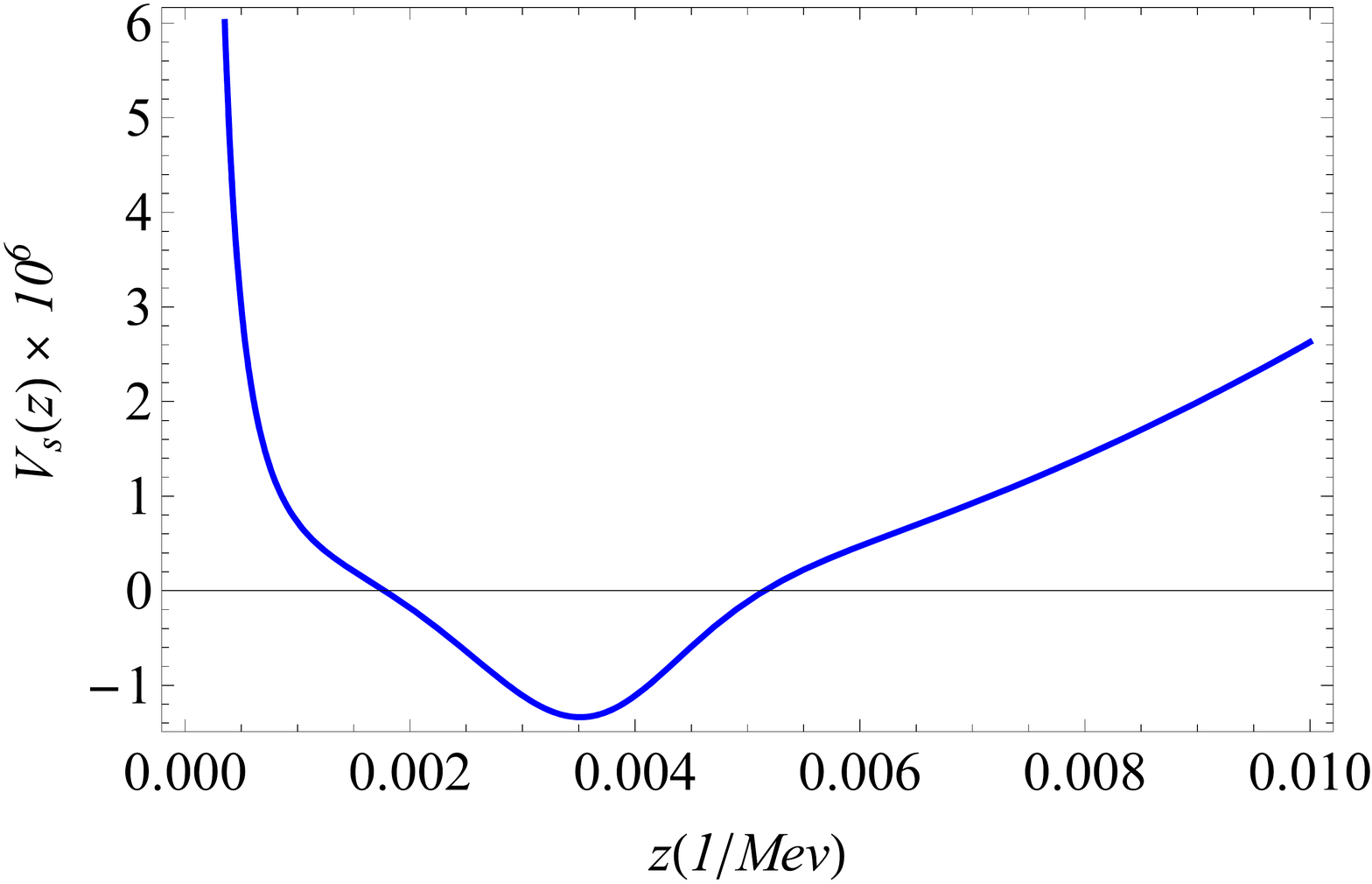}
\hfill
\includegraphics[width=7cm]{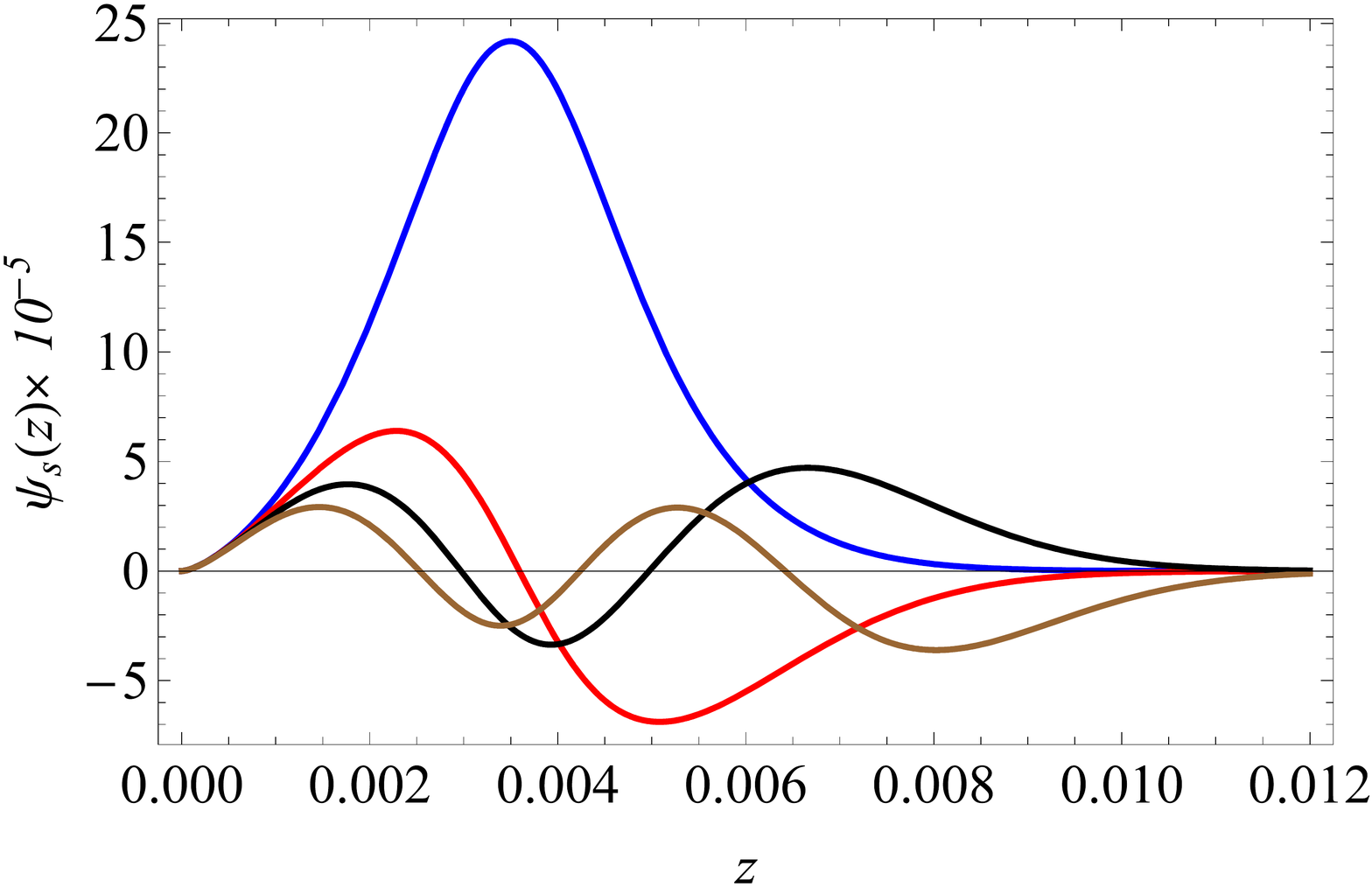}
\caption{Left: potential of the Schrödinger equation 
associated with the scalar mesons. 
Right: first fourth wave functions associated with the 
scalar mesons, see Table \ref{Taba:GKKScalar}. Both results 
were obtained setting $\kappa=15,$ $m_q=9.75$Mev, 
$\Sigma=(204.5\text{MeV})^3$ and 
$\phi_{\infty}=0.1831\text{GeV}^2$.}
\label{Fig:ScalarPotGKK}
\end{figure}

We have a special interest in the spectrum of the scalar sector of this model. Then, let us compute the spectrum. For doing that we must rewrite the perturbation equation in the Schrödinger form and solve it using a shooting method, for example. A plot of the potential is shown in the left panel of Fig.~\ref{Fig:ScalarPotGKK}. The results of the spectrum are displayed in the first column of Table~\ref{Taba:GKKScalar}, where we see that the first state has an imaginary mass, which means an instability, i.e., $m_s^2<0$. It is worth mentioning that this instability was reported previously in ~\cite{Sui:2009xe} (see also \cite{Li:2013oda}). To guarantee that this state is, in fact, a solution of the Schrödinger equation, we plot the wave functions of the corresponding first fourth eigenvalues in the right panel of Fig.~\ref{Fig:ScalarPotGKK}. All these results were obtained using the same parameters used in ~\cite{Gherghetta:2009ac}.

\begin{table}[ht]
\centering
\begin{tabular}{l |c|c|l}
\hline 
\hline
 $n$ & Model &
 GKK \cite{Gherghetta:2009ac}&
$f_0$ experimental \cite{Tanabashi:2018oca} \\
\hline 
 $1$ & 748\,i & 799   & $550^{+250}_{-150}$  \\
 $2$ & 799    & 1184  & $980\pm 10$  \\
 $3$ & 1184   & 1466  & $1350\pm 150$ \\
 $4$ & 1465   & 1699  & $1505\pm 6$  \\
 $5$ & 1698   & 1903  & $1724\pm 7$  \\
 $6$ & 1902   & 2087  & $1992\pm 16$ \\
 $7$ & 2087   & 2257  & $2103\pm 8$ \\
 $8$ & 2256   & 2414  & $2314\pm 25$  \\
\hline\hline
\end{tabular}
\caption{The masses (in MeV) obtained in the 
modified version of the soft wall model, including 
quartic interaction term, compared against the results 
of \cite{Gherghetta:2009ac} and experimental results of 
RPP \cite{Tanabashi:2018oca}. The value of the parameters used are: 
$\kappa=15,$ $m_q=9.75$Mev, 
$\Sigma=(404.5\text{MeV})^3$ and 
$\phi_{\infty}=0.1831\text{GeV}^2$.}
\label{Taba:GKKScalar}
\end{table}

\section{Numerical analysis: nonlinear soft wall model}
\label{Sec:Num}

\begin{figure}[ht]
\centering
\includegraphics[width=7cm]{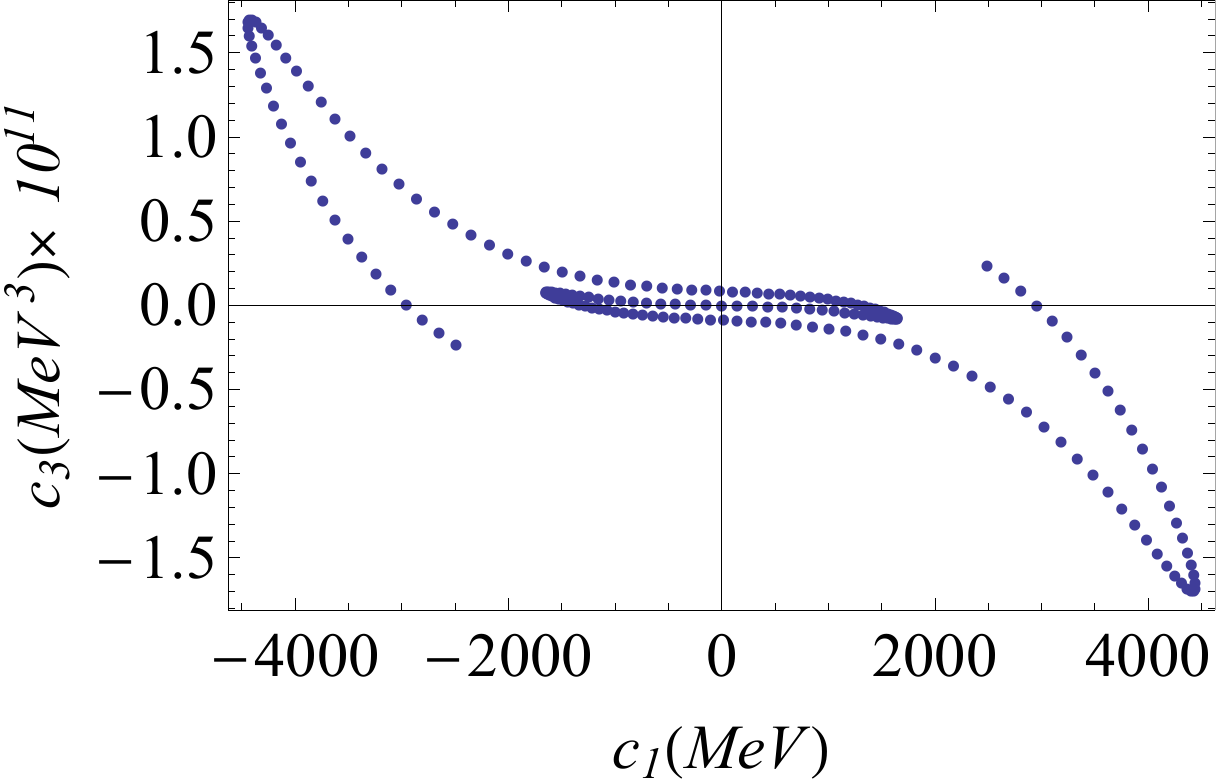}
\hfill
\includegraphics[width=7cm]{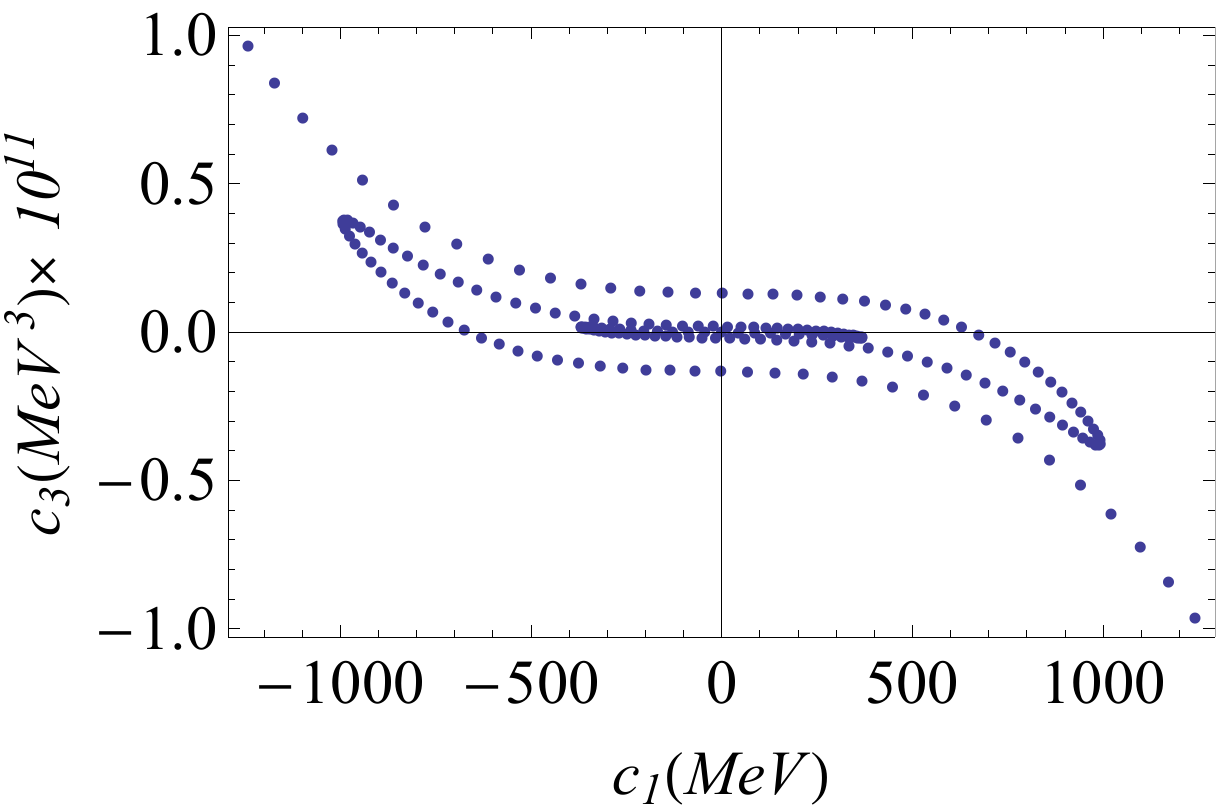}
\caption{
Numerical results of the nonlinear soft wall model. The corresponding parameters are: $\lambda=-1$ (left panel) and $\lambda=-20$ (right panel).}
\label{Fig:VEVC3C1}
\end{figure}

Here we write some details of our numerical results obtained investigating the tachyon field in the model for $\lambda<0$. We fix the parameter $\phi_{\infty}=(388\text{MeV})^2$. Then, we solve the differential equation \eqref{Eq:Tachyon} numerically, considering a family of two parametric solutions in the UV and a family of one parametric solution in the IR. We use as boundary condition the asymptotic solution in the IR \eqref{Eq:IRRegsol}. In the following analysis, we focus in the energy scale in the UV, such that the energy belongs to the interval $[10^{3}, 10^{6}]\text{MeV}$, which is equivalent to the interval of the holographic coordinate $z\in [10^{-6}, 10^{-3}]\text{MeV}^{-1}$, which lies close to the boundary. Thus, the problem was reduced to solve a one parameter family of solutions in the IR and two parameter family of solutions in the UV. What is expected is a non-trivial relationship between these parameters, which is obtained solving numerically Eq.~\eqref{Eq:Tachyon}. Our numerical results for $c_3$ as a function of $c_1$ are displayed in Fig.~\ref{Fig:VEVC3C1} for different values of $\lambda$. From this figure, we observe that besides the trivial solution there are solutions with $c_3\neq 0$ in the chiral limit, i.e., $c_1\to 0$, which corresponds to the limit of spontaneous chiral symmetry breaking.

\noindent

\begin{figure}[ht]
\centering
\includegraphics[width=7cm]{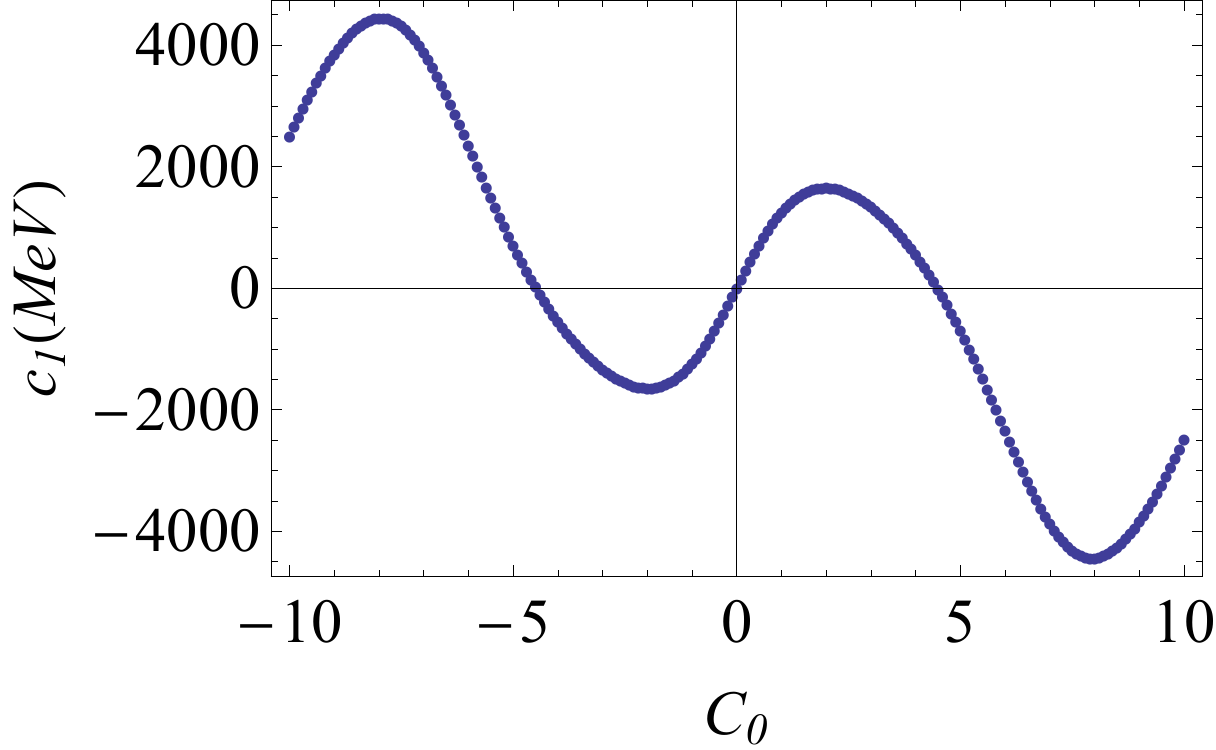}
\hfill
\includegraphics[width=7cm]{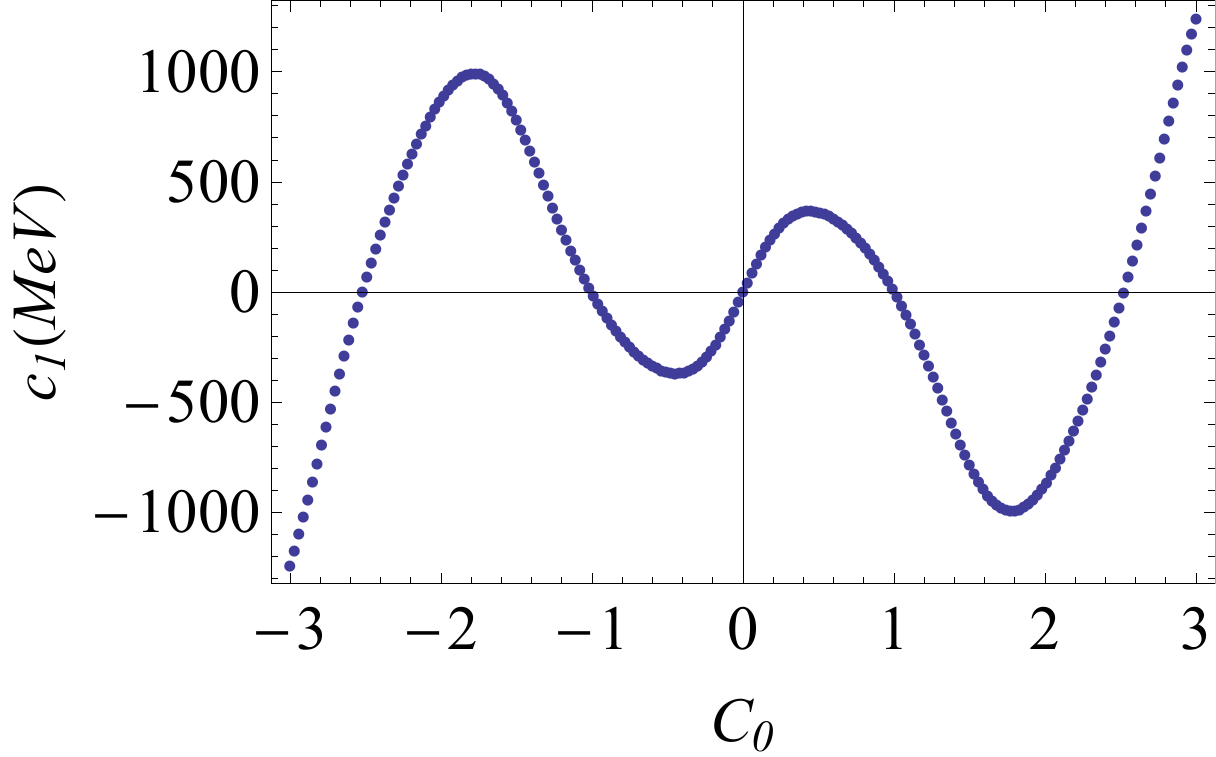}
\caption{
Numerical results of the nonlinear soft wall model. The corresponding parameters are: $\lambda=-1$ (left panel) and $\lambda=-20$ (right panel).}
\label{Fig:VEVC1C0}
\end{figure}

\begin{figure}[ht]
\centering
\includegraphics[width=7cm]{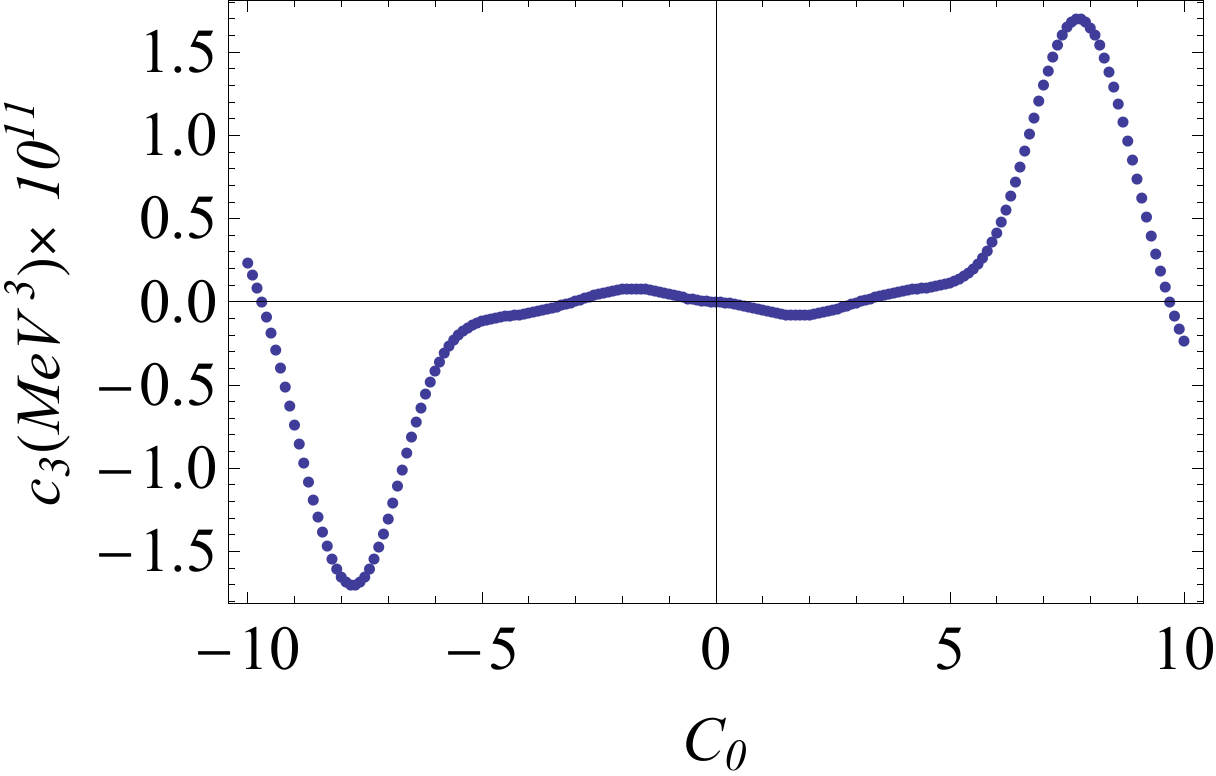}
\hfill
\includegraphics[width=7cm]{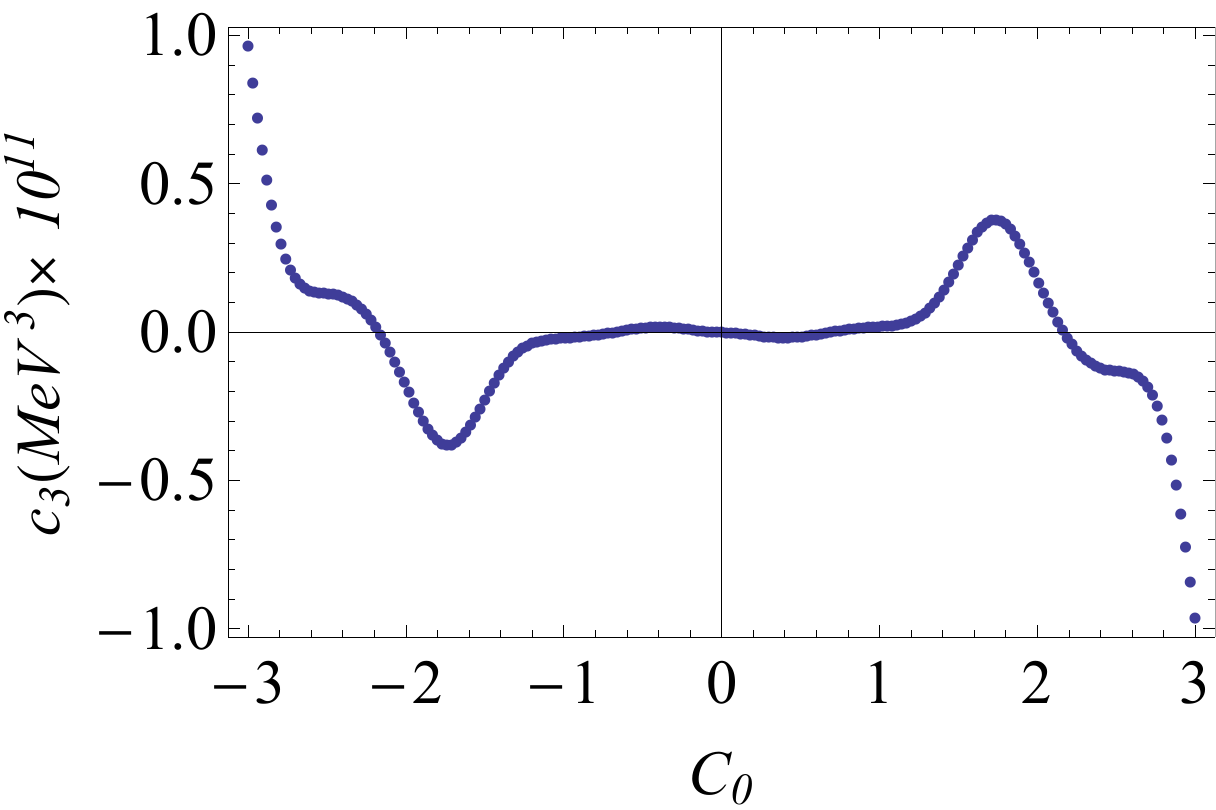}
\caption{
Numerical results of the nonlinear soft wall model. The corresponding parameters are: $\lambda=-1$ (left panel) and $\lambda=-20$ (right panel).}
\label{Fig:VEVC3C0}
\end{figure}

We point out that from the set of solutions showed in Fig.~\ref{Fig:VEVC3C1}, the physical solutions are those for what the tachyon field is a monotonic increasing function. The corresponding solutions for $c_1$ as a function of $C_0$ are displayed in Fig.~\ref{Fig:VEVC1C0}, whereas $c_3$ as a function of $C_0$ are displayed in Fig.~\ref{Fig:VEVC3C0}. However, when computing the spectrum of the vector mesons, for example, we obtain an inconsistence arising when we calculate the potential of the Schrödinger equation, which is given by 
\noindent
\begin{equation}
V_{V}=\frac{15}{4z^2}+\phi_{\infty}^2\,z^2+2\phi_{\infty}.
\end{equation}
\noindent
At $z_{min}=10^{-6}\text{MeV}^{-1}$, the first term of the last equation is leading, such that $V_{V}\sim (10^{10}\text{MeV})^2$. On the other hand, at $z_{max}=10^{-3}\text{MeV}^{-1}$, the first term is still the leading $V_{V}\sim (10^{3}\text{MeV})^2$, meaning that the potential is a monotonic decreasing function with no potential well. We also realised that the convergence of the asymptotic solution \eqref{Eq:IRRegsol} is not guaranteed.
\noindent

\noindent

In conclusion, the above analysis shows us that there is spontaneous chiral symmetry breaking in the nonlinear soft wall model for $\lambda<0$. However, it is arising in the UV. This conclusion may be justified because of the confinement scale introduced by the dilaton field is ``fake'', because the dilaton is introduced by hand and the backreaction on the metric neglected.


\begin{thebibliography}{99}

\bibitem{Karch:2006pv} 
  A.~Karch, E.~Katz, D.~T.~Son and M.~A.~Stephanov,
  Phys.\ Rev.\ D {\bf 74}, 015005 (2006).
  [hep-ph/0602229].
 
 \bibitem{Coleman:1980mx} 
  S.~R.~Coleman and E.~Witten,
  Phys.\ Rev.\ Lett.\  {\bf 45}, 100 (1980).

\bibitem{Kinar:1998vq} 
  Y.~Kinar, E.~Schreiber and J.~Sonnenschein,
  Nucl.\ Phys.\ B {\bf 566}, 103 (2000)
  [hep-th/9811192].
  
  \bibitem{Maldacena:1998im} 
  J.~M.~Maldacena,
  Phys.\ Rev.\ Lett.\  {\bf 80}, 4859 (1998)
  [hep-th/9803002].
  
 \bibitem{Erlich:2005qh} 
  J.~Erlich, E.~Katz, D.~T.~Son and M.~A.~Stephanov,
  Phys.\ Rev.\ Lett.\  {\bf 95}, 261602 (2005).
 [hep-ph/0501128].
 
 \bibitem{DaRold:2005mxj} 
  L.~Da Rold and A.~Pomarol,
  Nucl.\ Phys.\ B {\bf 721}, 79 (2005)
  [hep-ph/0501218].
  
\bibitem{Polchinski:2001tt} 
  J.~Polchinski and M.~J.~Strassler,
  Phys.\ Rev.\ Lett.\  {\bf 88}, 031601 (2002)
  [hep-th/0109174].
  
  \bibitem{McNeile:2012xh} 
  C.~McNeile, A.~Bazavov, C.~T.~H.~Davies, R.~J.~Dowdall, K.~Hornbostel, G.~P.~Lepage and H.~D.~Trottier,
  Phys.\ Rev.\ D {\bf 87}, no. 3, 034503 (2013)
  [arXiv:1211.6577 [hep-lat]].
  
  \bibitem{Chelabi:2015gpc} 
  K.~Chelabi, Z.~Fang, M.~Huang, D.~Li and Y.~L.~Wu,
  JHEP {\bf 1604}, 036 (2016)
  [arXiv:1512.06493 [hep-ph]]

  \bibitem{Iatrakis:2010jb} 
  I.~Iatrakis, E.~Kiritsis and A.~Paredes,
  JHEP {\bf 1011}, 123 (2010).
  [arXiv:1010.1364 [hep-ph]].
  
  \bibitem{Gherghetta:2009ac} 
  T.~Gherghetta, J.~I.~Kapusta and T.~M.~Kelley,
  Phys.\ Rev.\ D {\bf 79}, 076003 (2009).
 [arXiv:0902.1998 [hep-ph]].  
 
   \bibitem{BoschiFilho:2002vd} 
  H.~Boschi-Filho and N.~R.~F.~Braga,
  JHEP {\bf 0305}, 009 (2003)
  [hep-th/0212207].
  
\bibitem{Cherman:2008eh} 
  A.~Cherman, T.~D.~Cohen and E.~S.~Werbos,
  Phys.\ Rev.\ C {\bf 79}, 045203 (2009).
  [arXiv:0804.1096 [hep-ph]].

\bibitem{Csaki:2006ji} 
  C.~Csaki and M.~Reece,
  JHEP {\bf 0705}, 062 (2007)
  [hep-ph/0608266].
  
  \bibitem{Gursoy:2007er} 
  U.~Gursoy, E.~Kiritsis and F.~Nitti,
  JHEP {\bf 0802}, 019 (2008).
  [arXiv:0707.1349 [hep-th]].
  
  \bibitem{Ballon-Bayona:2017sxa} 
  A.~Ballon-Bayona, H.~Boschi-Filho, L.~A.~H.~Mamani, A.~S.~Miranda and V.~T.~Zanchin,
  Phys.\ Rev.\ D {\bf 97}, no. 4, 046001 (2018)
  [arXiv:1708.08968 [hep-th]].
  
  \bibitem{Ghoroku:2005vt} 
  K.~Ghoroku, N.~Maru, M.~Tachibana and M.~Yahiro,
  Phys.\ Lett.\ B {\bf 633}, 602 (2006)
  [hep-ph/0510334].
  
  
  \bibitem{Colangelo:2008us} 
  P.~Colangelo, F.~De Fazio, F.~Giannuzzi, F.~Jugeau and S.~Nicotri,
  Phys.\ Rev.\ D {\bf 78}, 055009 (2008)
  [arXiv:0807.1054 [hep-ph]].
  
  \bibitem{Brodsky:2014yha} 
  S.~J.~Brodsky, G.~F.~de Teramond, H.~G.~Dosch and J.~Erlich,
  Phys.\ Rept.\  {\bf 584}, 1 (2015)
  [arXiv:1407.8131 [hep-ph]].
 
 \bibitem{Bergman:2007pm} 
  O.~Bergman, S.~Seki and J.~Sonnenschein,
  JHEP {\bf 0712}, 037 (2007)
  [arXiv:0708.2839 [hep-th]].
 
 \bibitem{Casero:2007ae} 
  R.~Casero, E.~Kiritsis and A.~Paredes,
  Nucl.\ Phys.\ B {\bf 787}, 98 (2007).
  [hep-th/0702155 [HEP-TH]].
  
 \bibitem{Jarvinen:2011qe} 
  M.~Jarvinen and E.~Kiritsis,
  JHEP {\bf 1203}, 002 (2012)
  [arXiv:1112.1261 [hep-ph]].

\bibitem{Arean:2012mq} 
  D.~Arean, I.~Iatrakis, M.~Järvinen and E.~Kiritsis,
  Phys.\ Lett.\ B {\bf 720}, 219 (2013)
  [arXiv:1211.6125 [hep-ph]].
  
\bibitem{Jarvinen:2015ofa} 
  M.~Jarvinen,
  JHEP {\bf 1507}, 033 (2015)
  [arXiv:1501.07272 [hep-ph]].

\bibitem{Arean:2013tja} 
  D.~Areán, I.~Iatrakis, M.~Järvinen and E.~Kiritsis,
  JHEP {\bf 1311}, 068 (2013)
  [arXiv:1309.2286 [hep-ph]].
  
 
\bibitem{Karch:2005ms} 
  A.~Karch, A.~O'Bannon and K.~Skenderis,
  JHEP {\bf 0604}, 015 (2006)
  [hep-th/0512125].

\bibitem{Kelley:2010mu}
  T.~M.~Kelley, S.~P.~Bartz and J.~I.~Kapusta,
  Phys.\ Rev.\ D {\bf 83} (2011) 016002.
  [arXiv:1009.3009 [hep-ph]].
  
\bibitem{Herzog:2006ra} 
  C.~P.~Herzog,
  Phys.\ Rev.\ Lett.\  {\bf 98}, 091601 (2007).
  [hep-th/0608151].
  
\bibitem{Tanabashi:2018oca} 
  M.~Tanabashi {\it et al.} [Particle Data Group],
  Phys.\ Rev.\ D {\bf 98}, no. 3, 030001 (2018).

\bibitem{Kaplunovsky:2010eh} 
  V.~Kaplunovsky and J.~Sonnenschein,
  JHEP {\bf 1105}, 058 (2011)
  [arXiv:1003.2621 [hep-th]].
  
\bibitem{Ihl:2010zg} 
  M.~Ihl, M.~A.~C.~Torres, H.~Boschi-Filho and C.~A.~B.~Bayona,
  JHEP {\bf 1109}, 026 (2011)
  [arXiv:1010.0993 [hep-th]].
  
  
\bibitem{DaRold:2005vr} 
  L.~Da Rold and A.~Pomarol,
  JHEP {\bf 0601}, 157 (2006).
  [hep-ph/0510268].

\bibitem{Ballon-Bayona:2014oma}
  A.~Ballon-Bayona, G.~Krein and C.~Miller,
  Phys.\ Rev.\ D {\bf 91} (2015) 065024.
  [arXiv:1412.7505 [hep-ph]].



\bibitem{Ballon-Bayona:2017bwk} 
  A.~Ballon-Bayona, G.~Krein and C.~Miller,
  Phys.\ Rev.\ D {\bf 96}, no. 1, 014017 (2017).
  [arXiv:1702.08417 [hep-ph]].

\bibitem{Kwee:2007dd} 
  H.~J.~Kwee and R.~F.~Lebed,
  JHEP {\bf 0801}, 027 (2008).
  [arXiv:0708.4054 [hep-ph]].
  
\bibitem{Kwee:2007nq} 
  H.~J.~Kwee and R.~F.~Lebed,
  Phys.\ Rev.\ D {\bf 77}, 115007 (2008).
  [arXiv:0712.1811 [hep-ph]].

\bibitem{Holl:2004fr} 
  A.~Holl, A.~Krassnigg and C.~D.~Roberts,
  Phys.\ Rev.\ C {\bf 70}, 042203 (2004).
  [nucl-th/0406030].
  
\bibitem{Krassnigg:2006ps} 
  A.~Krassnigg, C.~D.~Roberts and S.~V.~Wright,
  Int.\ J.\ Mod.\ Phys.\ A {\bf 22}, 424 (2007).
 [nucl-th/0608039].
 
 \bibitem{Manohar:2000dt} 
  A.~V.~Manohar and M.~B.~Wise,
  Camb.\ Monogr.\ Part.\ Phys.\ Nucl.\ Phys.\ Cosmol.\  {\bf 10}, 1 (2000).
  
\bibitem{Maris:2005tt} 
  P.~Maris and P.~C.~Tandy,
  Nucl.\ Phys.\ Proc.\ Suppl.\  {\bf 161}, 136 (2006).
  [nucl-th/0511017].

\bibitem{Vega:2010ne} 
  A.~Vega and I.~Schmidt,
  Phys.\ Rev.\ D {\bf 82}, 115023 (2010).
  [arXiv:1005.3000 [hep-ph]].

\bibitem{Fang:2016nfj} 
  Z.~Fang, Y.~L.~Wu and L.~Zhang,
  Phys.\ Lett.\ B {\bf 762}, 86 (2016).
  [arXiv:1604.02571 [hep-ph]].
  
   
 
\bibitem{Sui:2009xe}
  Y.~Q.~Sui, Y.~L.~Wu, Z.~F.~Xie and Y.~B.~Yang,
  Phys.\ Rev.\ D {\bf 81} (2010) 014024.
  [arXiv:0909.3887 [hep-ph]].

\bibitem{Iatrakis:2010zf} 
  I.~Iatrakis, E.~Kiritsis and A.~Paredes,
  Phys.\ Rev.\ D {\bf 81}, 115004 (2010)
  [arXiv:1003.2377 [hep-ph]].


\bibitem{Cui:2013xva} 
  L.~X.~Cui, Z.~Fang and Y.~L.~Wu,
  Eur.\ Phys.\ J.\ C {\bf 76}, no. 1, 22 (2016)
  [arXiv:1310.6487 [hep-ph]].
  

\bibitem{Karch:2010eg} 
  A.~Karch, E.~Katz, D.~T.~Son and M.~A.~Stephanov,
  JHEP {\bf 1104}, 066 (2011)
  [arXiv:1012.4813 [hep-ph]].

\bibitem{Kiritsis:2006ua} 
  E.~Kiritsis and F.~Nitti,
  Nucl.\ Phys.\ B {\bf 772}, 67 (2007).
  [hep-th/0611344].   

\bibitem{Abidin:2009aj} 
  Z.~Abidin and C.~E.~Carlson,
  Phys.\ Rev.\ D {\bf 80}, 115010 (2009).
  [arXiv:0908.2452 [hep-ph]].
  
\bibitem{Li:2013oda} 
  D.~Li and M.~Huang,
  JHEP {\bf 1311}, 088 (2013).
  [arXiv:1303.6929 [hep-ph]].



  
 

\end{thebibliography}
\end{document}